\documentclass[aps,prb,amsmath,amssymb,nobibnotes,epsfig]{revtex4-1}

\usepackage{bm}
\usepackage{latexsym}
\usepackage{amsthm}
\usepackage{graphicx}

\newtheoremstyle{jmp}
{\topsep}{\topsep}
{\normalfont\itshape}
{\parindent}{\normalfont\bfseries}
{:}{.5em}
{}

\theoremstyle{jmp}

\newtheorem{prop}{Proposition}

\newcommand{\vect}[1]{\bm{#1}}

\makeatletter
\renewcommand\frontmatter@title@above{}
\renewcommand\frontmatter@title@format{\large\bfseries\raggedright}
\renewcommand\frontmatter@title@below{\addvspace{6\p@}}
\makeatother

\makeatletter
\renewcommand\frontmatter@authorformat
{\preprintsty@sw{\vskip0.5pc\relax}{}
\@tempskipa\@flushglue
\@flushglue\z@ plus50\p@\relax
\raggedright\advance\leftskip20mm\relax
\@flushglue\@tempskipa
\parskip\z@skip}
\makeatother

\makeatletter
\renewcommand\frontmatter@affiliationfont
{\small\slshape\selectfont\baselineskip10.5\p@\relax
\@tempskipa\@flushglue
\@flushglue\z@ plus50\p@\relax
\raggedright\advance\leftskip20mm\relax
\@flushglue\@tempskipa}
\makeatother

\makeatletter
\renewcommand{\section}{\@startsection{section}{1}{\z@}
{0.8cm \@plus1ex minus .2ex}{0.5cm}{\normalfont\small\bfseries}}
\renewcommand{\subsection}{\@startsection{subsection}{2}{-12pt}
{0.8cm \@plus1ex minus .2ex}{0.5cm}{\normalfont\small\bfseries}}
\renewcommand{\subsubsection}{\@startsection{subsubsection}{3}{-12pt}
{0.8cm \@plus1ex minus .2ex}{0.5cm}{\normalfont\small\itshape}}
\makeatother

\makeatletter
\renewcommand\bibsection
{\section*{}
\@nobreaktrue}
\makeatother

\begin{document}

\title{Trkalian fields: ray transforms and mini-twistors}

\author{K. Saygili}
\email{kamuran.saygili@istanbul.edu.tr}
\affiliation{Department of Mathematics, Istanbul University, 
             Beyazit Campus, 34134 Vezneciler Istanbul, Turkey}

\begin{abstract}
\noindent We study X-ray and Divergent beam transforms of Trkalian 
fields and their relation with Radon transform. We make use of four 
basic mathematical methods of tomography due to Grangeat, Smith, Tuy 
and Gelfand-Goncharov for an integral geometric view on them. We also 
make use of direct approaches which provide a faster but restricted 
view of the geometry of these transforms. These reduce to well known 
geometric integral transforms on a sphere of the Radon or the spherical 
Curl transform in Moses eigenbasis, which are members of an analytic 
family of integral operators. We also discuss their inversion. The X-ray 
(also Divergent beam) transform of a Trkalian field is Trkalian. Also the 
Trkalian subclass of X-ray transforms yields Trkalian fields in the physical 
space. The Riesz potential of a Trkalian field is proportional to the field. 
Hence, the spherical mean of the X-ray (also Divergent beam) transform of 
a Trkalian field over all lines passing through a point yields the field 
at this point. The pivotal point is the simplification of an intricate 
quantity: Hilbert transform of the derivative of Radon transform for 
a Trkalian field in the Moses basis. We also define the X-ray transform 
of the Riesz potential (of order $2$) and Biot-Savart integrals. Then, 
we discuss a mini-twistor respresentation, presenting a mini-twistor 
solution for the Trkalian fields equation. This is based on a time-harmonic 
reduction of wave equation to Helmholtz equation. A Trkalian field is given 
in terms of a null vector in $\mathbb{C}^{3}$ with an arbitrary function and 
an exponential factor resulting from this reduction. 
\end{abstract}

\maketitle

\setlength{\parindent}{4mm}

\section{INTRODUCTION}
\label{Intro}

  This is the second manuscript of a series\cite{KS} aimed at studying 
the mathematical structure of Trkalian\cite{VT, AL} class of Beltrami 
fields (eigenvectors of Curl operator with constant eigenvalue) in integral 
geometric and twistor terms. Integral geometry and Twistor theory provide 
new mathematical methods for studying the geometry of Trkalian fields. 
These lead to a deeper understanding of their physical 
aspects.\cite{KS, Ml1, Ml3, KS1}

  The Trkalian fields arise in different areas ranging from fluid dynamics 
and plasma physics to field theories. The field theoretic examples of 
Trkalian vectors are the force-free magnetic field and the Euclidean 
topologically massive Abelian gauge field.
  
  In plasma physics, a Trkalian field simply corresponds to a force-free 
equilibrium state of a plasma. The mathematical tomography is based on 
applications of integral geometric methods in tomography. It makes use 
of both pure and applied techniques. It provides the mathematical 
basis for modern tomography in the realistic sense. The tomographical 
study of an equilibrium state of a plasma is an active field of research 
in plasma tomography. 

  The topologically massive gauge theories\cite{DJTS1, DJTS2, DJTS3} are 
qualitatively different from Yang-Mills type gauge theories besides their 
mathematical elegance and consistency.\cite{ANS} These are introduced as 
an alternative to the mechanism of spontaneous symmetry breaking for  
generation of mass. The study of their physical and mathematical aspects 
is an active and exciting field of reseach today providing new insight 
into the relation of gauge theories and gravity in low dimensions. In this 
context,  Trkalian type solutions on $3$-sphere $\mathbb{S}^{3}$, anti-de 
Sitter space $\mathbb{H}^{3}$ and other spaces in connection with contact 
geometry are discussed in Refs. \onlinecite{ANS, KS2, KS3, KS1}.

  The Trkalian (Beltrami) fields equation also arises in connection 
with harmonic morphisms.\cite{BW} See Refs. \onlinecite{PW1, P1} 
and the refence therein for the solution on $\mathbb{S}^{3}$ 
(consisting of right/left-invariant $1$-forms) and its uniqueness 
(upto permutation). 

  A purpose of this manuscript is to develop physical insight for Trkalian 
fields in integral geometric and twistor terms. Intuitively speaking, from 
a higher point of view, we expect these to be related to the representations 
of the group underlying Trkalian fields. Basically, we are trying to 
investigate the aspects of functions invariant under the Curl operator 
generating rotations. These integral transforms naturally arise in 
geometric analysis (in Fourier sense) of physical systems.

  As a physical example of integral geometry in tomography, we 
shall frequently benefit the example of Lundquist\cite{SL} field which 
is used to model solar magnetic clouds, (see Refs. \onlinecite{Ml1}, 
\onlinecite{Ml3} and the references therein). We shall present 
a mathematical \textit{R\"{o}ntgen} of these clouds.

  Mathematically, we aim to expose the interrelations of the most basic 
transforms in integral geometry, for Trkalian fields. We are motivated 
by the geometric picture\cite{Ml1, Ml3} that is provided by the 
Radon\cite{JR} transform which is at a central place in integral geometry. 
This endows us with an intuition for Trkalian fields. These transforms 
which arise naturally in the study of integral geometric and tomographical 
aspects of Trkalian fields are intimately connected. The mathematical 
methods of tomography provide delicate ways for exposing their 
interrelations. This leads to an intuitive, integral understanding, 
besides its potential physical applications.         

  We also aim to discuss a mini-twistor representation for Trkalian fields 
presenting a mini-twistor solution for them. This leads to new mathematical 
challenges beside providing new physical insight into the Trkalian fields.

  In the first part of this manuscript we shall study X-ray (John\cite{John}) 
and Divergent beam transforms of Trkalian fields. Then we shall discuss 
the mini-twistor representation for Trkalian fields in the second part.

  The Radon transformation provides a geometric formulation 
of Trkalian fields.\cite{KS, KS1, Ml1, Ml3} Especially, Moses 
eigenfun\-ctions\cite{M1} of the Curl operator, which form a 
complete orthonormal basis, leads to a helicity decomposition of 
the Radon transform of Trkalian fields. The spherical Curl transform 
is a Radon probe transformation\cite{KS, Ml1, Ml3} in this basis. 
The Radon transform of a Trkalian field is tangent to a sphere in 
the transform space. It satisfies a corresponding eigenvalue equation 
on this sphere.

  Furthermore, we can associate an eigenvalue equation for 
Biot-Savart\cite{CDG, P} integral operator with Trkalian fields 
if they vanish at infinity. Meanwhile, the Riesz potential and 
Biot-Savart integrals naturally arise in Radon transform.\cite{KS} 
The Radon-Biot-Savart ($RBS$) integral is defined as the Radon 
transform of the Biot-Savart operator.\cite{KS} We can study Trkalian 
fields using the RBS operator in transform space.

  First we shall study X-ray and Divergent beam transforms 
of Trkalian fields. The field is to be taken from Schwartz 
space  $\mathbb{S}[\mathbb{R}^{3}]$ of rapidly decreasing 
functions on $\mathbb{R}^{3}$.\cite{GGG} 

  The X-ray and Divergent beam transforms are closely connected with 
the Radon transform. The mathematical methods of tomography (respectively 
Smith' s and Tuy' s methods discussed below) show that these transforms 
are basically in the form of a Minkowski-Funk\cite{M, PF} and a closely 
related (well known, but no specific name in the literature known to 
the author) integral transform of certain intricate quantities (Hilbert 
transform of the derivative of Radon transform). The Moses eigenbasis 
is especially efficient in exhibiting this connection for Trkalian fields. 
Then, these naturally reduce to well known geometric integral transforms 
on a sphere (mentioned above) of the Radon\cite{FBG} or the spherical 
Curl transforms. More precisely, the X-ray transform reduces to 
Minkowski-Funk transform of the Radon transform on this sphere. 
Meanwhile, the Divergent beam transform reduces to another closely 
related (an extension of Minkowski-Funk) integral transform of the 
spherical Curl transform. 

  Moreover, these geometric integral transforms are members of an analytic 
family of integral operators. See Refs. \onlinecite{Rubin3, Rubin4} and the 
references therein. This leads us to inverse transforms which also belong to
this family for the X-ray and Divergent beam transforms of Trkalian fields.

  The connections of X-ray and Divergent beam transforms of Trkalian 
fields to their Radon transforms can be directly obtained without further 
ado, (a bit difficult to see but) simply substituting the relevant expressions 
without any sense. We shall first present direct proofs without logical 
motivation. Because the direct approach is more appropriate for exposing 
the relations of X-ray and Divergent beam transforms of Trkalian fields 
to the Radon and other transforms. However the underlying geometric structure 
can be investigated by using the mathematical methods of tomography. Because 
these connections and their inversion deserve a separate discussion 
in its own right, we have interchanged the logical order for a clean 
presentation below. Yet, the mathematical methods of tomography provides 
a logically unified view for our motivations and also for the development 
of geometric intuition along the manuscript.   

  The X-ray (also the Divergent beam) transform and its inversion intertwine 
the Curl operator ($\vect{\nabla}\bm{\times}$) and also the Divergence 
($\vect{\nabla}\cdot$), Gradient ($\vect{\nabla}$) and Laplacian 
($\nabla^{2}$) operators. Thus the X-ray (Divergent beam) transform 
of a Trkalian field is Trkalian. Also, the Trkalian subclass of X-ray 
transforms $\vect{\mathcal{X}}\vect{F}$ (satisfying\cite{GGG} John' s 
differential equation)  yields Trkalian fields in the physical space. 
We shall also write John' s differential equation for Trkalian fields 
in an equivalent form. Thus, we can study Trkalian fields either in 
physical space or in the transform space.

  Another crucial quantity in integral geometry is the Riesz potential.
The Riesz potential, of order $\alpha$ where $0<\alpha<3$, of a  Trkalian 
field is proportional to the field. Hence, the spherical mean of X-ray 
(or Divergent beam) transform of a Trkalian field over all lines passing 
through a point yields the field itself at this point. This endows us with 
a new simple inversion formula for the X-ray (or Divergent beam) transform 
of Trkalian fields. This result is also logically implied by 
Gelfand-Goncharov' s mathematical approach to tomography below.

  Then we shall return back to the mathematical methods of tomography. 
The first purpose of this section is to provide a unified geometric 
view and motivation for the interrelations of integral transforms 
arising in our discussion, as mentioned above. The second purpose 
is to present a discussion of these mathematical methods with a view 
towards tomographical studies of Trkalian field models in nature. For 
this purpose, we shall study these mathematical methods using Trkalian 
fields. Especially for the sake of the second purpose and also for a 
clean presentation, this discussion will be postponed until the direct 
(but unmotivated) discussion of the geometry of X-ray and Divergent beam 
transforms finish. 

  We shall make use of four basic mathematical approaches of 
tomography due to Grangeat,\cite{G} Smith,\cite{S1} Tuy\cite{HKT} 
and Gelfand-Goncharov\cite{GG} for studying the X-ray and Divergent 
beam transforms of Trkalian fields and expressing their relations
with the other transforms. These relations are outflow of a formula 
essentially obtained in Ref. \onlinecite{HSSW}.\cite{NW} These 
methods basically make use of the Radon inversion (for tomographical 
reconstruction). They lead us to new inversion formulas for the X-ray 
and Divergent beam transforms of Trkalian fields with a view towards 
tomographical applications. They also provide the geometric motivation 
underlying the interrelations of the transforms mentioned. We shall 
adopt a mathematical approach rather than a tomographical implementation.  

  The Grangeat\cite{G} approach leads to another simple, direct inversion 
formula for the Divergent beam transform of Trkalian fields. 

  The Smith\cite{S1} method reveals that the X-ray transform is in the 
form of a Minkowski-Funk transform of an intricate quantity related to 
the Radon transform. This simply reduces to Minkowski-Funk transform 
of the Radon transform on a sphere, yielding the result mentioned 
above. In this approach, the inversion formula can be expressed in 
terms of the Radon transform of the field. 

  The Tuy\cite{HKT} method enables us to investigate the Divergent beam 
transform in detail. It reveals that the Divergent beam transform is in 
the form of another closely related integral transform of a quantity 
related to the Radon transform. This reduces to the above mentioned 
integral transform of the spherical Curl transform. In this case, 
the inversion formula can be expressed in terms of the spherical 
Curl transform of the field. 

  We calculate the Divergent beam transform of the Lundquist field using 
the Tuy method. This yields a mathematical \textit{R\"{o}ntgen} of 
solar magnetic clouds.

  Meanwhile, Gelfand-Goncharov' s\cite{GG} approach leads to a direct 
inversion through the spherical mean that is mentioned above. This naturally 
makes use of the inverse transform that belongs to the above family of 
integrals operators. 

  The basic simplification in these approaches are due to the same 
intricate quantity: Hilbert transform of the derivative of Radon 
transform in the Moses basis.

  The direct inversion formulas arising in Grangeat' s and 
Gelfand-Gonchorav' s approaches mathematically seem more 
feasible than the inversions in Smith' s and Tuy' s methods.

  These approaches provide different inversion formulas which may 
serve useful for designing reconstruction methods in tomographical 
studies of Trkalian field models in nature, depending on real physical 
situation. We shall not discuss tomographical implementations 
of these inversion formulas. 

  Furthermore, the Smith and Tuy methods mathematically enable us to 
define the X-ray and Divergent beam transforms of the Riesz potential 
(of order $2$) and Biot-Savart integrals. The X-ray transform of the 
Biot-Savart integral of a Trkalian field reduces to the X-ray transform 
of the field.

  In the second part of this manuscript we shall discuss Trkalian 
fields using (mini-)twistors.

  Twistor theory has been originally founded and developed by 
Penrose.\cite{RP4, RP3} It has led to a deeper understanding of 
nature. In simplest terms, this is based on writing contour integral 
solution for wave equation in ($3+1$) dimensional Minkowski space, 
using a holomorphic function. Similar formulas date back to Whittaker 
and Bateman.\cite{W, Bateman}

  The X-ray transform is a real analogue of the Penrose 
transform\cite{BEGM, BEGM1a} and a predecessor\cite{DW} 
of Twistor theory. See for example Refs. 
\onlinecite{BEGM, BEGM1a, DW, D1, S, BE, LJM3a} for 
relation of the X-ray transform and Twistor theory.

  Mini-twistor space as an intrinsic structure has been introduced 
by Hitchin.\cite{H1} The (mini-)twistor space of $\mathbb{R}^{3}$ 
is the space $\mathbb{TS}^{2}$ of oriented lines in $\mathbb{R}^{3}$. 
This can be identified with $\mathbb{TCP}^{1}$, the holomorphic tangent 
bundle of Riemann sphere $\mathbb{CP}^{1}$. This is also given by the 
quotient of twistor space ($\mathbb{CP}^{3}\backslash \mathbb{CP}^{1}$) 
of the  Minkowski space by the action of time translation.\cite{H1} 
The X-ray transform and the mini-twistors are both defined on the 
space $\mathbb{TS}^{2}\sim\mathbb{TCP}^{1}$ of oriented lines in 
$\mathbb{R}^{3}$.

  We shall discuss a mini-twistor representation, presenting a 
mini-twistor solution for the Trkalian fields equation. We shall make 
use of the solution\cite{WS} of (vector) Helmholtz equation which is 
based on a time-harmonic reduction of the wave equation. A Trkalian 
field is given in terms of a null vector in $\mathbb{C}^{3}$ with an 
arbitrary function and an exponential factor that results from the 
reduction.

  The exponential factor contains the spatial part of an integrating 
factor for the time-harmonicity condition. The solution is of the same 
form containing the spatial part of any choosen integrating factor. We 
shall also use the general solution\cite{WS} of this condition for writing 
the solution.

  This solution can also be derived as a time-harmonic reduction 
of the twistor solution\cite{RP3} for electromagnetic fields in 
$(3+1)$ dimensions.

  We are led to a time-harmonic extension of Trkalian fields implicitly 
keeping this condition. This can be interpreted as a time-harmonic 
electromagnetic field.

  We shall also present examples of Debye potentials for 
Chandrasekhar-Kendall\cite{CK} (CK) type solutions using 
the twistor solution of (scalar) Helmholtz equation. 

  The relation of (mini-)twistors and ray transforms for 
Trkalian fields is beyond the limitations of this manuscript.

\section{X-RAY AND DIVERGENT BEAM TRANSFORMS}
\label{Xraytrf}

\subsection{Trkalian fields: Radon transform} 

   Trkalian fields are eigenvectors of the curl operator

\begin{eqnarray} \label{Trkalianselfdual}
\vect{\nabla}\bm{\times}\vect{F}(\vect{x})-\nu\vect{F}(\vect{x})=0,
\end{eqnarray}

\noindent with constant eigenvalue $\nu$. The Radon transform 

\begin{eqnarray} \label{Radonvectortransform}
\vect{F}^{\mathcal{R}}(p, \vect{\kappa})
=\vect{\mathcal{R}}[\vect{F}(\vect{x})](p, \vect{\kappa})
=\int\vect{F}(\vect{x})\delta(p-\vect{\kappa}\cdot\vect{x})d^{3}x,
\end{eqnarray}

\noindent of a field $\vect{F}(\vect{x})$ (that belongs to Schwartz class) 
on $\mathbb{R}^{3}$  is defined as the integral of the field over a hyperplane 
at (orthogonal) distance $p$ to the origin, with unit normal vector 
$\vect{\kappa}$. The Radon transform $\vect{F}^{\mathcal{R}}(p, \vect{\kappa})$ 
of a Trkalian field satisfies

\begin{eqnarray} \label{transelfdualfield}
\vect{\Gamma} \bm{\times}  
\vect{F}^{\mathcal{R}}
(p, \vect{\kappa})
-\nu\vect{F}^{\mathcal{R}}
(p, \vect{\kappa})=0,
\end{eqnarray}

\noindent where $\vect{\Gamma}=\vect{\kappa}\partial/\partial p$.\cite{KS} 
We also have: $\vect{\kappa}\cdot\vect{F}^{\mathcal{R}}(p, \vect{\kappa})=0$ 
which leads to $\vect{\Gamma}\cdot\vect{F}^{\mathcal{R}}(p, \vect{\kappa})=0$. 
Because the Radon transform intertwines the operator $\vect{\nabla}$ with 
$\vect{\Gamma}$. We can write this equation as

\begin{equation} \label{transelfdualfield2}
\frac{\partial}{\partial p}
\vect{F}^{\mathcal{R}} 
(p, \vect{\kappa}) 
+\nu\kappa\bm{\times} 
\vect{F}^{\mathcal{R}}
(p, \vect{\kappa}) =0. 
\end{equation}

  The Curl transform\cite{M1} is based on decomposing a vector field 
into helical eigenfunctions $\vect{\chi}_{\lambda}(\vect{x}|\vect{k})
=(2\pi)^{-3/2}e^{i\vect{k}\cdot\vect{x}}\vect{Q}_{\lambda}(\vect{k})$ of the 
curl operator, which form an orthogonal and complete set, in the fashion 
of a Fourier transform refining the Helmholtz decomposition. This is a 
helicity ($\lambda=-1, \, 0, \, 1$) decomposition in the basis 
$\{ \vect{Q}_{\lambda}(\vect{k}) \}$. 

  A Trkalian field can be expressed as: $\vect{F}(\vect{x})
=\Sigma^{\prime}\vect{F}_{\lambda}(\vect{x})$ excluding the 
divergenceful component, where $\vect{F}_{\lambda}(\vect{x})
=(1/g)\int \vect{\chi}_{\lambda}(\vect{x}|\vect{k})
f_{\lambda}(\vect{k})d^{3}k$.\cite{Ml1, Ml3}  Then 
we find $f_{\lambda}(\vect{k})=\left[ \delta(k-\lambda\nu)/k^{2} \right] 
s_{\lambda}(\vect{k})$ relating the Curl transform $f_{\lambda}(\vect{k})$ 
and the spherical Curl transform $s_{\lambda}(\vect{k})$ of the field 
$\vect{F}(\vect{x})$. Thus an arbitrary solution is given entirely 
in terms of its transform on a sphere of radius $k=\lambda\nu=|\nu|$ 
in transform space. Further, only the eigenfunctions for which 
$\lambda=sgn(\nu)$ contribute to the field. The Radon transform 
of a Trkalian field is tangent to this sphere

\begin{eqnarray} \label{RadonMoses}
\vect{F}^{\mathcal{R}}_{\lambda} (p, \vect{\kappa}) 
=(2\pi)^{1/2} \frac{1}{g}\frac{1}{\nu^{2}} 
\left[ e^{i\lambda\nu p}\vect{Q}_{\lambda}(\vect{\kappa})
s_{\lambda}(\lambda\nu\vect{\kappa})
+ e^{-i\lambda\nu p}\vect{Q}_{\lambda}(-\vect{\kappa})
s_{\lambda}(-\lambda\nu\vect{\kappa}) \right].
\end{eqnarray}

\noindent The factor $1/g$ is introduced for the sake of a proper strength 
for the gauge potential in topologically massive gauge theory.\cite{KS, KS1} 
This can be taken as $1$ for general Trkalian fields. The inverse transform 
is given as 

\begin{eqnarray} \label{invRadtrf}
\vect{F}_{\lambda}(\vect{x})
&=& \frac{1}{8\pi^{2}}\nu^{2}\int_{S^{2}_{\vect{\kappa}}}
\vect{F}^{\mathcal{R}}_{\lambda} (\vect{\kappa}\cdot\vect{x}, \vect{\kappa}) 
d\Omega_{\vect{\kappa}} \\
&=& \frac{1}{(2\pi)^{3/2}} \frac{1}{g} \int_{S^{2}_{\vect{\kappa}}} 
e^{i\lambda\nu\vect{\kappa}
\cdot\vect{x}}\vect{Q}_{\lambda}(\vect{\kappa})
s_{\lambda}(\lambda\nu\vect{\kappa})
d\Omega_{\vect{\kappa}}, \nonumber
\end{eqnarray}

\noindent using the adjoint Radon transform $\vect{\mathcal{R}}^{\dagger}$, 
where $S^{2}_{\vect{\kappa}}$ is the unit sphere in the transform space.

  The simplest example of Trkalian fields is $\vect{F}(\vect{x})
= e^{i \vect{k}_{0}\cdot\vect{x}}\vect{F}_{0}$ where $\vect{k}_{0}
=k_{0}\vect{\kappa}_{0}$, $k_{0}=\lambda\nu>0$ and $\vect{F}_{0}
=\vect{Q}_{\lambda}(\vect{\kappa}_{0})$: 
$\vect{\kappa}_{0}\bm{\times}\vect{F}_{0}=-i\lambda\vect{F}_{0}$, 
$\vect{\kappa}_{0}\cdot\vect{F}_{0}=0$. Its Radon transform is 

\begin{eqnarray} \label{Radsimp}
\vect{F}^{\mathcal{R}}(p, \vect{\kappa}) 
=(2\pi)^{2} \frac{1}{k_{0}^{2}} \left[ 
e^{ik_{0}p}\delta(\vect{\kappa}-\vect{\kappa}_{0}) 
+e^{-ik_{0}p}\delta(\vect{\kappa}+\vect{\kappa}_{0}) 
\right] \vect{F}_{0}. 
\end{eqnarray}

\noindent The Lundquist\cite{SL} solution is given as 

\begin{eqnarray} \label{Lundquist}
\vect{F}_{L}(\vect{x})=F_{0} \left[\lambda J_{1}(\lambda\nu r)\vect{e}_{\phi} 
+J_{0}(\lambda\nu r)\vect{e}_{z} \right],
\end{eqnarray} 

\noindent in cylindrical coordinates, where 
$J_{n}$ is the $n$th order Bessel function. 
Its Radon transform, for $\lambda=1$ is 

\begin{eqnarray} \label{RadonLundquist}
\vect{F}^{\mathcal{R}}_{L(\lambda=1)}(p, \vect{\kappa}) 
=2\pi iF_{0}\frac{1}{\nu^{2}} \delta(\kappa_{z}) (e^{i\nu p}  
\vect{L}+e^{-i\nu p} \vect{L}^{\prime}), 
\end{eqnarray}

\noindent where $\vect{L}=\sin\psi\vect{e}_{x}
-\cos\psi\vect{e}_{y}-i\vect{e}_{z}$, 
$\vect{L}^{\prime}=-\sin\psi\vect{e}_{x}
+\cos\psi\vect{e}_{y}-i\vect{e}_{z}$ and
$\vect{k}=k_{r}\cos\psi\vect{e}_{x}
+k_{r}\sin\psi\vect{e}_{y}+k_{z}\vect{e}_{z}$.\cite{KS, Ml3}

  The Radon transform of the Riesz potential\cite{KS, RK} of order $2$: 
$\vect{\mathcal{I}}^{2}[\vect{F}](\vect{x})=(1/8\pi^{2})
\vect{\mathcal{R}}^{\dagger}
\vect{\mathcal{R}}[\vect{F}](\vect{x})$ is given as

\begin{eqnarray} \label{RadonRiesz}
\vect{\mathcal{R}}\{\vect{\mathcal{I}}^{2}[\vect{F}]\}(p, \vect{\kappa})
= \frac{1}{8\pi^{2}}\vect{\mathcal{R}}\vect{\mathcal{R}}^{\dagger}
\vect{\mathcal{R}}[\vect{F}](p, \vect{\kappa}) 
= \vect{{\mathcal{F}}}^{-1}\{\frac{1}{k^{2}}\vect{{\mathcal{F}}}
[\vect{F}^{\mathcal{R}}(q, \vect{\kappa})](k, \vect{\kappa})\} 
(p, \vect{\kappa}).
\end{eqnarray}

  We can associate an eigenvalue equation for Biot-Savart integral 
operator: $\vect{\mathcal{BS}}[\vect{F}](\vect{x})=\vect{\nabla}\bm{\times}
\vect{\mathcal{I}}^{2}[\vect{F}](\vect{x})$ with a Trkalian field, if its 
Radon transform exists (if the field is divergence-free and it vanishes at 
infinity): $\vect{\mathcal{BS}}[\vect{F}]=(1/\nu)\vect{F}$.\cite{KS, CDG, P} 
For example, the Lundquist field (\ref{Lundquist}) is an eigenvector of the 
$\vect{\mathcal{BS}}$ operator. 

  The Radon-Biot-Savart integral  

\begin{eqnarray} \label{RadonBiotSavart}
\vect{\mathcal{RBS}}[\vect{F}^{\mathcal{R}}(q, \vect{\kappa})]
(p, \vect{\kappa})= \vect{\Gamma}_{p}\bm{\times}
\vect{\mathcal{R}}\{I^{2}[\vect{F}]\}(p, \vect{\kappa}) 
= i\vect{\kappa} \bm{\times} 
\vect{{\mathcal{F}}}^{-1}\{\frac{1}{k}\vect{{\mathcal{F}}}
[\vect{F}^{\mathcal{R}}(q, \vect{\kappa})](k, \vect{\kappa})\} 
(p, \vect{\kappa}),
\end{eqnarray}

\noindent is defined as the Radon transform of the 
$\vect{\mathcal{BS}}$ integral.\cite{KS} The Radon transform 
(\ref{RadonMoses}) of a Trkalian field is an eigenvector 
of the $\vect{\mathcal{RBS}}$ integral operator.\cite{KS}

  The inverse spherical Curl transform expression (\ref{invRadtrf}) for 
a Trkalian field is in the form of Whittaker' s solution\cite{W} to the 
(scalar) Helmholtz equation (one for each Cartesian component or simply: 
$f\longrightarrow \vect{f}$ there and $\lambda\nu$ is normalized). 
If we substitute (the vectorial form of) Whittaker' s solution of 
the Helmholtz equation in (\ref{Trkalianselfdual}), we are led 
to a vector identity which is simply satisfied by the Moses\cite{M1} 
basis vectors: $\vect{\kappa}\bm{\times}\vect{Q}_{\lambda}(\vect{\kappa})
\sim-i\vect{Q}_{\lambda}(\vect{\kappa})$ (up to helicity factor 
$\lambda=\pm 1$).

  In Section \ref{mini-twistorsolution}, we shall follow another 
strand represented by Penrose and Hitchin which dates back to 
Whittaker and Bateman. The Twistor theory has been originally 
founded and developed by Penrose.

\subsection{X-ray and Divergent beam transforms}

  The X-ray transform of a vector-valued function $\vect{F}(\vect{x})$ 
(that belongs to Schwartz class) on $\mathbb{R}^{3}$ is defined as 

\begin{eqnarray} \label{xraytrf}
\vect{\mathcal{X}}\vect{F}(\vect{\theta}, \vect{x})
=\int_{-\infty}^{\infty} \vect{F}(\vect{x}+s\vect{\theta})ds,
\end{eqnarray} 

\noindent the integral of the field over line $L$ passing through 
point $\vect{x}$ in the direction determined by the unit vector 
$\vect{\theta}$.\cite{NW, N, GGG, RK} This is a componentwise 
generalization of the X-ray transform of scalar fields. The X-ray 
transform is defined on the space of oriented lines in $\mathbb{R}^{3}$. 
Note, $\vect{\mathcal{X}}\vect{F}(\vect{\theta}, \vect{x})$ is unchanged 
if $\vect{x}$ is translated in the direction of $\vect{\theta}$. Therefore 
we restrict $\vect{x}$ to $\theta^{\perp}$: $\vect{x}\cdot\vect{\theta}=0$. 
Hence $\vect{\mathcal{X}}\vect{F}(\vect{\theta}, \vect{x})$ is a function 
defined on the tangent bundle $\mathbb{TS}^{2}=\{ (\vect{\theta}, \vect{x}), 
\, \vect{\theta} \in \mathbb{S}^{2}, \, \vect{x} \in \theta^{\perp} \}$ of the 
sphere $\mathbb{S}^{2}$.\cite{NW, N} Also $\vect{\mathcal{X}}\vect{F}
(-\vect{\theta}, \vect{x})=\vect{\mathcal{X}}\vect{F}
(\vect{\theta}, \vect{x})$.

  The Divergent beam or Cone beam transform is defined as 

\begin{eqnarray} \label{draytrf}
\vect{\mathcal{D}}\vect{F}(\vect{\theta}, \vect{x})
=\int_{0}^{\infty} \vect{F}(\vect{x}+s\vect{\theta})ds, 
\end{eqnarray}

\noindent the integral of the field over the half-line. 
We have: $\vect{\mathcal{X}}\vect{F}(\vect{\theta}, \vect{x})
=\vect{\mathcal{D}}\vect{F}(\vect{\theta}, \vect{x})
+\vect{\mathcal{D}}\vect{F}(-\vect{\theta}, \vect{x})$.

  The X-ray transform (\ref{xraytrf}) [also the Divergent beam transform 
(\ref{draytrf})] satisfies John' s equation below. The inversion problem 
of the X-ray transform is overdetermined,\cite{GGG} that is the data of all 
line integrals are redundant.\cite{VP} There are various inversion methods 
for the X-ray (also Divergent beam) transform. For example, one can 
reconstruct $\vect{F}(\vect{x})$ knowing $\vect{\mathcal{X}}\vect{F}
(\vect{\theta}, \vect{x})$ where $\vect{\theta}\in \mathbb{S}^{2}$, 
$\vect{x}\in L$ and $L$ is a suitable curve, (see Ref. \onlinecite{RK}, 
p. 52, p. 276 and Ref. \onlinecite{VP1}). 

  The  X-ray transform of the Trkalian field $\vect{F}(\vect{x})
= e^{i \vect{k}_{0}\cdot\vect{x}}\vect{F}_{0}$ where $\vect{k}_{0}
=k_{0}\vect{\kappa}_{0}$, $k_{0}>0$ is 

\begin{eqnarray} \label{Xsimp}
\vect{\mathcal{X}}\vect{F}(\vect{\theta}, \vect{x})=2\pi \frac{1}{k_{0}} 
e^{i k_{0}\vect{\kappa}_{0}\cdot\vect{x}}
\delta(\vect{\kappa}_{0}\cdot\vect{\theta})\vect{F}_{0}.
\end{eqnarray}

\noindent Its Divergent beam transform is 

\begin{eqnarray} \label{Dsimp}
\vect{\mathcal{D}}\vect{F}(\vect{\theta}, \vect{x}) 
= 2\pi \frac{1}{k_{0}} e^{ik_{0}\vect{\kappa}_{0}\cdot\vect{x}} 
\delta^{+}(\vect{\kappa}_{0}\cdot\vect{\theta}) \vect{F}_{0}, 
\hspace*{15mm}
\vect{\mathcal{D}}\vect{F}(-\vect{\theta}, \vect{x}) 
= 2\pi \frac{1}{k_{0}} e^{ik_{0}\vect{\kappa}_{0}\cdot\vect{x}} 
\delta^{-}(\vect{\kappa}_{0}\cdot\vect{\theta}) \vect{F}_{0}, 
\end{eqnarray}

\noindent where $\delta^{\pm}(x)=(1/2) \left[ \delta(x)\mp1/(i\pi x) \right]$ 
is the socalled Heisenberg distribution\cite{GS, K} and $1/x$ is to be 
understood in the sense of Cauchy principal-value: $1/x=P(1/x)$. Here 
we have used Fourier transform of the Heaviside step function(al) $H(x)$: 
$\vect{{\mathcal{F}}}[H(\pm x)](k)=\sqrt{2\pi}\delta^{\mp}(k)$.

  The X-ray transform of the Lundquist 
field (\ref{Lundquist}) is 

\begin{eqnarray} \label{XLundquist}
\vect{\mathcal{X}}\vect{F}_{L}(\vect{\theta}, \vect{x}) 
= 2F_{0}\frac{1}{\lambda\nu}\frac{1}{v_{r}} \big\{ \sin 
\left[ \nu r\sin(\theta-\phi) \right] \vect{e}_{r}(\theta) 
+\cos \left[ \nu r\sin(\theta-\phi) \right] \vect{e}_{z} \big\}, 
\end{eqnarray}

\noindent where $\vect{x}=r\vect{e}_{r}(\phi)+z\vect{e}_{z}$, 
$\vect{\theta}=v_{r}\vect{e}_{r}(\theta)+v_{z}\vect{e}_{z}$ 
in cylindrical coordinates. See Appendix \ref{XrayLundquist}. 
We shall calculate its Divergent beam transform in Section 
\ref{Tuymethod}.

  The main results of this section are given in the following 
two propositions. Note that these results are logically motivated 
in a unified view by the mathematical methods in Sections 
\ref{Smithapproach}: Smith' s method and \ref{Tuymethod}: Tuy' s method. 
However, we have interchanged the logical order for a clean presentation, 
because the direct approach here is more appropriate for exposing the 
relations of X-ray and Divergent beam transforms of Trkalian fields to 
the Radon and other transforms. Also, the geometry of these results 
deserves a separate discussion in its own right. We shall discuss the 
mathematical point of view originating from tomography later on. We shall 
first present direct proofs without motivation below.

\begin{prop} \label{prop1} 
The X-ray transform of a Trkalian field is given as 

\begin{eqnarray} \label{XFR1}
\vect{\mathcal{X}}\vect{F}_{\lambda}(\vect{\theta}, \vect{x}) 
&=& \frac{1}{(2\pi)^{1/2}}\frac{1}{g}\frac{1}{\lambda\nu} 
\int_{S_{\vect{\kappa}}^{2}} e^{i\lambda\nu\vect{\kappa}\cdot\vect{x}}
\vect{Q}_{\lambda}(\vect{\kappa})s_{\lambda}(\lambda\nu\vect{\kappa})
\delta(\vect{\kappa}\cdot\vect{\theta}) d\Omega_{\vect{\kappa}} \\ 
&=& \frac{1}{4\pi}\lambda\nu \int_{S_{\vect{\kappa}}^{2}} 
\vect{F}_{\lambda}^{\mathcal{R}} (\vect{\kappa}\cdot\vect{x}, \vect{\kappa}) 
\delta(\vect{\kappa}\cdot\vect{\theta}) d\Omega_{\vect{\kappa}}, 
\nonumber
\end{eqnarray}

\noindent the Minkowski-Funk transform (in the transform space) 
of its Radon transform.

\end{prop} 

  {\bf{Proof:}\,} We substitute $\vect{F}_{\lambda}(\vect{x}+s\vect{\theta})$ 
using the second line of (\ref{invRadtrf}) in (\ref{xraytrf}). 
See Appendix \ref{reduceintegral}. Note that a result valid 
in a basis should be valid in any basis. \hfill {\boldmath $\Box$} 

  This proposition is based on the motivation in equation (\ref{Smithmine}) 
below which makes use of Smith' s formula (\ref{extxrayFourier21}) 
in Section \ref{Smithapproach}. See equation (\ref{identity7}) for 
the underlying intricate geometric quantity and its simplification.

  The Minkowski-Funk\cite{M, PF} transform $\vect{\mathcal{M}}$ of 
a continuous (even) function on a sphere is given by the integral 
of this function over great circles: geodesics in the sphere. This 
yields a function on the space of geodesics. This clearly annihilates 
odd functions.

  We symbolically write $\vect{\mathcal{X}}=[\lambda\nu/(4\pi)]
\vect{\mathcal{M}}\vect{\mathcal{R}}$ respectively relating the X-ray 
(John\cite{John}), Minkowski-Funk\cite{M, PF} and Radon\cite{JR} transforms 
for Trkalian fields. In $\mathbb{R}^{3}$, this corresponds to integral of the 
Radon transform of the field over a pencil of planes intersecting at a line 
(passing through the point $\vect{x}$ in the direction of $\vect{\theta}$) 
which are parametrized by a circle. Previously, Gonzales called this 
plane-to-line transform.\cite{FBG} See Ref. \onlinecite{BE} for 
a twistor view of this transform.

  There are various inversion methods for the Minkowski-Funk 
transform. We shall briefly discuss an inversion formula 
which is appropriate for our purpose below.

  If we use the example (\ref{Radsimp}) in the second line of 
(\ref{XFR1}) we find the result (\ref{Xsimp}). Also, we find 
(\ref{XLundquist}), ($\lambda=1$) substituting (\ref{RadonLundquist}) 
in the  second line of (\ref{XFR1}), where $\kappa_{r}=\sin \alpha
=\sqrt{1-\kappa_{z}^{2}}$, $\kappa_{z}=\cos \alpha$ and $\alpha$ 
is the polar angle on $S^{2}_{\vect{\kappa}}$ with spherical angles $\psi$, 
$\alpha$. We shall avoid the details of these straightforward 
calculations.

  We can arrive at the same result, as expressed in equation (\ref{XFR2}), 
using the Curl expansion for Trkalian fields in Fourier slice-projection 
theorem\cite{NW} 

\begin{eqnarray} \label{Fourierslicproj}
\vect{{\mathcal{F}}}[\vect{\mathcal{X}}\vect{F}(\vect{\theta}, \vect{x})]
(\vect{\theta}, \vect{\xi}) 
=(2\pi)^{1/2} \vect{{\mathcal{F}}}[\vect{F}(\vect{x})](\vect{\xi}), 
\hspace*{5mm} \vect{\xi}\in \theta^{\perp} 
\end{eqnarray} 

\noindent for the X-ray transform. See Appendix 
\ref{Fouriersliceprojection}. On the left-hand side 
$\vect{{\mathcal{F}}}$ stands for Fourier transform in 
$\theta^{\perp}$ plane whereas on the right-hand side 
it is a Fourier transform in three dimensions. 

\begin{prop} \label{prop2} 
The Divergent beam transform of a Trkalian field is given as 

\begin{eqnarray} \label{DFR1}
\vect{\mathcal{D}}\vect{F}_{\lambda}(\vect{\theta}, \vect{x}) 
&=& \frac{1}{(2\pi)^{1/2}}\frac{1}{g}\frac{1}{\lambda\nu} 
\int_{S_{\vect{\kappa}}^{2}} e^{i\lambda\nu\vect{\kappa}\cdot\vect{x}}
\vect{Q}_{\lambda}(\vect{\kappa})s_{\lambda}(\lambda\nu\vect{\kappa})
\delta^{+}(\vect{\kappa}\cdot\vect{\theta}) d\Omega_{\vect{\kappa}},
\end{eqnarray} 

\noindent in the Moses basis.

\end{prop}

  {\bf{Proof:}\,} We substitute $\vect{F}_{\lambda}
(\vect{x}+s\vect{\theta})$ using the second line of 
(\ref{invRadtrf}) in (\ref{draytrf}) and use
$\vect{{\mathcal{F}}}[H(\pm x)](k)
=\sqrt{2\pi}\delta^{\mp}(k)$. \hfill {\boldmath $\Box$} 

  This proposition is based on the motivation in equation 
(\ref{extdbeamtrfFourierinv}) below which makes use of 
(\ref{extdbeamFourier21}) in Section \ref{Tuymethod}: 
Tuy' s method. See equation (\ref{Tuycrucsimp}) for 
the underlying intricate geometric quantity and its 
simplification.

  This is another basic transform (with no specific name in the 
literature known by the author) in integral geometry containing 
the Heisenberg delta function $\delta^{\pm}$. The first term in 
$\delta^{\pm}$ is associated with the Minkowski-Funk transform 
mentioned above. Hence, this is an extension of the Minkowski-Funk 
transform. The second term is to be understood in the sense of 
Cauchy principal value (see above). Roughly speaking, it provides 
a description of the behaviour of a function on the sphere except 
the geodesics, [see (\ref{Y1}) below for the physical meaning]. 

  This leads to 

\begin{eqnarray} \label{DFR2}
\vect{\mathcal{D}}\vect{F}_{\lambda}(-\vect{\theta}, \vect{x}) 
&=& \frac{1}{(2\pi)^{1/2}}\frac{1}{g}\frac{1}{\lambda\nu} 
\int_{S_{\vect{\kappa}}^{2}} e^{i\lambda\nu\vect{\kappa}\cdot\vect{x}}
\vect{Q}_{\lambda}(\vect{\kappa})s_{\lambda}(\lambda\nu\vect{\kappa})
\delta^{-}(\vect{\kappa}\cdot\vect{\theta}) d\Omega_{\vect{\kappa}},
\end{eqnarray}

\noindent since $\delta^{+}(-\vect{\kappa}\cdot\vect{\theta})
=\delta^{-}(\vect{\kappa}\cdot\vect{\theta})$. 

  For $\vect{F}(\vect{x})= e^{i \vect{k}_{0}\cdot\vect{x}}\vect{F}_{0}$, 
[$\vect{k}_{0}=k_{0}\vect{\kappa}_{0}$, $k_{0}=\lambda\nu>0$, $\vect{F}_{0}
=\vect{Q}_{\lambda}(\vect{\kappa}_{0})$], we find the result (\ref{Dsimp}) 
substituting $s_{\lambda}(\lambda\nu\vect{\kappa})=(2\pi)^{3/2}
g\delta(\vect{\kappa}-\vect{\kappa}_{0})$\cite{KS, Ml1, Ml3} 
in (\ref{DFR1}, \ref{DFR2}).

  In analogy with the X-ray transform (\ref{xraytrf}), 
the difference

\begin{eqnarray} \label{Y1}
\vect{\mathcal{Y}}\vect{F}(\vect{\theta}, \vect{x}) 
= \vect{\mathcal{D}}\vect{F}(\vect{\theta}, \vect{x}) 
-\vect{\mathcal{D}}\vect{F}(-\vect{\theta}, \vect{x})
= \int_{-\infty}^{\infty} \vect{F}
(\vect{x}+s\vect{\theta}) sgn(s) ds,
\end{eqnarray}

\noindent where $sgn(s)=H(s)-H(-s)$ is the signum function, 
reduces to
 
\begin{eqnarray} \label{Y2}
\vect{\mathcal{Y}}\vect{F}_{\lambda}(\vect{\theta}, \vect{x}) 
=i \frac{2}{(2\pi)^{3/2}} \frac{1}{g} 
\frac{1}{\lambda\nu} \int_{S_{\vect{\kappa}}^{2}} 
e^{i\lambda\nu\vect{\kappa}\cdot\vect{x}} \vect{Q}_{\lambda}(\vect{\kappa}) 
s_{\lambda}(\lambda\nu\vect{\kappa}) \frac{1}{\vect{\kappa}\cdot\vect{\theta}}
d\Omega_{\vect{\kappa}}, 
\end{eqnarray}

\noindent the difference of equations (\ref{DFR1}) and (\ref{DFR2}), 
for a Trkalian field (\ref{invRadtrf}) in the Moses basis 
using: $\vect{{\mathcal{F}}}[sgn(x)](k)=\sqrt{2\pi}(1/i\pi)(1/k)$. 

  For the Trkalian field $\vect{F}(\vect{x})= e^{i \vect{k}_{0}\cdot\vect{x}}
\vect{F}_{0}$, [$\vect{k}_{0}=k_{0}\vect{\kappa}_{0}$, $k_{0}=\lambda\nu>0$, 
$\vect{F}_{0}=\vect{Q}_{\lambda}(\vect{\kappa}_{0})$], we find 

\begin{eqnarray} \label{Ytrfsimp}
\vect{\mathcal{Y}}\vect{F}_{\lambda}(\vect{\theta}, \vect{x}) 
=2i \frac{1}{k_{0}}
e^{ik_{0}\vect{\kappa}_{0}\cdot\vect{x}}
\frac{1}{\vect{\kappa}_{0}\cdot\vect{\theta}}
\vect{F}_{0}.
\end{eqnarray}

  We can write the equation (\ref{DFR1}) as

\begin{eqnarray} \label{Atrf}
D\vect{F}_{\lambda}(\vect{\theta}, \vect{x}) 
=\frac{1}{2^{1/2}}\frac{1}{g}\frac{1}{\lambda\nu} 
\vect{\mathcal{A}}^{0}[\vect{G}]
=\frac{1}{2^{1/2}}\frac{1}{g}\frac{1}{\lambda\nu} 
\left( \vect{\mathcal{U}}^{0}[\vect{G}]
+i\vect{\mathcal{V}}^{0}[\vect{G}] \right),
\end{eqnarray}

\noindent where $\vect{G}(\vect{\kappa}, 
\vect{x})=e^{i\lambda\nu\vect{\kappa}\cdot\vect{x}}
\vect{Q}_{\lambda}(\vect{\kappa})s_{\lambda}(\lambda\nu\vect{\kappa})$, 
$\vect{\mathcal{A}}^{0}=\vect{\mathcal{U}}^{0}+i\vect{\mathcal{V}}^{0}$ 
and 

\begin{eqnarray} \label{UV}
\vect{\mathcal{U}}^{0}[\vect{G}]
&=& \frac{1}{2}\frac{1}{\pi^{1/2}}
\int_{S_{\vect{\kappa}}^{2}} \vect{G}(\vect{\kappa}, \vect{x}) 
\delta(\vect{\kappa}\cdot\vect{\theta}) d\Omega_{\vect{\kappa}}
\hspace*{10mm} 
\vect{\mathcal{V}}^{0}[\vect{G}]=\frac{1}{2}\frac{1}{\pi^{3/2}}
\int_{S_{\vect{\kappa}}^{2}} \vect{G}(\vect{\kappa}, \vect{x}) 
\frac{1}{\vect{\kappa}\cdot\vect{\theta}} d\Omega_{\vect{\kappa}}, \\
&=& \frac{1}{2}\frac{1}{\pi^{1/2}}
\int_{\vect{\kappa}\cdot\vect{\theta}=0} 
\vect{G}(\vect{\kappa}, \vect{x}) d_{\vect{\theta}}\vect{\kappa}, 
\nonumber
\end{eqnarray}

\noindent with [a slight change of notation: 
$\vect{e}_{r}(\phi) \longrightarrow \vect{\kappa}$ in (\ref{XFR2})]
the integration measure $d_{\vect{\theta}}\vect{\kappa}$ on the great 
circle $C$ determined by $\vect{\theta}$. The integrals in 
$\vect{\mathcal{U}}^{0}$ and $\vect{\mathcal{V}}^{0}$ respectively 
correspond to the Minkowski-Funk transform in (\ref{XFR1}, \ref{XFR2}) 
and the transform in (\ref{Y2}). The Minkowski-Funk transform 
$\vect{\mathcal{U}}^{0}$ describes behaviour of the 
function on great circles in the $2$-sphere. Roughly, the transform 
$\vect{\mathcal{V}}^{0}$ describes behaviour of the function 
on the $2$-sphere except the great circles. 

  The transform $\vect{\mathcal{A}}^{0}$ is a member (via analytic 
continuation) of an analytic family of integral operators 
$\vect{\mathcal{A}}^{\alpha}=\vect{\mathcal{U}}^{\alpha}
+i\vect{\mathcal{V}}^{\alpha}$ which arise in the study 
of Fourier transforms of homogeneous functions. Recently, 
these have been studied by Rubin.\cite{Rubin3, Rubin4}
For a brief summary of these transforms see Appendix \ref{transforms}.

  The equations (\ref{XFR1}, \ref{XFR2}), (\ref{DFR1}) and 
(\ref{Atrf}, \ref{UV}) respectively suggest their inversion methods 
for the X-ray and Divergent beam transforms of Trkalian fields. For 
the sake of motivation, we shall briefly discuss the inversion of  
Minkowski-Funk transform $\vect{\mathcal{U}}^{0}$ using a member of this 
family. This was established by Semyanistyi.\cite{VIS, VIS1} Because a 
detailed study of these would be distracting. 

  We have $\vect{\mathcal{X}}\vect{F}=[\lambda\nu/(4\pi)]\vect{\mathcal{M}}
[\vect{F}^{R}]$, (\ref{XFR1}) and $\vect{\mathcal{U}}^{0}=[1/(2\pi^{1/2})]
\vect{\mathcal{M}}$, (\ref{UV}), hence $\vect{\mathcal{X}}\vect{F}
=[\lambda\nu/(2\pi^{1/2})]\vect{\mathcal{U}}^{0}[\vect{F}^{R}]$. 
The inverse of $\vect{\mathcal{U}}^{0}$ is $(\vect{\mathcal{U}}^{0})^{-1}
=\vect{\mathcal{U}}^{-1}$ (Appendix \ref{transforms}), hence 
$\vect{\mathcal{M}}^{-1}=[1/(2\pi^{1/2})]\vect{\mathcal{U}}^{-1}$. 
Thus

\begin{eqnarray} \label{InvFunkSem}
\vect{F}_{\lambda}^{\mathcal{R}} 
(\vect{\kappa}\cdot\vect{x}, \vect{\kappa}) 
=-\frac{1}{2\pi}\frac{1}{\lambda\nu} 
\int_{S_{\vect{\theta}}^{2}}\vect{\mathcal{X}}
\vect{F}_{\lambda}(\vect{\theta}, \vect{x}) 
\frac{1}{|\vect{\kappa}\cdot\vect{\theta}|^{2}} 
d\Omega_{\vect{\theta}},
\end{eqnarray}

\noindent which is to be understood in a regularized 
sense.\cite{VIS, VIS1} See equation (\ref{GelfandGoncharovMoses}) 
and the following discussion in Section \ref{GelfandGoncharovmethod} 
for a derivation of this inversion formula from the mathematical methods 
of tomography.

  If we substitute (\ref{XFR1}) in (\ref{InvFunkSem}) 
and interchange the order of integrations, we find

\begin{eqnarray} \label{InvFunkSemTrkal}
\vect{F}_{\lambda}^{\mathcal{R}} 
(\vect{\kappa}\cdot\vect{x}, \vect{\kappa}) 
&=& -\frac{1}{(2\pi)^{3/2}}\frac{1}{g}\frac{1}{\nu^{2}}
\int_{S_{\vect{\kappa}}^{2}} 
e^{i\lambda\nu\vect{\kappa}^{\prime}\cdot\vect{x}}
\vect{Q}_{\lambda}(\vect{\kappa}^{\prime})
s_{\lambda}(\lambda\nu\vect{\kappa}^{\prime})
I(\vect{\kappa}, \vect{\kappa}^{\prime}) 
d\Omega_{\vect{\kappa}^{\prime}}.
\end{eqnarray}

\noindent The integral

\begin{eqnarray} \label{integralimp}
I(\vect{\kappa}, \vect{\kappa}^{\prime}) 
= \int_{S_{\vect{\theta}}^{2}} 
\frac{\delta(\vect{\kappa}^{\prime}\cdot\vect{\theta})}
{|\vect{\kappa}\cdot\vect{\theta}|^{2}} 
d\Omega_{\vect{\theta}} 
= -4\pi^{2}[\delta(\vect{\kappa}-\vect{\kappa}^{\prime}) 
+\delta(\vect{\kappa}+\vect{\kappa}^{\prime})], 
\end{eqnarray}

\noindent can be evaluated using the planewave decomposition\cite{K, GS} 
of Dirac delta function. See Appendix \ref{evaluationintegral}. Then the 
equation (\ref{InvFunkSemTrkal}) yields  
$\vect{F}_{\lambda}^{\mathcal{R}} (\vect{\kappa}\cdot\vect{x}, \vect{\kappa})$, 
(\ref{RadonMoses}). We can find $\vect{F}_{\lambda}(\vect{x})$
using the inverse Radon transform (\ref{invRadtrf}). See Section 
\ref{GelfandGoncharovmethod} for this.

  For $\vect{F}(\vect{x})= e^{i \vect{k}_{0}\cdot\vect{x}}\vect{F}_{0}$, 
[$\vect{k}_{0}=k_{0}\vect{\kappa}_{0}$, $k_{0}=\lambda\nu>0$, 
$\vect{F}_{0}=\vect{Q}_{\lambda}(\vect{\kappa}_{0})$],
if we substitute (\ref{Xsimp}) in (\ref{InvFunkSem}), we find
$\vect{F}_{\lambda}^{\mathcal{R}}(\vect{\kappa}\cdot\vect{x}, \vect{\kappa})$, 
(\ref{Radsimp}) with a similar reasoning as in (\ref{integralimp}), 
($\kappa^{\prime}\longrightarrow \kappa_{0}$).  For the Lundquist 
solution (\ref{Lundquist}), if we substitute (\ref{XLundquist}), 
[that is (\ref{XLe})] in (\ref{InvFunkSem}) and use the planewave 
decomposition of Dirac delta function, then we find 
$\vect{F}^{\mathcal{R}}_{L(\lambda=1)}(\vect{\kappa}\cdot\vect{x}, 
\vect{\kappa})$, (\ref{RadonLundquist}) for $\lambda=1$. 
See Appendix \ref{InvFunkSemLunquist}.

  If we substitute (\ref{InvFunkSem}) in (\ref{invRadtrf}), we find 
$\vect{F}_{\lambda}(\vect{x})$ is given by the spherical mean of 
$\vect{\mathcal{X}}\vect{F}_{\lambda}(\vect{\theta}, \vect{x})$. 
We shall derive this in a more direct way below, 
[see (\ref{Sphmean}), (\ref{contint})].

  As a different alternative, one can try expressing $\vect{F}^{\mathcal{R}}$ 
in terms of $\vect{\mathcal{X}}\vect{F}$: $\vect{F}^{\mathcal{R}}
(p, \vect{\kappa})=\left[ \vect{\mathcal{R}}_{2} \vect{\mathcal{X}}\vect{F}
(\vect{\theta}, \vect{x}) \right] (p, \vect{\kappa})$ using a Radon 
transform: $\vect{\mathcal{R}}_{2}$  in the plane 
$\theta^{\perp}$.\cite{N, T, PM}

  We can also write the Divergent beam transform 
(\ref{DFR1}, \ref{Atrf}) as 

\begin{eqnarray} \label{Divbeamalter}
\vect{\mathcal{D}}\vect{F}_{\lambda}(\vect{\theta}, \vect{x}) 
&=& \frac{1}{(2\pi)^{3/2}}\frac{1}{g}\frac{1}{\lambda\nu} \eta 
\int_{R_{\vect{k}}^{3}} \frac{1}{k^{2}} 
\vect{G}(\vect{\kappa}, \vect{x})
e^{i\vect{k}\cdot\vect{\eta}} d^{3}k,  
\end{eqnarray}

\noindent the equation (\ref{Fourierhomogeneous}) with 
$Re(\alpha)=0$, where $\vect{k}=k\vect{\kappa}$, $k=|\vect{k}|$ 
and $\vect{\eta}=\eta\vect{\theta}$, $\eta=|\vect{\eta}|$. 
It is straightforward to verify this for the field 
$\vect{F}(\vect{x})= e^{i \vect{k}_{0}\cdot\vect{x}}\vect{F}_{0}$, 
[$\vect{k}_{0}=k_{0}\vect{\kappa}_{0}$, $k_{0}=\lambda\nu>0$, 
$\vect{F}_{0}=\vect{Q}_{\lambda}(\vect{\kappa}_{0})$
and $s_{\lambda}(\lambda\nu\vect{\kappa})=(2\pi)^{3/2}
g\delta(\vect{\kappa}-\vect{\kappa}_{0})$], using 
(\ref{Dsimp}).

\section{JOHN' S EQUATION}
\label{Johneqn}
 
   The X-ray\cite{GGG} transform (\ref{xraytrf}) [also the Divergent 
beam transform (\ref{draytrf})] satisfies John' s equation\cite{John} 

\begin{eqnarray} \label{John}
\frac{\partial^{2}}{\partial x^{i} \partial \theta^{j}} 
\vect{\mathcal{X}}\vect{F}(\vect{\theta}, \vect{x}) 
=\frac{\partial^{2}}{\partial x^{j} \partial \theta^{i}} 
\vect{\mathcal{X}}\vect{F}(\vect{\theta}, \vect{x}).
\end{eqnarray} 

\noindent The John' s equation is the necessary and sufficient condition 
for a function to be expressible as an X-ray transform.\cite{GGG} This 
can be written as an ultrahyperbolic wave equation. 

  We can easily check (\ref{XFR1}) [also (\ref{DFR1})] satisfies the 
equation (\ref{John}). It is also easy to check (\ref{Xsimp}) [also 
(\ref{Dsimp})].

\begin{prop} \label{prop3}
The X-ray (also Divergent beam) transform intertwines the Curl operator 
($\vect{\nabla}\bm{\times}$): 

\begin{eqnarray} 
\vect{\mathcal{X}}[\vect{\nabla}\bm{\times}
\vect{F}](\vect{\theta}, \vect{x})=\vect{\nabla}_{\vect{x}}\bm{\times}
\vect{\mathcal{X}}\vect{F}(\vect{\theta}, \vect{x}), 
\end{eqnarray}

\noindent and also the Divergence ($\vect{\nabla}\cdot$), Gradient 
($\vect{\nabla}$) and Laplacian ($\nabla^{2}$) operators. 

\end{prop}

  {\bf{Proof:}\,} Let $\vect{x}^{\prime}=\vect{x}+s\vect{\theta}$, 
then $\vect{\nabla}_{\vect{x}^{\prime}}=\vect{\nabla}_{\vect{x}}$  
and we have $\vect{\nabla}_{\vect{x}^{\prime}}\bm{\times}\vect{F}
(\vect{x}^{\prime})=\vect{\nabla}_{\vect{x}}\bm{\times}\vect{F}
(\vect{x}+s\vect{\theta})$. Then we find, the X-ray (Divergent beam) 
transform intertwines the Curl operator: $\vect{\mathcal{X}}
[\vect{\nabla}\bm{\times}\vect{F}](\vect{\theta}, \vect{x})
=\vect{\nabla}_{\vect{x}}\bm{\times}\vect{\mathcal{X}}\vect{F}
(\vect{\theta}, \vect{x})$, using the definition (\ref{xraytrf}), 
[(\ref{draytrf})]. The proofs for the Divergence: $\vect{\mathcal{X}}
[\vect{\nabla}\cdot\vect{F}](\vect{\theta}, \vect{x})=\vect{\nabla}_{\vect{x}}
\cdot \vect{\mathcal{X}}\vect{F}(\vect{\theta}, \vect{x})$, Gradient: 
$\vect{\mathcal{X}}[\vect{\nabla}f](\vect{\theta}, \vect{x})
=\vect{\nabla}_{\vect{x}}\vect{\mathcal{X}}f(\vect{\theta}, \vect{x})$ 
and hence the Laplacian: $\vect{\mathcal{X}}[\nabla^{2}f]
(\vect{\theta}, \vect{x})=\nabla^{2}_{\vect{x}} \vect{\mathcal{X}}
f(\vect{\theta}, \vect{x})$ operators work with a similar 
reasoning. \hfill {\boldmath $\Box$}

\begin{prop} \label{prop4}
The X-ray (also Divergent beam) transform intertwines the operator 
$(\vect{\nabla}\bm{\times})-\nu=0$ for constant $\nu$. 

\end{prop} 

  {\bf{Proof:}\,} The X-ray (Divergent beam) 
transform is linear. \hfill {\boldmath $\Box$}

  Thus the X-ray (Divergent beam) transform $\vect{\mathcal{X}}\vect{F}
(\vect{\theta}, \vect{x})$ of a Trkalian field $\vect{F}(\vect{x^{\prime}})$, 
(\ref{Trkalianselfdual}) is also Trkalian

\begin{eqnarray} \label{XrayTrkalian}
\vect{\nabla}_{\vect{x}}\bm{\times}\vect{\mathcal{X}}\vect{F}
(\vect{\theta}, \vect{x})-\nu \vect{\mathcal{X}}\vect{F}
(\vect{\theta}, \vect{x})=0,
\end{eqnarray}

\noindent [ $\vect{\nabla}_{\vect{x}}\bm{\times}\vect{\mathcal{D}}\vect{F}
(\vect{\theta}, \vect{x})-\nu \vect{\mathcal{D}}\vect{F}
(\vect{\theta}, \vect{x})=0$ ] and $\vect{\nabla}_{\vect{x}}\cdot
\vect{\mathcal{X}}\vect{F}(\vect{\theta}, \vect{x})=0$. 

  We also find $\vect{\nabla}_{\vect{\theta}} 
\cdot \vect{\mathcal{X}}\vect{F}(\vect{\theta}, \vect{x})=0$,  
($\vect{\nabla}_{\vect{\theta}}=s \vect{\nabla}_{\vect{x}^{\prime}} 
=s\vect{\nabla}_{\vect{x}}$) for a Trkalian field. 

  The X-ray transform (\ref{XFR1}) [also the Divergent beam 
transform (\ref{DFR1})] satisfies (\ref{XrayTrkalian}) and 
$\vect{\nabla}_{\vect{\theta}}\cdot\vect{\mathcal{X}}\vect{F}
(\vect{\theta}, \vect{x})=0$. We can easily check the X-ray 
transform $\vect{\mathcal{X}}\vect{F}(\vect{\theta}, \vect{x})$, 
(\ref{Xsimp}) [also the Divergent beam transform 
$\vect{\mathcal{D}}\vect{F}(\vect{\theta}, \vect{x})$, (\ref{Dsimp})]
of the field $\vect{F}(\vect{x})= e^{i \vect{k}_{0}\cdot\vect{x}}\vect{F}_{0}$, 
[$\vect{k}_{0}=k_{0}\vect{\kappa}_{0}$, $k_{0}=\lambda\nu>0$, $\vect{F}_{0}
=\vect{Q}_{\lambda}(\vect{\kappa}_{0})$] is Trkalian and 
$\vect{\nabla}_{\vect{\theta}} \cdot \vect{\mathcal{X}}\vect{F}
(\vect{\theta}, \vect{x})=0$. It is also straightforward to 
show that the X-ray transform $\vect{\mathcal{X}}\vect{F}_{L}
(\vect{\theta}, \vect{x})$, (\ref{XLundquist}) of the Lundquist 
field is Trkalian and $\vect{\nabla}_{\vect{\theta}}\cdot 
\vect{\mathcal{X}}\vect{F}_{L}(\vect{\theta}, \vect{x})=0$.

  We immediately find 

\begin{eqnarray} \label{identityTrkalian}
\vect{\nabla}_{\vect{\theta}} \cdot ( \vect{\nabla}_{\vect{x}}
\bm{\times} \vect{\mathcal{X}}\vect{F} )= ( \vect{\nabla}_{\vect{\theta}} 
\bm{\times} \vect{\nabla}_{\vect{x}} )\cdot\vect{\mathcal{X}}\vect{F}
=- \vect{\nabla}_{\vect{x}}\cdot( \vect{\nabla}_{\vect{\theta}} \bm{\times} 
\vect{\mathcal{X}}\vect{F} )=\nu \vect{\nabla}_{\vect{\theta}}\cdot 
\vect{\mathcal{X}}\vect{F}=0,
\end{eqnarray}

\noindent using $\vect{\nabla}_{\vect{\theta}}\cdot\vect{\mathcal{X}}
\vect{F}(\vect{\theta}, \vect{x})=0$, for a Trkalian field.

  We find

\begin{eqnarray} \label{nablaXFTrkalian}
\vect{\nabla}_{\vect{x}} \bm{\times} \left( \vect{\nabla}_{\vect{\theta}} 
\bm{\times}\vect{\mathcal{X}}\vect{F} \right)-\nu \vect{\nabla}_{\vect{\theta}} 
\bm{\times}\vect{\mathcal{X}}\vect{F}=0, 
\end{eqnarray}

\noindent that is $\vect{\nabla}_{\vect{\theta}}\bm{\times}\vect{\mathcal{X}}
\vect{F}$ is also Trkalian, using a straightforward reasoning similar to that 
above. This yields $\vect{\nabla}_{\vect{x}}\cdot 
\left( \vect{\nabla}_{\vect{\theta}}\bm{\times}\vect{\mathcal{X}}
\vect{F} \right)=0$ as we expect, (\ref{identityTrkalian}). We 
can also prove the equation (\ref{nablaXFTrkalian}) using John' s 
equation (\ref{John}). 

  It is straightforward to verify (\ref{nablaXFTrkalian}) 
respectively for the X-ray transforms (\ref{Xsimp}) of the 
field $\vect{F}(\vect{x})=e^{i \vect{k}_{0}\cdot\vect{x}}\vect{F}_{0}$, 
[$\vect{k}_{0}=k_{0}\vect{\kappa}_{0}$, $k_{0}=\lambda\nu>0$, 
$\vect{F}_{0}=\vect{Q}_{\lambda}(\vect{\kappa}_{0})$] and 
(\ref{XLundquist}) of the Lundquist field.

\begin{prop} \label{prop5}
For a Trkalian field the John' s equation (\ref{John}) is equivalent 
(both necessary and sufficient) to $\partial_{x^{m}} \epsilon_{ijk}
\partial_{\theta^{i}} (  \vect{\mathcal{X}}\vect{F} )_{j}
=\nu \partial_{\theta^{m}} (  \vect{\mathcal{X}}\vect{F} )_{k}$ 
that is $\partial_{x^{m}}(\vect{\nabla}_{\vect{\theta}}\bm{\times}
\vect{\mathcal{X}}\vect{F})=\nu\partial_{\theta^{m}}\vect{\mathcal{X}}\vect{F}$. 

\end{prop}

  {\bf{Proof:}\,} We can write (\ref{John}) as $\vect{\nabla}_{\vect{\theta}}
\bm{\times}\vect{\nabla}_{\vect{x}}\vect{\mathcal{X}}A=0$, 
$\vect{\nabla}_{\vect{\theta}}\bm{\times}\vect{\nabla}_{\vect{x}}
\vect{\mathcal{X}}B=0$, $\vect{\nabla}_{\vect{\theta}}\bm{\times}
\vect{\nabla}_{\vect{x}}\vect{\mathcal{X}}C=0$ where $\vect{\mathcal{X}}
\vect{F}=(\vect{\mathcal{X}}A,\, \vect{\mathcal{X}}B,\, \vect{\mathcal{X}}C)$. 
The proof is straightforward writing these and (\ref{XrayTrkalian}) and 
$\vect{\nabla}_{\vect{\theta}}\cdot\vect{\mathcal{X}}\vect{F}
(\vect{\theta}, \vect{x})=0$ in components. \hfill {\boldmath $\Box$} 

  The difference of symmetric ($m \leftrightarrow k$) equations in this 
yields (\ref{nablaXFTrkalian}). The sum of diagonal equations ($m=k$) 
yields $\vect{\nabla}_{\vect{x}}\cdot(\vect{\nabla}_{\vect{\theta}}
\bm{\times}\vect{\mathcal{X}}\vect{F} )=\nu\vect{\nabla}_{\vect{\theta}}
\cdot\vect{\mathcal{X}}\vect{F}$ which identically vanishes, 
(\ref{identityTrkalian}).

  We also have $\vect{\nabla}_{\vect{\theta}}\bm{\times} 
\left( \vect{\nabla}_{\vect{x}} \bm{\times}\vect{\mathcal{X}}\vect{F} \right)
-\nu \vect{\nabla}_{\vect{\theta}} \bm{\times}\vect{\mathcal{X}}\vect{F}=0$, 
(\ref{nablaXFTrkalian}). Then $\vect{\nabla}_{\vect{\theta}}\cdot
\left( \vect{\nabla}_{\vect{\theta}}\bm{\times}\vect{\mathcal{X}}
\vect{F} \right)=0$ trivially.

  The X-ray transform and its formal adjoint

\begin{eqnarray} \label{AdjXray}
\vect{\mathcal{X}}^{\dagger}[\vect{G}](\vect{x})=\int_{S^{2}_{\vect{\theta}}} 
\vect{G}(\vect{\theta}, E_{\theta}\vect{x}) d\Omega_{\vect{\theta}}, 
\hspace*{5mm} E_{\theta}\vect{x}=\vect{x}
-(\vect{x}\cdot\vect{\theta})\vect{\theta}
\end{eqnarray} 

\noindent where [$\vect{G}(\vect{\theta}, \vect{x})$ is to be taken 
in the range of X-ray transform:] $\vect{G}(\vect{\theta}, \vect{x})
=\vect{\mathcal{X}}[\vect{F}](\vect{\theta}, \vect{x})$, are related 
to the Riesz potential of order $\alpha$ through $\vect{\mathcal{I}}^\alpha
[\vect{F}](\vect{x})=[1/(2\pi)^{2}]\vect{\mathcal{X}}^{\dagger}
\vect{\mathcal{I}}^{\alpha-1}\vect{\mathcal{X}}[\vect{F}]
(\vect{x})$.\cite{NW, N} Here $\vect{\mathcal{I}}^{\alpha-1}$ 
is the Riesz potential on $\mathbb{TS}^{2}$ acting on the second variable 
$\vect{x}$ of $\vect{\mathcal{X}}[\vect{F}](\vect{\theta}, \vect{x})$ and 
$0<\alpha<n=3$. For $\alpha=1$ this yields 

\begin{eqnarray} \label{xrayRiesz}
\vect{\mathcal{I}}^{1}[\vect{F}](\vect{x})= \frac{1}{(2\pi)^{2}} 
\vect{\mathcal{X}}^{\dagger}\vect{\mathcal{X}}[\vect{F}](\vect{x}),
\end{eqnarray}

\noindent where 

\begin{eqnarray} \label{Riesz}
\vect{\mathcal{I}}^{1}[\vect{F}](\vect{x})=\frac{1}{2\pi^{2}} \int 
\frac{\vect{F}(\vect{y})}{|\vect{x}-\vect{y}|^{2}} d^{3}y.
\end{eqnarray}

  It is straightforward to prove 

\begin{eqnarray} \label{SphmeanDivbeam}
\int_{S^{2}_{\vect{\theta}}} \vect{\mathcal{D}}
\vect{F}(\vect{\theta}, \vect{x}) d\Omega_{\vect{\theta}} 
=2\pi^{2} \vect{\mathcal{I}}^{1}[\vect{F}](\vect{x}),
\end{eqnarray}

\noindent see Ref. \onlinecite{Markoe}, p. 283.
This leads to 

\begin{eqnarray} \label{Riesz1simp}
\vect{\mathcal{I}}^{1}[\vect{F}](\vect{x}) 
=\frac{1}{4\pi^{2}} \int_{S^{2}_{\vect{\theta}}} 
\vect{\mathcal{X}}\vect{F}(\vect{\theta}, \vect{x}) 
d\Omega_{\vect{\theta}},  
\end{eqnarray}

\noindent using\cite{Markoe} $\vect{\theta}\longrightarrow -\vect{\theta}$ 
in (\ref{SphmeanDivbeam}). We could infer this from (\ref{xrayRiesz}) using 
invariance of $\vect{\mathcal{X}}\vect{F}(\vect{\theta}, \vect{x})$ under 
translation of $\vect{x}$ in the direction $\vect{\theta}$: 
$\vect{\mathcal{X}}\vect{F}(\vect{\theta}, E_{\theta}\vect{x})
=\vect{\mathcal{X}}\vect{F}(\vect{\theta}, \vect{x})$. We can 
use equation (\ref{Riesz1simp}) for inverting the X-ray transform.
 
\begin{prop} \label{prop6}
The inversion of X-ray transform intertwines the Curl 
$(\vect{\nabla}\bm{\times})$ and also the Divergence 
$(\vect{\nabla}\cdot)$, Gradient $(\vect{\nabla})$ 
and Laplacian $(\nabla^{2})$ operators. 

\end{prop}

  {\bf{Proof:}\,} 
We can easily show: $\vect{\nabla}_{\vect{x}}\bm{\times} 
\vect{\mathcal{I}}^{1}[\vect{F}](\vect{x})=\vect{\mathcal{I}}^{1}
[\vect{\nabla}\bm{\times}\vect{F}](\vect{x})$ for $\vect{F}(\vect{x})$ 
in the Schwartz class, using the Fourier transform of Riesz 
potential $\vect{{\mathcal{F}}}\{\vect{\mathcal{I}}^{\alpha}[\vect{F}]\}
(\vect{\xi})=|\vect{\xi}|^{-\alpha}\vect{{\mathcal{F}}}[\vect{F}]
(\vect{\xi})$, $\alpha<n=3$.\cite{NW, N, Markoe} We also have 
$\vect{\mathcal{I}}^{-\alpha}\vect{\mathcal{I}}^{\alpha}=1$.\cite{N}  
Thus 

\begin{eqnarray}
\vect{\mathcal{I}}^{-1} \Big[ \frac{1}{4\pi^{2}} 
\int_{S^{2}_{\vect{\theta}}} \vect{\nabla}_{\vect{x}}
\bm{\times} \vect{\mathcal{X}}\vect{F}
(\vect{\theta}, \vect{x}) 
d\Omega_{\vect{\theta}} \Big]
=\vect{\nabla}\bm{\times}\vect{F},
\end{eqnarray}

\noindent (\ref{Riesz1simp}). $\vect{\nabla}_{\vect{x}}\bm{\times} 
\vect{\mathcal{X}}\vect{F}(\vect{\theta}, \vect{x})$ satisfies 
(\ref{John}), if $\vect{\mathcal{X}}\vect{F}(\vect{\theta}, \vect{x})$ 
does. The inversion of X-ray transform also intertwines the Divergence 
$(\vect{\nabla}\cdot)$, Gradient $(\vect{\nabla})$ and hence the Laplacian 
$(\nabla^{2})$ operators with a similar reasoning. \hfill {\boldmath $\Box$}

\begin{prop} \label{prop7}
The inversion of X-ray transform intertwines the operator 
$(\vect{\nabla}\bm{\times})-\nu=0$ for constant $\nu$. 

\end{prop}

  {\bf{Proof:}\,} The transforms $\vect{\mathcal{I}}^{-1}$ 
and $\vect{\mathcal{X}}^{\dagger}$ are linear. \hfill {\boldmath $\Box$}

  Thus the Trkalian subclass (\ref{XrayTrkalian}) of functions 
$\vect{\mathcal{X}}\vect{F}(\vect{\theta}, \vect{x})$ satisfying 
John' s equation (\ref{John}) yields Trkalian fields in the physical 
space. 

  The propositions \ref{prop3}, \ref{prop4}, \ref{prop5}, \ref{prop6} 
and \ref{prop7} enable us to study Trkalian fields either in physical 
space or in the transform space. 

\begin{prop} \label{prop8} 
The Riesz potential for a Trkalian field is given by 

\begin{eqnarray} 
\vect{\mathcal{I}}^{\alpha}[\vect{F}_{\lambda}](\vect{x})
=(\lambda\nu)^{-\alpha}\vect{F}_{\lambda}(\vect{x}),
\end{eqnarray}

\noindent where $\alpha<3$.

\end{prop}

  {\bf{Proof:}\,} We are led to 

\begin{eqnarray}
\vect{{\mathcal{F}}}\{\vect{\mathcal{I}}^{\alpha}[\vect{F}]\}(\vect{\xi})
= \frac{1}{g}\frac{\delta(\xi-\lambda\nu)}{\xi^{\alpha+2}}
\vect{Q}_{\lambda}(\vect{u})s_{\lambda}(\lambda\nu\vect{u}),
\end{eqnarray}

\noindent substituting the inverse spherical Curl  transform 
(\ref{invRadtrf}) into: $\vect{{\mathcal{F}}}\{\vect{\mathcal{I}}^{\alpha}
[\vect{F}]\}(\vect{\xi})=|\vect{\xi}|^{-\alpha}\vect{{\mathcal{F}}}[\vect{F}]
(\vect{\xi})$, $\alpha<n=3$ and using\cite{Ml3} $\delta(\lambda\nu
\vect{\kappa}-\vect{\xi})=[\delta(\xi-\lambda\nu)/\xi^{2}]
\delta(\vect{\kappa}-\vect{u})$, [$\vect{\xi}=\xi\vect{u}$,  
$\xi=|\vect{\xi}|$, $|\vect{u}|=1$]. The result follows by 
inverting this. \hfill {\boldmath $\Box$}

  The Riesz potential (\ref{Riesz}) 
for $\vect{F}(\vect{x})= e^{i \vect{k}_{0}\cdot\vect{x}}\vect{F}_{0}$, 
[$\vect{k}_{0}=k_{0}\vect{\kappa}_{0}$, $k_{0}=\lambda\nu>0$] and the 
Lundquist field (\ref{Lundquist}) respectively lead to these fields 
themselves. We shall not present the details of these calculations here.

\begin{prop} \label{prop9}
The spherical mean of the X-ray (or Divergent beam) transform 
of a Trkalian field over all lines passing through a point 
yields the field at this point

\begin{eqnarray} \label{Sphmean}
\int_{S^{2}_{\vect{\theta}}} 
X\vect{F}_{\lambda}(\vect{\theta}, \vect{x}) 
d\Omega_{\vect{\theta}} 
=4\pi^{2}\frac{1}{\lambda\nu}
\vect{F}_{\lambda}(\vect{x}).
\end{eqnarray}

\end{prop}

  {\bf{Proof:}\,} The result follows from (\ref{Riesz1simp}) 
and Proposition \ref{prop8}. [Divergent beam transform: 
$\vect{\mathcal{X}}\vect{F}(\vect{\theta}, \vect{x})
=\vect{\mathcal{D}}\vect{F}(\vect{\theta}, \vect{x})
+\vect{\mathcal{D}}\vect{F}(-\vect{\theta}, \vect{x})$, 
the integrals of Divergent beam transforms in opposite 
directions are equal. See equation (\ref{SphmeanDiv}) 
below.] \hfill {\boldmath $\Box$}

  This provides us a simple inversion formula for Trkalian fields. 
This proposition is geometrically motivated by equation 
(\ref{GelfandGoncharoveqn}) below in Section \ref{GelfandGoncharovmethod}: 
Gelfand-Goncharov' s method. The underlying geometry is again based on the 
simplification in the same intricate geometric quantity: Hilbert transform 
of the derivative of Radon transform in the Moses basis.

  The spherical mean of the X-ray transform (\ref{Xsimp}) yields the field 
$\vect{F}(\vect{x})= e^{i \vect{k}_{0}\cdot\vect{x}}\vect{F}_{0}$, decomposing 
$\vect{\theta}$ into components which are respectively parallel and orthogonal 
to $\vect{\kappa}_{0}$. If we substitute (\ref{XFR1}) in (\ref{Sphmean}), 
we find (\ref{invRadtrf}) with a similar reasoning. 

  An analogous result for the Radon transform of Trkalian fields:
$\vect{\mathcal{I}}^{2}[\vect{F}_{\lambda}]=[1/(8\pi^{2})]
\vect{\mathcal{R}}^{\dagger}\vect{\mathcal{R}}[\vect{F}_{\lambda}]
=(1/\nu^{2})\vect{F}_{\lambda}$ also follows from equations (34, 40) 
in Ref. \onlinecite{KS}. 

\section{MATHEMATICAL METHODS OF TOMOGRAPHY}
\label{Tomographymethod}

  We shall use four basic mathematical approaches of tomography 
due to Grangeat,\cite{G} Smith,\cite{S1} Tuy\cite{HKT} 
and Gelfand\cite{GG}-Goncharov. These are based on relations 
of the X-ray and Divergent beam transforms to Hilbert transform 
of the derivative of Radon transform. Mathematically, these 
relations are outflow of the formula 

\begin{eqnarray} \label{essential}
\int_{S^{2}_{\vect{\theta}}} \vect{\mathcal{D}}\vect{F}(\vect{\theta}, \vect{x})
h(\vect{\theta}\cdot\vect{b}) d\Omega_{\vec{\theta}} 
=\int \vect{F}^{\mathcal{R}}(p, \vect{b}) h(p-\vect{b}\cdot\vect{x})dp,
\end{eqnarray}

\noindent essentially obtained in Ref. \onlinecite{HSSW}.\cite{NW} 
Here we have introduced the normalization: $\vect{\beta}=\beta\vect{b}$, 
$\beta=|\vect{\beta}|$, $s=\beta p$, $\vect{F}^{\mathcal{R}}(s, \vect{\beta})
=(1/\beta)\vect{F}^{\mathcal{R}}(p, \vect{b})$ for the sake of our 
conventions.\cite{NW} The distribution $h$ satisfies $h(ap)=(1/a^{2})h(p)$, 
$a>0$. 

   We provide the mathematical derivations of these formulas for the sake 
of a self-contained presentation with a unique, consistent convention. See 
Appendix \ref{Identity} for a derivation of (\ref{essential}) following 
Ref. \onlinecite{RK}, p. 276 and also Ref. \onlinecite{CD}. For a unified 
discussion of these approaches see Refs. \onlinecite{CD}, 
\onlinecite{XY}. 

  For the attentive reader, we remark that history has followed a path
different from the mathematically logical one. These mathematical methods 
were developed independent of equation (\ref{essential}), only using 
practical tomographical ideas. We shall adopt a mathematical approach 
rather than a tomographical implementation.

  The first purpose of this section is to provide a unified geometric 
motivation and intuition for the previous results in Sections \ref{Xraytrf} 
and \ref{Johneqn}. The second purpose is to present a discussion of these
mathematical methods with a view towards tomographical studies of Trkalian 
field models in nature. For this purpose, we shall study these mathematical 
methods using Trkalian fields. Especially for the sake of the second 
purpose and also for a clean presentation, this discussion was postponed 
until the direct (but unmotivated) discussion of the geometry of X-ray 
and Divergent beam transforms in Sections \ref{Xraytrf} and 
\ref{Johneqn} finished. 

  These methods basically make use of the Radon inversion (for 
tomographical reconstruction). They lead us to new inversion 
formulas for the X-ray and Divergent beam transforms of Trkalian 
fields with a view towards tomographical applications. They also  
provide the geometric motivation underlying the interrelations 
of the transforms mentioned.

\subsection{Grangeat' s method} 
\label{Grangeatformula}

  We obtain Grangeat' s formula\cite{G} 

\begin{eqnarray} \label{Grangeat}
\frac{\partial}{\partial p}\vect{F}^{\mathcal{R}}
(p, \vect{\kappa}) \Big|_{p=\vect{\kappa}\cdot\vect{x}} 
=-\int_{S^{2}_{\vect{\theta}}}
\vect{\mathcal{D}}\vect{F}(\vect{\theta}, \vect{x})
\delta^{\prime} (\vect{\kappa}\cdot\vect{\theta}) 
d\Omega_{\theta},
\end{eqnarray}

\noindent using $h(p)=\delta^{\prime}(p)$ in (\ref{essential}). 
This is a componentwise\cite{NW} generalization of the formula 
for scalar fields. For a derivation of this formula see Appendix 
\ref{ProofGrangeat' s}.\cite{CD, TZG} Thus we find

\begin{eqnarray} \label{Grangeat2}
\frac{\partial}{\partial p}\vect{F}^{\mathcal{R}}
(p, \vect{\kappa}) \Big|_{p=\vect{\kappa}\cdot\vect{x}} 
= \int_{S^{2}_{\vect{\theta}}\cap\kappa^{\perp}} 
\frac{\partial}{\partial \vect{\kappa}} 
\vect{\mathcal{D}}\vect{F}(\vect{\theta}, \vect{x}) d\vect{\theta},
\end{eqnarray}

\noindent using the identity (\ref{identity1}).\cite{NW} 
Here ${\partial}/{\partial \vect{\kappa}}$ denotes directional 
derivative along $\vect{\kappa}$.

  We find 

\begin{eqnarray} \label{GrangeatTrkalianfield}
\vect{F}^{\mathcal{R}} (\vect{\kappa}\cdot\vect{x}, \vect{\kappa}) 
&=& -\frac{1}{\nu}\vect{\kappa} \bm{\times} \int_{S^{2}_{\vect{\theta}}}
\vect{\mathcal{D}}\vect{F}(\vect{\theta}, \vect{x})
\delta^{\prime} (\vect{\kappa}\cdot\vect{\theta}) 
d\Omega_{\theta},
\end{eqnarray}

\noindent for a Trkalian field using 
(\ref{transelfdualfield}, \ref{Grangeat}). 
We also have $\vect{\kappa} \cdot \int_{S^{2}_{\vect{\theta}}}
\vect{\mathcal{D}}\vect{F}(\vect{\theta}, \vect{x})
\delta^{\prime} (\vect{\kappa}\cdot\vect{\theta}) 
d\Omega_{\theta} =0$, (\ref{transelfdualfield2}, \ref{Grangeat}).
The equation (\ref{GrangeatTrkalianfield}) leads to

\begin{eqnarray} \label{invY}
\vect{F}^{\mathcal{R}} (\vect{\kappa}\cdot\vect{x}, \vect{\kappa}) 
= -\frac{1}{2}\frac{1}{\nu}\vect{\kappa} \bm{\times} 
\int_{S^{2}_{\vect{\theta}}} \vect{\mathcal{Y}}\vect{F}(\vect{\theta}, \vect{x})
\delta^{\prime} (\vect{\kappa}\cdot\vect{\theta}) 
d\Omega_{\theta},
\end{eqnarray}

\noindent using $\vect{\theta}\longrightarrow -\vect{\theta}$ 
and rearranging.

  We can check (\ref{Grangeat}) for $\vect{F}(\vect{x})=e^{i \vect{k}_{0}
\cdot\vect{x}}\vect{F}_{0}$, [$\vect{k}_{0}=k_{0}\vect{\kappa}_{0}$, 
$k_{0}>0$]. See Appendix \ref{GrangeatTrkalian}. If we substitute 
$\vect{\mathcal{Y}}\vect{F}_{\lambda}(\vect{\theta}, \vect{x})$, 
(\ref{Ytrfsimp}) in (\ref{invY}), we are similarly led to 
$\vect{F}^{\mathcal{R}}(\vect{\kappa}\cdot\vect{x}, \vect{\kappa})$, 
(\ref{Radsimp}).

  If we substitute (\ref{GrangeatTrkalianfield}) [or (\ref{invY})] 
into the first line of (\ref{invRadtrf}), we find 

\begin{eqnarray} \label{Grangeatinv}
\vect{F}(\vect{x}) 
= -\frac{1}{8\pi^{2}} \nu \int_{S^{2}_{\vect{\kappa}}} 
\vect{\kappa} \bm{\times} \int_{S^{2}_{\vect{\theta}}}
\vect{\mathcal{D}}\vect{F}(\vect{\theta}, \vect{x})
\delta^{\prime} (\vect{\kappa}\cdot\vect{\theta}) 
d\Omega_{\theta} d\Omega_{\vect{\kappa}}.
\end{eqnarray}

\noindent This leads to

\begin{eqnarray} \label{GrangetTrkalianinv}
\vect{F}(\vect{x}) 
= \pm\frac{1}{4\pi} \nu \int_{S^{2}_{\vect{\theta}}} 
\vect{\theta} \bm{\times} 
\vect{\mathcal{D}}\vect{F}(\pm\vect{\theta}, \vect{x}) d\Omega_{\vect{\theta}},
\end{eqnarray} 

\noindent interchanging the order of integrations 
and using the identity (\ref{identity1}), 
(also $\vect{\theta}\longrightarrow -\vect{\theta}$). 

  This is another simple, direct inversion (reconstruction) formula 
for the Divergent beam transform of Trkalian fields. Note that the 
simplification is basically due to the eigenvalue equation for the 
Radon transform of Trkalian fields. We shall not discuss the tomographical 
implementation of this inversion formula.

  We can verify the formula (\ref{GrangetTrkalianinv}) using
(\ref{DFR1},\ref {DFR2}). See Appendix \ref{GrangetTrkalianinversion}. 
If we substitute $\vect{\mathcal{D}}\vect{F}(\pm\vect{\theta}, \vect{x})$, 
(\ref{Dsimp}) in (\ref{GrangetTrkalianinv}), we similarly find 
$\vect{F}(\vect{x})= e^{i \vect{k}_{0}\cdot\vect{x}}\vect{F}_{0}$, 
($\vect{\kappa}\longrightarrow\vect{\kappa}_{0}$).

\subsection{Smith' s method} 
\label{Smithapproach}

   Smith' s formula\cite{S1} can be written in various 
ways.\cite{NW, CD, YL} We shall follow Ref. \onlinecite{NW}, p. 24 
(correcting misprints there) and Ref. \onlinecite{YL}. We extend the 
X-ray transform (\ref{xraytrf}) as a function 

\begin{eqnarray} \label{extxraytrf}
\vect{\mathfrak{g}}\vect{F}(\vect{\alpha}, \vect{x}) 
= \int_{-\infty}^{\infty} 
\vect{F}(\vect{x}+t\vect{\alpha})dt
= \frac{1}{\alpha} 
\vect{\mathcal{X}}\vect{F}(\vect{\theta}, \vect{x}), 
\end{eqnarray}

\noindent homogeneous of degree $-1$, using a non-unit vector 
$\vect{\alpha}=\alpha\vect{\theta}$, $\alpha=|\vect{\alpha}|$, 
$s=\alpha t$. Then its Fourier transform is 

\begin{eqnarray} \label{extxrayFourier}
\vect{\mathcal{G}}\vect{F}(\vect{\beta}, \vect{x}) 
&=& \vect{\mathcal{F}}[\vect{\mathfrak{g}}\vect{F}
(\vect{\alpha}, \vect{x})](\vect{\beta}, \vect{x})
= \frac{1}{(2\pi)^{3/2}} \frac{1}{\beta^{2}} 
\int_{S^{2}_{\vect{\theta}}} \vect{\mathcal{D}}\vect{F}
(\vect{\theta}, \vect{x})h(\vect{\theta}\cdot\vect{b}) 
d\Omega_{\vect{\theta}}. 
\end{eqnarray}

\noindent See Appendix \ref{Smith}. Here

\begin{eqnarray} \label{Smithdist}
h(p) = \int_{\alpha=-\infty}^{\infty} |\alpha|e^{-i\alpha p}d\alpha, 
\hspace*{5mm} (p=\vect{\theta}\cdot\vect{b})
\end{eqnarray}

\noindent where $h(ap)=(1/a^{2})h(p)$, $a>0$ and $h(-p)=h(p)$.

  Thus, we find

\begin{eqnarray} \label{extxrayFourier2}
\vect{\mathcal{G}}\vect{F}(\vect{\beta}, \vect{x}) 
= \frac{1}{(2\pi)^{3/2}} \frac{1}{\beta^{2}}
\int \vect{F}^{\mathcal{R}}(p, \vect{b}) 
h(p-\vect{b}\cdot\vect{x})dp,
\end{eqnarray}

\noindent using (\ref{essential}). 
This leads to Smith' s formula

\begin{eqnarray} \label{extxrayFourier21}
\vect{\mathcal{G}}\vect{F}(\vect{\beta}, \vect{x}) 
= \frac{1}{(2\pi)^{1/2}} \frac{1}{\beta^{2}}
[\vect{\mathcal{H}}\partial_{p} \vect{F}^{\mathcal{R}}(p, \vect{b})] 
(\vect{b}\cdot\vect{x}, \vect{b}),
\end{eqnarray}

\noindent for vector fields, using the identity

\begin{eqnarray} \label{identity10}
\int \vect{F}^{\mathcal{R}}(p, \vect{b}) h(p-p_{0}) dp 
=2\pi [\vect{\mathcal{H}}\partial_{p}\vect{F}^{\mathcal{R}}(p, \vect{b})]
(p_{0}, \vect{b}), \hspace*{5mm} (p_{0}=\vect{b}\cdot\vect{x}).
\end{eqnarray} 

\noindent Here $\vect{\mathcal{H}}$ is the Hilbert transform\cite{NW} 
defined (on $\mathbb{R}$) by the principal-value integral

\begin{eqnarray}
\vect{\mathcal{H}}[r(y)](x)=\frac{1}{\pi} \int \frac{r(y)}{x-y} dy.
\end{eqnarray}

\noindent We can prove this identity noting 
$h(p-p_{0})=2 \partial_{p} [1/(p-p_{0})]$. 
We also have 

\begin{eqnarray} \label{Hilbertidentity}
[\vect{\mathcal{H}}\partial_{y} r(y)](x)
=\frac{1}{\pi} 
\int \frac{\partial_{y}r(y)}{x-y} dy
=-\frac{1}{\pi} 
\int \frac{r(y)}{(x-y)^{2}} dy. 
\end{eqnarray}

  If we invert (\ref{extxrayFourier}) using (\ref{extxrayFourier21}) 
and (\ref{extxraytrf}), we find 

\begin{eqnarray} \label{Smithmine}
\vect{\mathcal{X}}\vect{F}(\vect{\theta}, \vect{x}) 
&=& \alpha \vect{\mathcal{F}}^{-1}[\vect{\mathcal{G}}
\vect{F}(\vect{\beta}, \vect{x})](\vect{\alpha}, \vect{x}) 
= \frac{1}{4\pi} \int_{S^{2}_{\vect{b}}}
[\vect{\mathcal{H}}\partial_{p} \vect{F}^{\mathcal{R}}(p, \vect{b})] 
(\vect{b}\cdot\vect{x}, \vect{b}) \delta(\vect{b}\cdot\vect{\theta}) 
d\Omega_{\vect{b}}, 
\end{eqnarray}

\noindent using $\vect{\beta}=\beta\vect{b}$ $\Rightarrow$ 
$d^{3}\beta=\beta^{2} d\beta d\Omega_{\vect{b}}$, $\vect{\alpha}
=\alpha\vect{\theta}$ and $[\vect{\mathcal{H}}\partial_{p} 
\vect{F}^{\mathcal{R}}(p, \vect{b})] (\vect{b}\cdot\vect{x}, \vect{b})]$ 
is even under $\vect{b}\longrightarrow -\vect{b}$. 
Hence, the X-ray transform is in the form of a Minkowski-Funk 
transform: $X\vect{F}(\vect{\theta}, \vect{x})=[1/(2\pi^{1/2})]
\vect{\mathcal{U}}^{0}\{[\vect{\mathcal{H}}\partial_{p} 
\vect{F}^{\mathcal{R}}(p, \vect{b})] (\vect{b}\cdot\vect{x}, \vect{b})\}
(\vect{\theta}, \vect{x})$, (\ref{UV}) of $\vect{\mathcal{H}}\partial_{p} 
\vect{F}^{\mathcal{R}}(p, \vect{b})$.

  For a Trkalian field (\ref{transelfdualfield}), 
$\vect{\mathcal{G}}\vect{F}(\vect{\beta}, \vect{x})$, 
(\ref{extxrayFourier2}) satisfies: $\vect{\nabla}_{\vect{x}}
\bm{\times}\vect{\mathcal{G}}\vect{F}(\vect{\beta}, \vect{x})
-\nu \vect{\mathcal{G}}\vect{F}(\vect{\beta}, \vect{x})=0$, 
[equivalently from (\ref{extxrayFourier}, \ref{extxraytrf}) 
and (\ref{XrayTrkalian})]. The equation (\ref{Smithmine}) 
reduces to

\begin{eqnarray} \label{SmithTrkalian}
\vect{\mathcal{X}}\vect{F}(\vect{\theta}, \vect{x}) 
= -\frac{1}{4\pi} \nu \int_{S^{2}_{\vect{b}}}
\vect{b} \bm{\times}
[\vect{\mathcal{H}} \vect{F}^{\mathcal{R}}(p, \vect{b})] 
(\vect{b}\cdot\vect{x}, \vect{b}) \delta(\vect{b}\cdot\vect{\theta}) 
d\Omega_{\vect{b}},
\end{eqnarray}

\noindent using  (\ref{transelfdualfield2}). We find 

\begin{eqnarray} \label{identity7}
[\vect{\mathcal{H}}\partial_{p} \vect{F}_{\lambda}^{\mathcal{R}}(p, \vect{b})] 
(p_{0}, \vect{b}) 
= -\nu \vect{b} \bm{\times} 
[\vect{\mathcal{H}} \vect{F}_{\lambda}^{\mathcal{R}}(p, \vect{b})] 
(p_{0}, \vect{b})=\lambda\nu \vect{F}_{\lambda}^{\mathcal{R}}(p_{0}, \vect{b}),
\hspace*{5mm} (p_{0}=\vect{b}\cdot\vect{x})
\end{eqnarray}

\noindent using the Moses basis, (\ref{RadonMoses}).
Thus (\ref{Smithmine}), (\ref{SmithTrkalian}) reduce to 
(\ref{XFR1}), ($\vect{b} \longrightarrow \vect{\kappa}$) 
which can be directly proven as in Proposition \ref{prop1} 
in Section \ref{Xraytrf}. This can be inverted, for example 
using (\ref{InvFunkSem}). The simplification here is basically 
due to the eigenvalue equation and the Hilbert transform of the 
derivative for the Radon transform of Trkalian fields in the Moses basis.

  Smith' s inversion method\cite{YL} 
makes use of the intermediate function

\begin{eqnarray} \label{intermediate}
\vect{K}(\omega, \vect{\beta})
&=& \int \vect{{\mathcal{F}}}[\vect{F}(\vect{x})]
(\tau\vect{\beta})
|\tau|e^{i\omega\tau}d\tau
= \frac{1}{(2\pi)^{3/2}} \int 
\vect{F}^{\mathcal{R}}(s, \vect{\beta}) 
h(s-\omega) ds,
\end{eqnarray}

\noindent $\vect{\mathcal{G}}\vect{F}(\vect{\beta}, \vect{x}) 
=\vect{K}(\vect{\beta}\cdot\vect{x}, \vect{\beta})$, (see also 
Ref. \onlinecite{CD}). Here we have used the Fourier slice 
theorem: $\vect{{\mathcal{F}}}[\vect{F}^{R}(s, \vect{\beta})]
(\tau, \vect{\beta})=2\pi \vect{{\mathcal{F}}}[\vect{F}(\vect{x})]
(\tau\beta)$.\cite{KS} This leads to $\vect{K}(\omega, \vect{\beta})
=[1/(2\pi)^{1/2}] (1/\beta^{2}) [\vect{\mathcal{H}}\partial_{p} 
\vect{F}^{\mathcal{R}} (p, \vect{b})] (\omega/\beta, \vect{b})$ 
using (\ref{identity10}) and $\vect{K}_{\lambda}(\omega, \vect{\beta}) 
=[1/(2\pi)^{1/2}](1/\beta^{2})\lambda\nu\vect{F}^{\mathcal{R}}_{\lambda} 
(\omega/\beta, \vect{b})$ using (\ref{identity7}), [or appropriately 
using (\ref{invRadtrf}) in (\ref{intermediate})] for a Trkalian field. 
Hence $\vect{K}_{\lambda}(\vect{\beta}\cdot\vect{x}, \vect{\beta})
=[1/(2\pi)^{1/2}] (1/\beta^{2})\lambda\nu \vect{F}^{\mathcal{R}}_{\lambda} 
(\vect{b}\cdot\vect{x}, \vect{b})$. Then the inversion formula\cite{YL} 

\begin{eqnarray} \label{Smithinversion}
\vect{F}_{\lambda}(\vect{x}) 
=\frac{1}{(2\pi)^{5/2}} \int \int \vect{K}_{\lambda}(\omega, \vect{\beta}) 
e^{-i(\omega-\vect{\beta}\cdot\vect{x})} d\omega d^{3}\beta,
\end{eqnarray}

\noindent can be expressed in terms of the Radon transform of the field. 
This reduces to (\ref{invRadtrf}). We shall not discuss its tomographical 
implementation.

\subsection{Tuy' s method} 
\label{Tuymethod}

  We shall follow Ref. \onlinecite{CD} for Tuy' s approach.\cite{HKT}
We extend the Divergent beam transform (\ref{draytrf}) as a function   

\begin{eqnarray} \label{extdraytrf}
\vect{\mathfrak{g}}\vect{F}(\vect{\alpha}, \vect{x}) 
= \int_{0}^{\infty} 
\vect{F}(\vect{x}+t\vect{\alpha})dt
= \frac{1}{\alpha} 
\vect{\mathcal{D}}\vect{F}(\vect{\theta}, \vect{x}),
\end{eqnarray}

\noindent homogeneous of degree $-1$, using a non-unit vector 
$\vect{\alpha}=\alpha\vect{\theta}$, $\alpha=|\vect{\alpha}|$, 
$s=\alpha t$. Then its Fourier transform is 

\begin{eqnarray} \label{extdbeamFourier}
\vect{\mathcal{G}}\vect{F}(\vect{\beta}, \vect{x}) 
= \vect{\mathcal{F}}[\vect{\mathfrak{g}}
\vect{F}(\vect{\alpha}, \vect{x})](\vect{\beta}, \vect{x}) 
= \frac{1}{(2\pi)^{3/2}} \frac{1}{\beta^{2}}
\int_{S^{2}_{\vect{\theta}}} \vect{\mathcal{D}}
\vect{F}(\vect{\theta}, \vect{x})
f(\vect{\theta}\cdot\vect{b}) d\Omega_{\vect{\theta}}, 
\end{eqnarray}

\noindent See Appendix \ref{Tuy}.
Here $f(p)=2\pi i \partial_{p} \delta^{-}(p)$, 
($p=\vect{\theta}\cdot\vect{b}$) where
$f(ap)=(1/a^{2})f(p)$, $a>0$ and 
$f(-p)=-2\pi i \partial_{p} \delta^{+}(p)$. 

  Thus we find 

\begin{eqnarray} \label{extdbeamFourier2}
\vect{\mathcal{G}}\vect{F}(\vect{\beta}, \vect{x})  
=\frac{1}{(2\pi)^{3/2}}\frac{1}{\beta^{2}} 
\int \vect{F}^{\mathcal{R}}(p, \vect{b}) 
f(p-\vect{b}\cdot\vect{x}) dp,
\end{eqnarray}

\noindent using (\ref{essential}). This leads to

\begin{eqnarray} \label{extdbeamFourier21}
\vect{\mathcal{G}}\vect{F}(\vect{\beta}, \vect{x})  
=\frac{\pi}{(2\pi)^{3/2}}\frac{1}{\beta^{2}} 
\Big\{ \left[ \vect{\mathcal{H}}\partial_{p} 
\vect{F}^{\mathcal{R}}(p, \vect{b}) \right] (\vect{b}\cdot\vect{x}, \vect{b})
-i \left[ \partial_{p} \vect{F}^{\mathcal{R}}(p, \vect{b}) \right] 
(\vect{b}\cdot\vect{x}, \vect{b}) \Big\},
\end{eqnarray}

\noindent using the identity

\begin{eqnarray} \label{I}
\int \vect{F}^{\mathcal{R}}(p, \vect{b}) f(p-p_{0}) dp, 
&=& \pi \Big\{ \left[ (\vect{\mathcal{H}}-i)\partial_{p} \right] 
\vect{F}^{\mathcal{R}}(p, \vect{b}) \Big\} (p_{0}, \vect{b}),
\hspace*{5mm} (p_{0}=\vect{b}\cdot\vect{x}).
\end{eqnarray}

\noindent We can prove this identity noting 
$f(p-p_{0})=2\pi i \partial_{p} \delta^{-}(p-p_{0})$,
upon simple manipulations. 

  If we invert (\ref{extdbeamFourier}) using (\ref{extdbeamFourier21})
and (\ref{extdraytrf}), we find

\begin{eqnarray} \label{extdbeamtrfFourierinv}
\vect{\mathcal{D}}\vect{F}(\vect{\theta}, \vect{x}) 
&=& \alpha \vect{\mathcal{F}}^{-1}[\vect{\mathcal{G}}
\vect{F}(\vect{\beta}, \vect{x})](\vect{\alpha}, \vect{x}) 
= \frac{\pi}{(2\pi)^{2}} \int_{S_{\vect{b}}^{2}} 
\Big\{ \left[ (\vect{\mathcal{H}}-i)\partial_{p} \right] 
\vect{F}^{\mathcal{R}}(p, \vect{b}) \Big\} 
(\vect{b}\cdot\vect{x}, \vect{b})
\delta^{+}(\vect{b}\cdot\vect{\theta}) d\Omega_{\vect{b}}, 
\end{eqnarray}

\noindent using $\vect{\beta}=\beta\vect{b}$ $\Rightarrow$ 
$d^{3}\beta=\beta^{2} d\beta d\Omega_{\vect{b}}$, 
$\vect{\alpha}=\alpha\vect{\theta}$ and 
$\vect{{\mathcal{F}}}[H(-\beta^{\prime})](p)
=\sqrt{2\pi}\delta^{+}(p)$, $\beta^{\prime}=\alpha\beta$. Hence, the 
Divergent beam transform is in the form of the integral transform:
$\vect{\mathcal{D}}\vect{F}(\vect{\theta}, \vect{x})=[1/(4\pi^{1/2})] 
\vect{\mathcal{A}}^{0} \big\{ \{ \left[ (\vect{\mathcal{H}}-i)
\partial_{p} \right] \vect{F}^{\mathcal{R}}(p, \vect{b}) \} 
(\vect{b}\cdot\vect{x}, \vect{b}) \big\} 
(\vect{\theta}, \vect{x})$, (\ref{UV}) of 
$\left[ (\vect{\mathcal{H}}-i)\partial_{p} \right] 
\vect{F}^{\mathcal{R}}(p, \vect{b})$.
An easy check reveals $\left[ \partial_{p}\vect{F}^{\mathcal{R}}
(p, \vect{b}) \right] (\vect{b}\cdot\vect{x}, \vect{b})$ is odd 
under $\vect{b}\longrightarrow -\vect{b}$ while 
$\left[ \vect{\mathcal{H}}\partial_{p} \vect{F}^{\mathcal{R}}(p, \vect{b}) 
\right] (\vect{b}\cdot\vect{x}, \vect{b})$ is even.
Therefore only the even terms survive the integration.
Thus we find

\begin{eqnarray} \label{Tuymine}
\vect{\mathcal{D}}\vect{F}(\vect{\theta}, \vect{x}) 
= \frac{1}{2} \vect{\mathcal{X}}\vect{F}(\vect{\theta}, \vect{x}) 
+\frac{1}{2(2\pi)^{2}} \int_{S_{\vect{b}}^{2}} 
\left[ \partial_{p} \vect{F}^{\mathcal{R}}(p, \vect{b}) \right] 
(\vect{b}\cdot\vect{x}, \vect{b})
\frac{1}{\vect{b}\cdot\vect{\theta}} d\Omega_{\vect{b}},
\end{eqnarray}

\noindent using (\ref{Smithmine}). 
The second term is associated 
with the difference

\begin{eqnarray} \label{Tuymine3}
\vect{\mathcal{Y}}\vect{F}(\vect{\theta}, \vect{x}) 
=\frac{1}{(2\pi)^{2}} \int_{S_{\vect{b}}^{2}} 
\left[ \partial_{p} \vect{F}^{\mathcal{R}}(p, \vect{b}) \right] 
(\vect{b}\cdot\vect{x}, \vect{b})
\frac{1}{\vect{b}\cdot\vect{\theta}} d\Omega_{\vect{b}},
\end{eqnarray}

\noindent (\ref{Y1}).

  For a Trkalian field (\ref{transelfdualfield}), 
$\vect{\mathcal{G}}\vect{F}(\vect{\beta}, \vect{x})$, 
(\ref{extdbeamFourier2}) satisfies: $\vect{\nabla}_{\vect{x}}
\bm{\times} \vect{\mathcal{G}}\vect{F}(\vect{\beta}, \vect{x})
-\nu \vect{\mathcal{G}}\vect{F}(\vect{\beta}, \vect{x})=0$,
[equivalently from (\ref{extdbeamFourier}, \ref{extdraytrf}) and 
(\ref{XrayTrkalian})]. The equation (\ref{Tuymine}) reduces to

\begin{eqnarray} \label{TuyTrkalian}
\vect{\mathcal{D}}\vect{F}(\vect{\theta}, \vect{x}) 
= \frac{1}{2} \vect{\mathcal{X}}\vect{F}(\vect{\theta}, \vect{x}) 
-\frac{1}{2(2\pi)^{2}} \nu \int_{S_{\vect{b}}^{2}} 
\vect{b} \bm{\times} \vect{F}^{\mathcal{R}}
(\vect{b}\cdot\vect{x}, \vect{b})
\frac{1}{\vect{b}\cdot\vect{\theta}} d\Omega_{\vect{b}},
\end{eqnarray}

\noindent using (\ref{transelfdualfield2}). 
This leads to 

\begin{eqnarray} \label{Tuymine4}
\vect{\mathcal{Y}}\vect{F}(\vect{\theta}, \vect{x}) 
=-\frac{1}{(2\pi)^{2}} \nu \int_{S_{\vect{b}}^{2}} 
\vect{b} \bm{\times} \vect{F}^{\mathcal{R}}(\vect{b}\cdot\vect{x}, \vect{b})
\frac{1}{\vect{b}\cdot\vect{\theta}} d\Omega_{\vect{b}}.
\end{eqnarray}

\noindent The equations (\ref{Tuymine}, \ref{TuyTrkalian}) reduce 
to (\ref{DFR1}), ($\vect{b} \longrightarrow \vect{\kappa}$) 
using the Moses basis (\ref{RadonMoses}), [also (\ref{XFR1})] 
which can be directly proven as in Proposition \ref{prop2} 
in Section \ref{Xraytrf}. Similarly (\ref{Tuymine3}, \ref{Tuymine4}) 
reduce to (\ref{Y2}). Note 

\begin{eqnarray} \label{Tuycrucsimp}
\{ \left[ (\vect{\mathcal{H}}-i)\partial_{p} 
\right] \vect{F}_{\lambda}^{\mathcal{R}}(p, \vect{b}) \} 
(p_{0}, \vect{b}) 
&=& -\nu \vect{b} \bm{\times} 
[(\vect{\mathcal{H}}-i) 
\vect{F}_{\lambda}^{\mathcal{R}}(p, \vect{b})] 
(p_{0}, \vect{b}) \\ 
&=& 2(2\pi)^{1/2}(1/g)(1/\lambda\nu)
e^{i\lambda\nu p_{0}}\vect{Q}_{\lambda}(\vect{b})
s_{\lambda}(\lambda\nu\vect{b}),  
\hspace*{5mm} (p_{0}=\vect{b}\cdot\vect{x}) 
\nonumber
\end{eqnarray}

\noindent in the Moses basis. The simplification here is also 
due to the eigenvalue equation and the Hilbert transform of the 
derivative for the Radon transform of Trkalian fields in the Moses basis.

  It is straightforward to find the difference for 
Lundquist field ($\lambda=1$)

\begin{eqnarray} \label{Lundquistdiftrf}
\vect{\mathcal{Y}}\vect{F}_{L(\lambda=1)}(\vect{\theta}, \vect{x})
&=& -2 F_{0}\frac{1}{\nu}\frac{1}{v_{r}}
\left\{ 2\sum_{k=1}^{\infty} \sin [2k(\theta-\phi)]J_{2k}(\nu r)
\vect{e}_{r}(\theta)-J_{0}(\nu r)\vect{e}_{\theta} \right. \\
& & \hspace{68mm} + \left. 2\sum_{k=0}^{\infty} 
\cos [(2k+1)(\theta-\phi)]J_{2k+1}(\nu r)\vect{e}_{z} \right\}, 
\nonumber
\end{eqnarray}

\noindent substituting the Radon transform (\ref{RadonLundquist}) 
into (\ref{Tuymine3}). See Appendix \ref{LundquistDivbeamTuy}.
Then, we find the Divergent beam transform 

\begin{eqnarray} \label{DivbeamLundquist}
\vect{\mathcal{D}}\vect{F}_{L(\lambda=1)}(\vect{\theta}, \vect{x}) 
&=& F_{0}\frac{1}{\nu}\frac{1}{v_{r}}
\left\{ -2\sum_{n=1}^{\infty} (-1)^{n} \sin[n(\theta-\phi)] J_{n}(\nu r)
\vect{e}_{r}(\theta)+J_{0}(\nu r)\vect{e}_{\theta} \right. \\ 
& & \hspace*{70mm} + \left. \left[ J_{0}(\nu r)+ 
2\sum_{n=1}^{\infty} (-1)^{n} \cos[n(\theta-\phi)] J_{n}(\nu r) \right] 
\vect{e}_{z} \right\}, \nonumber
\end{eqnarray}

\noindent of the Lundquist field using (\ref{expansion}) 
in (\ref{XLundquist}) and then in (\ref{Tuymine}). See 
p. 23 and p. 538 in Ref. \onlinecite{GNW} for these series. 

  Tuy' s inversion method\cite{CD} makes use 
of the intermediate function

\begin{eqnarray} \label{Tuyintermediate}
\vect{K}(\omega, \vect{\beta})
&=& \frac{1}{2} \int \vect{{\mathcal{F}}}[\vect{F}(\vect{x})]
(\tau\vect{\beta})
(|\tau|+\tau) e^{i\omega\tau} d\tau
= \frac{1}{(2\pi)^{3/2}} \int \vect{F}^{\mathcal{R}}(s, \vect{\beta}) 
f(s-\omega) ds,
\end{eqnarray}

\noindent $\vect{\mathcal{G}}\vect{F}(\vect{\beta}, \vect{x}) 
=\vect{K}(\vect{\beta}\cdot\vect{x}, \vect{\beta})$. 
This yields $\vect{K}(\omega, \vect{\beta})
=[1/(2\pi)^{3/2}](1/\beta^{2})I(\omega/\beta, \vect{b})$
where $I(p, \vect{b})=\pi \left[ (\vect{\mathcal{H}}-i)
\partial_{q} \vect{F}^{\mathcal{R}}(q, \vect{b}) \right]$ 
$(p, \vect{b})$, [see (\ref{extdbeamFourier21})]. 
For a Trkalian field in the Moses basis
$\vect{\mathcal{G}}\vect{F}(\vect{\beta}, \vect{x}) 
=\vect{K}(\vect{\beta}\cdot\vect{x}, \vect{\beta})
=[1/(2\pi)^{3/2}](1/\beta^{2})I(\vect{b}\cdot\vect{x}, \vect{b})
=[\pi/(2\pi)^{3/2}](1/\beta^{2})\{ \left[ (\vect{\mathcal{H}}-i)
\partial_{p} \right] \vect{F}^{\mathcal{R}}(p, \vect{b}) \} 
(\vect{b}\cdot\vect{x}, \vect{b})=(1/g)(1/\lambda\nu)(1/\beta^{2})
e^{i\lambda\nu\vect{b}\cdot\vect{x}}\vect{Q}_{\lambda}(\vect{b})s_{\lambda}
(\lambda\nu\vect{b})$. The inversion formula\cite{CD} 

\begin{eqnarray} \label{TuyinversionI}
\vect{F}(\vect{x}) 
=-\frac{1}{(2\pi)^{3}} \frac{1}{i} \int_{S^{2}_{\vect{b}}} 
\int I(p, \vect{b}) \delta^{\prime}(p-\vect{b}\cdot\vect{x}) 
dp d\Omega_{\vect{b}},
\end{eqnarray}

\noindent can be expressed in terms of the spherical Curl transform 
of the field. This reduces to (\ref{invRadtrf}). We shall not discuss 
its tomographical implementation.

\subsection{Gelfand-Goncharov' s method} 
\label{GelfandGoncharovmethod}

  If we use $h(p)=1/p^{2}$, $p=\vect{\theta}\cdot\vect{b}$, 
then (\ref{essential}) leads\cite{GG} to

\begin{eqnarray} \label{GelfandGoncharoveqn}
\left[ \vect{\mathcal{H}}\partial_{p} 
\vect{F}^{\mathcal{R}}(p, \vect{b}) \right] 
(\vect{b}\cdot\vect{x}, \vect{b}) 
=-\frac{1}{\pi} \int_{S^{2}_{\vect{\theta}}} 
\frac{\vect{\mathcal{D}}\vect{F}(\vect{\theta}, \vect{x})}
{(\vect{\theta}\cdot\vect{b})^{2}} 
d\Omega_{\vect{\theta}}, 
\end{eqnarray}

\noindent see (\ref{Hilbertidentity}). This can also be inferred from 
(\ref{extxrayFourier}, \ref{extxrayFourier21}) and (\ref{identity3}).

  For a Trkalian field (\ref{transelfdualfield2}), 
this yields 

\begin{eqnarray} \label{GelfandGoncharoveTrkalqn}
\vect{b} \bm{\times} \left[ \vect{\mathcal{H}}
\vect{F}^{\mathcal{R}}(p, \vect{b}) \right] 
(\vect{b}\cdot\vect{x}, \vect{b}) 
=\frac{1}{\pi} \frac{1}{\nu} \int_{S^{2}_{\vect{\theta}}} 
\frac{\vect{\mathcal{D}}\vect{F}(\vect{\theta}, \vect{x})}
{(\vect{\theta}\cdot\vect{b})^{2}} 
d\Omega_{\vect{\theta}}.
\end{eqnarray}

\noindent This reduces to

\begin{eqnarray} \label{GelfandGoncharovMoses}
\vect{F}_{\lambda}^{\mathcal{R}} (\vect{b}\cdot\vect{x}, \vect{b}) 
= -\frac{1}{\pi} \frac{1}{\lambda\nu} \int_{S^{2}_{\vect{\theta}}} 
\frac{\vect{\mathcal{D}}\vect{F}_{\lambda} (\vect{\theta}, \vect{x})}
{(\vect{\theta}\cdot\vect{b})^{2}} d\Omega_{\vect{\theta}},
\end{eqnarray}

\noindent using the Moses basis, (\ref{identity7}). The simplification 
here is again due to the eigenvalue equation and the Hilbert transform 
of the derivative for the Radon transform of Trkalian fields in the 
Moses basis. 

  We can write similar formulas with $X\vect{F} (\vect{\theta}, \vect{x})$
using the substitution $\vect{\theta}\longrightarrow -\vect{\theta}$ 
through the equations (\ref{GelfandGoncharoveqn}, 
\ref{GelfandGoncharoveTrkalqn}, \ref {GelfandGoncharovMoses}).
This leads to Semyanistyi' s inversion formula (\ref{InvFunkSem}) 
in Section \ref{Xraytrf}.

  If we substitute the equation (\ref{GelfandGoncharovMoses}), 
($\vect{b}\longrightarrow\vect{\kappa}$) in (\ref{invRadtrf}), 
we find 

\begin{eqnarray} \label{SphmeanDiv}
\int_{S^{2}_{\vect{\theta}}} 
\vect{\mathcal{D}}\vect{F}_{\lambda}(\vect{\theta}, \vect{x}) 
d\Omega_{\vect{\theta}} 
=2\pi^{2}\frac{1}{\lambda\nu}
\vect{F}_{\lambda}(\vect{x}),
\end{eqnarray}

\noindent which can be directly proved using the Riesz potential 
as in Proposition \ref{prop9} in Section \ref{Johneqn}. See Appendix 
\ref{GelfandGoncharov}. We shall not discuss its tomographical implementation.

  We also see that the substitution of Semyanistyi' s inversion formula 
(\ref{InvFunkSem}) in the Radon inversion (\ref{invRadtrf}) leads to the 
inversion through spherical mean in Proposition \ref{prop9}.

  If we use (\ref{Dsimp}) in (\ref{GelfandGoncharovMoses}), 
the second term vanishes and we find

\begin{eqnarray} \label{GelfandGoncharovexample}
\vect{F}_{\lambda}^{\mathcal{R}} (\vect{b}\cdot\vect{x}, \vect{b}) 
= - \frac{1}{\nu^{2}} e^{ik_{0}\vect{\kappa}_{0}\cdot\vect{x}}
I(\vect{b}, \vect{\kappa}_{0})
\vect{F}_{0},
\end{eqnarray}

\noindent where the integral $I(\vect{b}, \vect{\kappa}_{0})$ is given 
in (\ref{integralimp}), ($\vect{\kappa}\longrightarrow\vect{b}$, 
$\vect{\kappa}^{\prime}\longrightarrow\vect{\kappa}_{0}$). Then we 
find $\vect{F}_{\lambda}^{\mathcal{R}} (\vect{b}\cdot\vect{x}, \vect{b})$, 
(\ref{Radsimp}). Similarly, we are led to (\ref{RadonMoses}) 
substituting (\ref{DFR1}) in (\ref{GelfandGoncharovMoses}).

  If we substitute (\ref{DivbeamLundquist}) in (\ref{SphmeanDiv}) 
we are led to (\ref{Lundquist}), ($\lambda=1$).

\section{RIESZ POTENTIAL AND BIOT-SAVART INTEGRALS}
\label{XrayRieszBiotSavart}

  We can write the X-ray transform of Riesz potential (of order $2$) 
and Biot-Savart ($\vect{\mathcal{BS}}$) integrals using (\ref{Smithmine}) 
in terms of their\cite{KS} Radon transform.

  If we replace $\vect{F}\longrightarrow \vect{\mathcal{I}}^{2}[\vect{F}]
=1/(8\pi^{2})\vect{\mathcal{R}}^{\dagger}\vect{\mathcal{R}}[\vect{F}]$ 
$\Rightarrow$ $\vect{F}^{\mathcal{R}}\longrightarrow \vect{\mathcal{R}}
\{\vect{\mathcal{I}}^{2}[\vect{F}]\}$ in (\ref{Smithmine}) using 
(\ref{RadonRiesz}), we find 

\begin{eqnarray} \label{JohnRiesz}
\vect{\mathcal{X}}\vect{\mathcal{I}}^{2}[\vect{F}](\vect{\theta}, \vect{x}) 
= \frac{1}{8\pi^{2}} \vect{\mathcal{X}}\vect{\mathcal{R}}^{\dagger}
\vect{\mathcal{R}}[\vect{F}](\vect{\theta}, \vect{x})= \frac{1}{4\pi} i 
\int_{S^{2}_{\vect{\kappa}}}\big\{ \vect{\mathcal{H}}\vect{{\mathcal{F}}}^{-1}
\{\frac{1}{k}\vect{{\mathcal{F}}}[\vect{F}^{\mathcal{R}}(q, \vect{\kappa})]
(k, \vect{\kappa})\} (p, \vect{\kappa}) \big\} 
(\vect{\kappa}\cdot\vect{x}, \vect{\kappa})
\delta(\vect{\kappa}\cdot\vect{\theta}) 
d\Omega_{\vect{\kappa}}.
\end{eqnarray}

  For Trkalian fields $\vect{\mathcal{I}}^{2}[\vect{F}]
=1/(8\pi^{2})\vect{\mathcal{R}}^{\dagger}\vect{\mathcal{R}}
[\vect{F}]=(1/\nu^{2})\vect{F}$, (Proposition \ref{prop8}) 
and hence

\begin{eqnarray} 
\vect{\mathcal{X}}\vect{\mathcal{I}}^{2}[\vect{F}](\vect{\theta}, \vect{x}) 
=\frac{1}{8\pi^{2}} \vect{\mathcal{X}}\vect{\mathcal{R}}^{\dagger}
\vect{\mathcal{R}}[\vect{F}](\vect{\theta}, \vect{x})
= \frac{1}{\nu^{2}} \vect{\mathcal{X}}\vect{F} (\vect{\theta}, \vect{x}).
\end{eqnarray} 

\noindent We can easily verify this substituting (\ref{RadonMoses}) 
in (\ref{JohnRiesz}) and comparing with (\ref{XFR1}).

  The equation (\ref{RadonBiotSavart}) yields $\partial_{p} 
\vect{\mathcal{RBS}}[\vect{F}^{\mathcal{R}}(q, \vect{\kappa})](p, \vect{\kappa})
=-\vect{\kappa} \bm{\times} \vect{F}^{\mathcal{R}}(p, \vect{\kappa})$.
If we replace $\vect{F}\longrightarrow \vect{\mathcal{BS}}[\vect{F}]
=\vect{\nabla}\bm{\times}\vect{\mathcal{I}}^{2}[\vect{F}]$ $\Rightarrow$ 
$\vect{F}^{\mathcal{R}}\longrightarrow \vect{\mathcal{RBS}}
[\vect{F}^{\mathcal{R}}]$ in (\ref{Smithmine}) using this, 
we find 

\begin{eqnarray} \label{XBS}
\vect{\mathcal{XBS}} \left[ \vect{F} \right] (\vect{\theta}, \vect{x}) 
= -\frac{1}{4\pi} \int_{S^{2}_{\vect{\kappa}}} 
\vect{\kappa} \bm{\times} \left[ \vect{\mathcal{H}} \vect{F}^{\mathcal{R}}
(p, \vect{\kappa}) \right] (\vect{\kappa}\cdot\vect{x}, \vect{\kappa})
\delta(\vect{\kappa}\cdot\vect{\theta}) d\Omega_{\vect{\kappa}}.
\end{eqnarray} 

\noindent Thus $\vect{\mathcal{XBS}}\left[ \vect{F} \right]
= (1/4\pi) \vect{\mathcal{M}} \big\{ \vect{\mathcal{H}}\partial_{p} 
\vect{\mathcal{RBS}} \left[ \vect{F}^{\mathcal{R}}(q, \vect{\kappa}) \right] 
(p, \vect{\kappa}) \big\}= -(1/4\pi) \vect{\mathcal{M}}\big\{ \vect{\kappa} 
\bm{\times} \left[ \vect{\mathcal{H}}\vect{F}^{\mathcal{R}}
(\vect{p, \kappa}) \right] \big\}$. This is again in the form 
of a Minkowski-Funk transform. We call this John-Biot-Savart 
integral. Further, one can express $\vect{F}^{\mathcal{R}}$ in terms 
of $\vect{\mathcal{X}}\vect{F}$.\cite{N, T, PM} The equation (\ref{XBS}) 
also follows from (\ref{JohnRiesz}): $\vect{\mathcal{XBS}} \left[ \vect{F} 
\right]=\vect{\mathcal{X}}\{\vect{\nabla}\bm{\times}\vect{\mathcal{I}}^{2}
[\vect{F}]\}=\vect{\nabla}\bm{\times}\vect{\mathcal{X}}\vect{\mathcal{I}}^{2}
[\vect{F}]$, (Proposition \ref{prop3}).

  For Trkalian fields (\ref{transelfdualfield2}), 
we find 

\begin{eqnarray} \label{XBSTrkalian}
\vect{\mathcal{XBS}} \left[ \vect{F} \right] (\vect{\theta}, \vect{x}) 
&=& \frac{1}{4\pi} \frac{1}{\nu} \int_{S^{2}_{\vect{\kappa}}}
\left[ \vect{\mathcal{H}}\partial_{p} 
\vect{F}^{\mathcal{R}}(p, \vect{\kappa}) \right] 
(\vect{\kappa}\cdot\vect{x}, \vect{\kappa})
\delta(\vect{\kappa}\cdot\vect{\theta}) d\Omega_{\vect{\kappa}}
= \frac{1}{\nu} \vect{\mathcal{X}}\vect{F} (\vect{\theta}, \vect{x}),
\end{eqnarray} 

\noindent (\ref{Smithmine}) as we expect, since: 
$\vect{\mathcal{BS}}[\vect{F}]=(1/\nu)\vect{F}$. 

  We can write $\vect{\mathcal{DBS}}[\vect{F}](\vect{\theta}, \vect{x})$ 
and $\vect{\mathcal{YBS}}[\vect{F}](\vect{\theta}, \vect{x})$ integrals 
respectively using (\ref{extdbeamtrfFourierinv}, \ref{Tuymine}) and 
(\ref{Tuymine3}), with a similar reasoning. For example, we find 

\begin{eqnarray} \label{YBS}
\vect{\mathcal{YBS}} \left[ \vect{F} \right] (\vect{\theta}, \vect{x}) 
= -\frac{1}{(2\pi)^{2}} \int_{S^{2}_{\vect{\kappa}}} 
\vect{\kappa} \bm{\times} \vect{F}^{\mathcal{R}}
(\vect{\kappa}\cdot\vect{x}, \vect{\kappa})
\frac{1}{\vect{\kappa}\cdot\vect{\theta}} d\Omega_{\vect{\kappa}}.
\end{eqnarray} 

\noindent We have $\vect{\mathcal{YBS}} \left[ \vect{F} \right]
(\vect{\theta}, \vect{x})=(1/\nu) \vect{\mathcal{Y}}\vect{F}
(\vect{\theta}, \vect{x})$, (\ref{Tuymine4}) for Trkalian fields.

  These, together with the Radon transform,\cite{KS} lead to an 
integral geometric understanding of these integrals. However, a physical 
or tomographical discussion of these integrals is beyond the scope of this 
manuscript.

\section{MINI-TWISTOR REPRESENTATION} 
\label{mini-twistorsolution}

  The mini-twistor space as an intrinsic structure have been introduced 
by Hitchin.\cite{H1} The Twistor theory, in simplest terms, is based on 
writing contour integral solutions for the wave equation in ($3+1$) 
dimensional Minkowski space, using holomorphic functions.\cite{RP4, RP3} 
We can write the solution of Helmholtz equation as a time-harmonic reduction 
of this, using mini-twistor space variables.\cite{WS} See Appendix 
\ref{Helmholtztwistor}. The quotient of twistor space ($\mathbb{CP}^{3}
\backslash \mathbb{CP}^{1}$) of the Minkowski space by the action of time 
translation yields the mini-twistor space $\mathbb{TCP}^{1}$.\cite{H1}

  The (mini-)twistor space of $\mathbb{R}^{3}$ is the space 
$\mathbb{TS}^{2}=\big\{ (\vect{u}, \vect{v})\in \mathbb{S}^{2} \bm{\times} 
\mathbb{R}^{3} , \, \vect{u},\vect{v} \in \mathbb{R}^{3}, \, |\vect{u}|=1, \, 
\vect{v}\cdot\vect{u}=0 \big\}\subset \mathbb{S}^{2}\bm{\times}\mathbb{R}^{3}$
of oriented lines $l$: $\vect{p}=\vect{v}+t\vect{u}$ where $\vect{u}$ is the 
direction vector, $\vect{v}$ is the position (shortest) vector of $l$ and 
$\vect{p}$ denotes a point of this line.\cite{H1} This has a natural complex 
structure which can be identified with the holomorphic tangent bundle 
$\mathbb{TCP}^{1}$ of projective line: the Riemann sphere $\mathbb{CP}^{1}$, 
with local coordinates $(\eta, \omega)$, [on $\mathbb{S}^{2}-\{N\}$, 
$\vect{u}\neq(0, 0, 1)$].\cite{D1, H1, MKM} Here $\vect{u}\in \mathbb{S}^{2}
\sim \mathbb{CP}^{1}$, $\omega$ is the coordinate on the base 
$\mathbb{CP}^{1}$ and $\eta$ denotes the fiber coordinate which 
decribes a holomorphic section. 

  We can regard a point $\vect{p}$ in $\mathbb{R}^{3}$ as the 
intersection of all oriented straight lines through it\cite{WS1} 
which are parametrised by a $2$-sphere $\mathbb{S}_{p}^{2}$ 
in $\mathbb{TS}^{2}\sim\mathbb{TCP}^{1}$. More precisely, each 
point $\vect{p}$ corresponds to a holomorphic section of 
$\mathbb{TCP}^{1}$.\cite{H1, MKM} These are fixed\cite{MKM} by an 
involutive map $\tau$ on $\mathbb{TS}^{2}$, $\tau^{2}=1$ reversing 
the orientation of lines which is called the \textit{real} 
structure.\cite{H1} The incidence relation\cite{MKM} between a point 
$\vect{p}(x, y, z)$ and a twistor $(\eta, \omega)$ that
defines this section is given by $\eta=(1/2) \left[ \left( x+iy \right) 
+2z\omega-\left( x-iy \right)\omega^{2} \right]$. The set of twistors 
incident with a given point (the set of lines passing through this point) 
form a copy of $\mathbb{CP}^{1}$ which lies as a \textit{real} section of 
$\mathbb{TCP}^{1}$.\cite{H1, PB, PB1} If we hold $(\eta, \omega)$ fixed, 
then $(x, y, z)$ satisfying the incidence relation  defines a line in 
$\mathbb{R}^{3}$. If we hold $(x, y, z)$ fixed, then $(\eta, \omega)$ 
satisfying the incidence relation parametrises the set of all lines 
through the point $\vect{p}(x, y, z)$.\cite{MKM1,CGHKM} The local 
coordinates on $\mathbb{S}^{2}-\{S\}$, $[\vect{u} \neq (0, 0, -1)]$ 
are given by $\omega^{\prime}=1/\omega$, $\eta^{\prime}=-\eta/\omega^{2}
=(1/2) \left[ \left( x-iy \right) -2z\omega^{\prime} -\left( x+iy \right) 
\omega^{\prime 2} \right]$.\cite{WS, WS1, D1} We shall ignore the factor 
$1/2$ in $\eta$. 

  We shall restrict a function defined on a domain of the mini-twistor 
space to a (projective) line and then integrate along a closed contour 
contained in this $\mathbb{CP}^{1}$.\cite{D1} 

  We shall use (mini-)twistor solution of the Helmholtz equation 
for finding Trkalian fields. A Trkalian field (\ref{Trkalianselfdual}) 
also satisfies the vector Helmholtz equation 
 
\begin{eqnarray} \label{vectorHelmholtz}
\nabla^{2}\vect{F}(\vect{x}) =-k^{2}\vect{F}(\vect{x}), 
\hspace*{5mm} k=\nu
\end{eqnarray} 

\noindent but the converse is not necessarily true. 
We can write a solution of this equation as

\begin{eqnarray} \label{vectorHelmholtzsol1}
\vect{F}(\vect{x}) 
= \left[ A(\vect{x}), \, B(\vect{x}), \, C(\vect{x}) \right]
= \int_{C} e^{-ikf} \left( L, \, M, \, N \right) d\omega,
\end{eqnarray}

\noindent where $L \left( \eta_{\vect{x}}(\omega), 
\omega \right)$, $M \left( \eta_{\vect{x}}(\omega), 
\omega \right)$, $N \left( \eta_{\vect{x}}(\omega), 
\omega \right)$ are holomorphic functions of $\eta_{\vect{x}}(\omega)
=x+iy+2z\omega-(x-iy)\omega^{2}$ and $\omega$. Here $f=\omega(x-iy)-z$ 
is, for example chosen as the spatial part of integrating factor 
(\ref{intfac1}, \ref{scalarHelmholtzSol}) for the time-harmonicity 
condition (\ref{eqn1}).

  If we substitute (\ref{vectorHelmholtzsol1}) 
in (\ref{Trkalianselfdual}), we find

\begin{eqnarray} \label{Trkalianselfdual1twistor}
& & i \left( 1+\omega^{2} \right) N_{\eta}-2\omega M_{\eta} 
=k (L+iM+\omega N), \nonumber \\
& & 2\omega L_{\eta}- \left( 1-\omega^{2} \right) N_{\eta} 
=-ik (L+iM+\omega N), \\
& & \left( 1-\omega^{2} \right) M_{\eta}-i \left( 1+\omega^{2} \right) L_{\eta} 
=-k \left[ \omega(L-iM)-N \right]. \nonumber 
\end{eqnarray}

\noindent We shall avoid further considerations such as modifying this 
equation by introducing arbitrary holomorphic functions or modifying 
the contour $C$ which calls for sheaf cohomology here.\cite{D1, PB}

   The first and second equations in (\ref{Trkalianselfdual1twistor}) 
yields $-2\omega \left[ \left( 1-\omega^{2} \right) M_{\eta} 
-i \left( 1+\omega^{2} \right) L_{\eta} \right]=2k (L+iM+\omega N)$.
This leads to

\begin{eqnarray} \label{incidence1}
\left( 1-\omega^{2} \right) L + i \left( 1+\omega^{2} \right) M 
+ 2\omega N =0,
\end{eqnarray}

\noindent that is $L+iM+2N\omega-(L-iM)\omega^{2}=0$ 
using the third equation. If we substitute $N$, (\ref{incidence1}) 
in (\ref{Trkalianselfdual1twistor}), we find

\begin{eqnarray} \label{condition} 
-i \left( 1+\omega^{2} \right) L
+ \left( 1-\omega^{2} \right) M=0. 
\end{eqnarray}

\noindent The equations (\ref{incidence1}, \ref{condition}) lead 
to: $L(\eta, \omega) = \left( 1-w^{2} \right) u(\eta, \omega)$, 
$M(\eta, \omega) = i \left( 1+w^{2} \right) u(\eta, \omega)$,  
$N(\eta, \omega) = 2w u(\eta, \omega)$ where $u(\eta, \omega)$ 
is an arbitrary holomorphic function (except some poles) of 
$\eta$ and $\omega$. 

  Thus a Trkalian field (\ref{Trkalianselfdual})
is given by 

\begin{eqnarray} \label{Trkaliantwistor1}
\vect{F}(\vect{x}) 
= \left[ A(\vect{x}), \, B(\vect{x}), \, C(\vect{x}) \right]
= \int_{C} \left[ \left( 1-\omega^{2} \right) , 
\, i \left( 1+\omega^{2} \right) , \, 2\omega \right] 
e^{-ikf} u(\eta, \, \omega) d\omega.
\end{eqnarray}

\noindent Note $\left[ \left( 1-\omega^{2} \right) , 
i \left( 1+\omega^{2} \right) , 2\omega \right]$ 
is a null vector in $\mathbb{C}^{3}$.  

  This solution is in the form of twistor solution to Maxwell 
equations in $(3+1)$ dimensions, as we expect (since Trkalian 
fields correspond to the spatial part of  time-harmonic 
electromagnetic fields with no source). We can also derive this 
solution from the twistor solution of Maxwell equations (see 
Ref. \onlinecite{RP3}, p. 33, pp. 206-207), using a similar 
time-harmonic reduction (with minor changes of conventions). 

    For example, we choose $u(\eta, \omega)
=g(\eta_{\vect{x}}(\omega))/h(\omega)$ with $h(\omega)
=(\omega-\omega_{0})^{m}$. If $g(\eta_{\vect{x}}(\omega))
= \eta^{n}_{\vect{x}}(\omega)$ where $n$ is positive, $m=1$ 
and $\omega_{0}=0$, that is $u(\eta, \omega)
=\eta^{n}_{\vect{x}}(\omega)/\omega$, we find 

\begin{eqnarray}
\vect{F}(\vect{x})=2\pi i e^{i\nu z} \zeta^{n} (1, i, 0), 
\hspace*{3mm} \zeta=x+iy.
\end{eqnarray}

\noindent If $\omega_{0}\neq 0$, that is 
$u(\eta, \omega)=\eta^{n}_{\vect{x}}(\omega)/(\omega-\omega_{0})$, 
we find 

\begin{eqnarray}
\vect{F}(\vect{x})=2\pi i e^{-i\nu[\omega_{0}(x-iy)-z]}
\eta^{n}_{\vect{x}}(\omega_{0}) 
\left( (1-\omega^{2}_{0}), i(1+\omega^{2}_{0}), 2\omega_{0} \right). 
\end{eqnarray}

\noindent Hence for a holomorphic function 
$g(\eta)=\sum_{n=0}^{\infty} a_{n} \eta^{n}$ and $\omega_{0}=0$ 
that is $u(\eta, \omega)=g(\eta_{\vect{x}}(\omega))/\omega$ we find 

\begin{eqnarray}
\vect{F}(\vect{x})=2\pi i e^{i\nu z} g(\zeta) (1, i, 0). 
\end{eqnarray}

\noindent The orthogonality of real contact structures arising in 
case $g(\zeta)$ is given by a derivative: $g\longrightarrow g^{\prime}$ 
is discussed in Ref. \onlinecite{KS1}. If we choose 
$u(\eta, \omega)=g(\eta_{\vect{x}}(\omega))/\omega^{2}$, 
we find 

\begin{eqnarray} \label{sol}
\vect{F}(\vect{x})=2\pi i e^{i\nu z} \{ \left[ 
-i\nu\overline{\zeta} g(\zeta)+2zg^{\prime}(\zeta) \right] (1, i, 0) 
+2g(\zeta)(0, 0, 1) \} .
\end{eqnarray}

  If we choose  $u(\eta, \omega)=(1/\omega^{2})e^{-i(\nu/2)\omega^{-1}\eta}$, 
[$-(1/2)\omega^{-1}\eta=f-g$, $g=x(\omega+w^{-1})/2-iy(\omega-w^{-1})/2$] 
where $\vect{x}=(x, y, z)=(r\cos \varphi, r\sin \varphi, z)$ in cylindrical 
coordinates and $\omega=e^{i\theta}$, $C$ is a circle of unit radius about 
the origin, then we find the Lundquist solution (\ref{Lundquist}) with 
$F_{0}=4\pi i$ and $\lambda=1$, using the integrals (\ref{Besselintegral3}) 
in Appendix \ref{Otherformulas}.

\subsection{Arbitrary integrating factor} 
\label{arbitraryintegratingfactor}

  We can choose different integrating factors for the time-harmonicity 
condition (\ref{eqn1}). In fact, we do not have to choose an integrating 
factor initially. We can see this in a time-harmonic extension of the 
solution above. The integrating factor in (\ref{intfac1}) yields a 
time-harmonic extension of Trkalian fields: $\vect{F}\longrightarrow 
e^{-ikt}\vect{F}$. This satisfies (vector) wave equation which reduces 
to (\ref{vectorHelmholtz}). 

  If we use an arbitrary integrating factor

\begin{eqnarray} 
g(p, q, \omega) &=& e^{-ik\tilde{f}(p, q, \omega)} h(p, q, \omega) 
= e^{-ik\tilde{f}(p, q, \omega)} H(\eta_{\vect{x}}(\omega), \omega),
\end{eqnarray}

\noindent in (\ref{eqn1}), we find

\begin{eqnarray} \label{condition1}
\omega \frac{\partial \tilde{f}}{\partial p} 
+ \frac{\partial \tilde{f}}{\partial q}=1, 
\end{eqnarray}

\noindent that is $\partial \tilde{f}/\partial t=1$ 
as a condition on the integrating factor. Then the field 
(\ref{vectorHelmholtzsol1}), (with $f\longrightarrow\tilde{f}$
containing both temporal and spatial pieces) satisfies 
the wave equation which reduces to (\ref{vectorHelmholtz}) 
upon imposing  the condition (\ref{condition1}). If we 
substitute this field in (\ref{Trkalianselfdual}), we find

\begin{eqnarray} \label{arbintfac}
& & i \left( 1+\omega^{2} \right) N_{\eta}-2\omega M_{\eta} 
=kL-k(i\omega M+N)\tilde{f}_{p}+k(iM+\omega N)\tilde{f}_{q}, \nonumber \\
& & 2\omega L_{\eta}- \left( 1-\omega^{2} \right) N_{\eta} 
=kM+ik(\omega L-N)\tilde{f}_{p}-ik(L+\omega N)\tilde{f}_{q}, \\
& & \left( 1-\omega^{2} \right) M_{\eta}-i \left( 1+\omega^{2} \right) L_{\eta} 
=kN+k(L+iM)\tilde{f}_{p}-k\omega (L-iM)\tilde{f}_{q}. \nonumber 
\end{eqnarray}

\noindent This reduces to (\ref{Trkalianselfdual1twistor}) for 
$\tilde{f}=q$. The equations (\ref{arbintfac}) lead to the same 
equations (\ref{incidence1}, \ref{condition}) using a similar 
reasoning. These yield the solution (\ref{Trkaliantwistor1}), 
[$f\longrightarrow\tilde{f}$ satisfying (\ref{condition1})] 
with a harmonic time dependence now.

  Thus the solution is of the same form containing the spatial 
part of the chosen integrating factor. 

  Any solution of the time-harmonicity condition (\ref{eqn1}) can 
be written using $\tilde{f}=(1/2)(p/\omega+q)$, (\ref{intfac2a}), 
see (\ref{gensol}). This (excluding the temporal piece) leads to 
the solution (\ref{Trkaliantwistor1}) with $f=(1/2)
[\omega(x-iy)+(x+iy)/\omega]$, see (\ref{intfac2b}).

  In this case, the Lundquist solution is simply given 
by $u(\eta, \omega)=1/\omega^{2}$, ($\omega=e^{i\theta}$, 
$C$: unit circle about the origin). If we use 
$u=h(\omega^{\prime})$ which has a Laurent series: 
$h(\omega^{\prime})=1/{\omega^{\prime}}^{(n+1)}$ 
with $\omega=i\omega^{\prime}$, ( $k=\nu$), we find 

\begin{eqnarray} \label{twistorCKz}
\vect{F}(\vect{x})= 4\pi i e^{-im\varphi}  
\left[ im\frac{1}{\nu r} J_{m}(\nu r) \vect{e}_{r} 
+ J_{m}^{\prime}(\nu r) \vect{e}_{\varphi} 
- J_{m}(\nu r) \vect{e}_{z} \right],  
\hspace*{5mm} m=n-1
\end{eqnarray}

\noindent in cylindrical coordinates. See Appendix 
\ref{SolutionLaurent}.\cite{WS} This is a circular 
cylindrical CK\cite{CK} solution\cite{Y, TC} with 
no $z$ dependence, upto conventions.

\subsection{Chandrasekhar-Kendall type solutions: Debye potentials}

  We can use the solution 

\begin{eqnarray} \label{scalarHelmholtzSolTrkallian1}
\phi(x, y, z)= \int_{C} e^{-i\sigma f} 
H(\eta_{\vect{x}}(\omega), \omega) d\omega, 
\hspace*{5mm} f=\frac{1}{2}[\omega(x-iy)+\frac{1}{\omega}(x+iy)]
\end{eqnarray}

\noindent of the scalar Helmholtz equation $\nabla^{2}\phi=-k^{2}\phi$, 
($k=\sigma$) as Debye potential for CK\cite{CK} type Trkalian 
fields

\begin{eqnarray} \label{CK}
\vect{F}(\vect{x})=- \left[ \sigma \vect{\nabla} \bm{\times} 
\left( \phi\vect{w} \right) 
+ \vect{\nabla} \bm{\times} \vect{\nabla} \bm{\times} 
\left( \phi\vect{w} \right) \right], 
\end{eqnarray}

\noindent where $\vect{\omega}$ is a fixed vector and $\vect{\nabla}
\bm{\times}\vect{F}(\vect{x})-\sigma\vect{F}(\vect{x})=0$. It is 
straightforward to write a time-harmonic extension of the CK solution.

  The potential for the circular cylindrical CK solution\cite{Y, TC} is 
effectively (apart from $z$ coordinate) a 2 dimensional solution\cite{WS} 
(satisfying the Helmholtz equation: $\nabla^{2}\phi+\nu^{2}\phi=0$ in 2 
dimensions). We can express this as 

\begin{eqnarray} \label{CKcyl}
\phi(r, \varphi, z) = e^{-ikz} \int_{C}  e^{-i\nu f} H d\omega
= 2\pi i \frac{1}{i^{m}} J_{m}(\nu r) e^{im\varphi-ikz}, 
\hspace*{5mm} \sigma^{2}=\nu^{2}+k^{2}
\end{eqnarray}

\noindent where $H=\omega^{m-1}$, $\omega=e^{i\theta}$, $C$:
unit circle about the origin and $\vect{w}=\vect{e}_{z}$. 
Here we use (\ref{Besselintegral1}, \ref{Besselintegral2}). 
This reduces to 

\begin{eqnarray} \label{Lundquisttwistor}
\phi(r, \varphi, z) = \int_{C}  e^{-i\nu f} H d\omega
= 2\pi i J_{0}(\nu r), 
\end{eqnarray}

\noindent the potential for the Lundquist solution, 
for $m=0$, $k=0$, ($\sigma=\nu$), $H=\omega^{-1}$. 

   An interesting case is the class of axially symmetric potentials. 
If we assume axial symmetry about $z$-azis, then a rotation in $xy$-plane 
is given by $\omega\longrightarrow e^{i\psi}\omega$ (treated as a spinor 
coordinate) which induces the rotation $x+iy \longrightarrow e^{i\psi}(x+iy)$ 
(see Refs. \onlinecite{WS} and also \onlinecite{H1}, \onlinecite{WS1}) and 
$\eta_{\vect{x}}(\omega)\longrightarrow  e^{i\psi}\eta_{\vect{x}}(\omega)$,
$d\omega \longrightarrow e^{i\psi} d\omega$ while 
$f=(1/2)[\omega(x-iy)+(x+iy)/\omega]\longrightarrow f$. 
Hence we consider fields of the form

\begin{eqnarray} \label{scalarHelmholtzSolaxsym}
\phi(x, y, z)= \int_{C} e^{-i\sigma f} 
G \left( \frac{\eta_{\vect{x}}(\omega)}{\omega} \right) \frac{1}{\omega} d\omega,
\end{eqnarray}

\noindent for some holomorphic $G$. We assume 
that $G$ has a Laurent series about the origin: 
$G=[\eta_{\vect{x}}(\omega)/\omega]^{n}$, then 

\begin{eqnarray} \label{scalarHelmholtzSolaxsym1}
\phi(x, y, z)= \int_{C} e^{-i\sigma f} 
\left[ \frac{\eta_{\vect{x}}(\omega)}{\omega} \right]^{n} 
\frac{1}{\omega} d\omega.
\end{eqnarray}

\noindent For $n=0$ ($\sigma=\nu$), this reduces
to the potential (\ref{Lundquisttwistor}). 

  We can use $\omega=-ie^{iu}$ for parametrizing 
(\ref{scalarHelmholtzSolaxsym1}). This yields

\begin{eqnarray} \label{scalarHelmholtzSolaxsym2}
\phi(x, y, z)= 2^{n} i \int_{C} e^{-i\sigma f} 
\left( z+ix\cos u+iy\sin u \right)^{n} du,
\end{eqnarray}

\noindent where $f=x\sin u-y\cos u$, with
$\eta/\omega=2(z+ix\cos u+iy\sin u)$. 
For $n=1$, we find  

\begin{eqnarray} \label{scalarHelmholtzSolaxsym3}
\phi &=& 2 i R \int_{u=0}^{u=2\pi} 
e^{-i\sigma R \sin \theta\sin(u-\varphi)} 
\left[ \cos\theta+i\sin\theta\cos(u-\varphi) \right] du \\
&=& 2 i R \int_{u=0}^{u=2\pi} e^{-i\sigma R \sin \theta\sin u} 
\left( \cos\theta+i\sin\theta\cos u \right) du, \nonumber
\end{eqnarray}

\noindent using spherical coordinates: $x=R\sin\theta\cos \varphi$, 
$y=R\sin\theta\sin \varphi$, $z=R\cos\theta$. The integral of the 
second term vanishes and the first term yields 

\begin{eqnarray} \label{Debye}
\phi=4\pi i z J_{0}(\sigma r), 
\hspace*{5mm} 
r=R\sin\theta,
\end{eqnarray}

\noindent in cylindrical coordinates, using (\ref{Besselintegral4}): 
($\beta=\sigma r$). The potential (\ref{Debye}), with $\vect{\omega}
=\vect{e}_{z}$, leads to 

\begin{eqnarray}
\vect{F}(\vect{x})=-4\pi i \sigma^{2} \Big\{ 
-\frac{1}{\sigma} J_{1}(\sigma r) \vect{e}_{r}
+z \left[ J_{1}(\sigma r)\vect{e}_{\varphi} 
+J_{0}(\sigma r)\vect{e}_{z} \right] \Big\} , 
\end{eqnarray} 

\noindent a generalization of the Lundquist field (\ref{Lundquist}), 
($\sigma=\nu$).

  In case $n=-1$, we use $f=\omega(x-iy)-z$ for 
the sake of simplicity.\cite{WS1, PB} The denominator 
in (\ref{scalarHelmholtzSolaxsym1}) can be factored as 
$\eta_{\vect{x}}(\omega)=-(x-iy)(\omega-\omega_{1})(\omega-\omega_{2})$ 
where $\omega_{1}=(z-|\vect{x}|)/(x-iy)=-e^{i\varphi} \tan (\theta/2)$,  
$\omega_{2}=(z+|\vect{x}|)/(x-iy)=e^{i\varphi} \cot (\theta/2)$. 
The integral branches depending on the sign of $z$.\cite{WS1}. 
If $z>0$, $|\omega_{1}|<1$, then the single residue inside the 
unit circle $C$ yields 

\begin{eqnarray} \label{Helmholtzfunsol}
\phi= \frac{1}{2} \frac{e^{i\sigma|\vect{x}|}}{|\vect{x}|}.
\end{eqnarray} 

\noindent This is the fundamental solution of the scalar Helmholtz 
equation: $\nabla^{2}\phi+\sigma^{2}\phi=-2\pi\delta(\vect{x})$.
It is beyond the scope of this manuscript to provide a complete 
treatment of this case.\cite{WS1, PB}

   The classical spheromak equilibrium solution\cite{CK, RB1}
(see also Ref. \onlinecite{Ml3} and the references therein) is 
given by 

\begin{eqnarray} \label{classSpheromakequisol}
\vect{F}=F_{0} \Big\{ 2\frac{j_{1}(kR)}{kR}\cos\theta\vect{e}_{R} 
+\frac{1}{kR}\left[ j_{1}(kR)-\sin(kR) \right] \sin\theta \vect{e}_{\theta} 
+j_{1}(kR)\sin\theta \vect{e}_{\phi} \Big\}, 
\end{eqnarray} 

\noindent in spherical coordinates, where $j_{1}(kR)$ is the spherical 
Bessel function, $\vect{w}=\vect{R}$, (with the conventions of Ref. 
\onlinecite{RB1}) and $\sigma=k$, (\ref{CK}). We need to consider 
an expression of the form (\ref{scalarHelmholtzSolTrkallian1}) 
which reduces to\cite{JAS}

\begin{eqnarray} \label{classSpheromakequisoltwis}
\phi = - \frac{i}{2} \frac{F_{0}}{k} 
\int^{\pi}_{0} e^{-ikR\cos \theta \cos \alpha} 
J_{0}(kR\sin \theta \sin \alpha) P^{0}_{1}(\cos \alpha) 
\sin\alpha d\alpha
= - \frac{F_{0}}{k} j_{1}(kR)P^{0}_{1}(\cos \theta).
\end{eqnarray}

   An integral of this type was recently reconsidered by 
various authors,\cite{NPFRCBC, VD, CS, SK} refering to Refs. 
\onlinecite{GR, GNW, E} and including alternative proofs. 
To the knowledge of the author, the simplest derivation of 
this integral expression is given in Ref. \onlinecite{JAS}, 
p. 411 (with a misprint of coefficient).

\section{CONCLUSION}
\label{Conc}

  We have studied the X-ray and Divergent beam transforms of Trkalian 
fields in connection with their Radon transform. We remind that the Radon 
transform of a Trkalian field is defined on a sphere in the transform space 
and it satisfies a corresponding eigenvalue equation there. The mathematical 
methods of tomography (respectively Smith' s and Tuy' s methods) show that 
these transforms are basically in the form of a Minkowski-Funk and a closely 
related integral transform of certain intricate quantities (Hilbert transform 
of the derivative of Radon transform). The Moses eigenbasis is especially 
efficient in exhibiting this connection for Trkalian fields. These naturally 
reduce to well known geometric integral transforms on a sphere of the Radon 
or the spherical Curl transform. 

  More precisely, the X-ray transform of a Trkalian field is given by 
the Minkowski-Funk transform of its Radon transform on this sphere. In 
$\mathbb{R}^{3}$, this corresponds to the integral of the Radon 
transform of the field over a pencil of planes intersecting at 
a line. Previously, this transform was introduced by Gonzalez\cite{FBG}, 
called the plane-to-line transform, as an elementary geometric 
transform in integral geometry. This transform naturally arises 
for Trkalian fields. We refer the interested reader to Ref. 
\onlinecite{BE} for a twistor approach to this transform. 

  Meanwhile the Divergent beam transform is given by another closely 
related (an extension of Minkowski-Funk) geometric integral transform 
of the spherical Curl transform of the field on the sphere. This also 
provides an extension over the plane-to-line transform. This seems a 
natural extension from the point of view of generalized functions.

  We remark that these transforms are naturally defined on the sphere 
in the transform space. Geometrically, we are endowed with a simple 
picture showing the interrelations of these transforms for Trkalian 
fields on this sphere. This is made possible with the Moses basis.

  Intuitively speaking, the X-ray or Divergent beam transform
of a Trkalian field respectively integrates the field on a whole line 
or a half-line which is determined by a direction vector. This direction 
vector determines a great circle on the sphere in the transform space. 
Then these transforms are given by integrals of the Radon transform 
on this great circle. The X-ray: whole-line transform is given by a 
whole (in the distributional sense: Dirac delta) integral. This leads 
to the plane-to-line picture in $\mathbb{R}^{3}$. Meanwhile, the 
Divergent beam: half-line transform is given by a half (in the 
distributional sense: Heisenberg delta) integral, depending on 
the orientation. The picture in $\mathbb{R}^{3}$ for this is left 
to the imagination of the reader.

  We can logically derive these results starting from the fundamental 
relation (\ref{essential}) of mathematical tomography in a unified manner. 

  However, we have postponed the mathematical discussion originating 
from tomography until the basic investigation of X-ray and Divergent 
beam transforms of Trkalian fields had finished. This gave us the 
opportunity to consider the mathematical basis for tomographical 
studies of Trkalian field models in nature, for its own sake.

  These transforms are members (via analytic continuation) of a well known 
analytic family of integral operators which arise in the study of Fourier 
transforms of homogeneous functions. Recently, these integral operators 
have been studied by Rubin.\cite{Rubin3, Rubin4} We have inverted the 
X-ray transform of a Trkalian field using Semyanistyi' s 
formula\cite{VIS, VIS1} which also belongs to this family, 
so as to yield its Radon transform. This leads to an inversion 
through the spherical mean of the X-ray transform of Trkalian fields.  

  The X-ray (also the Divergent beam) transform and its inversion 
intertwine the Curl operator and also the Divergence, Gradient and 
Laplacian operators. Thus the X-ray (Divergent beam) transform 
of a Trkalian field is Trkalian. Also, the Trkalian subclass of X-ray 
transforms $\vect{\mathcal{X}}\vect{F}$ yields Trkalian fields in the 
physical space. We have also written the John' s equation for Trkalian fields 
in an equivalent form. Thus, we can study Trkalian fields either in physical 
space or in the transform space. 

  Another crucial quantity in integral geometry is the Riesz potential.
The Riesz potential, of order $\alpha$ where $0<\alpha<3$, of a Trkalian field 
is proportional to the field. Hence, the spherical mean of the X-ray (or 
Divergent beam) transform of a Trkalian field over all lines passing 
through a point yields the field at this point. This provides a new 
simple inversion formula for the X-ray (or Divergent beam) transform 
of Trkalian fields. This result is also logically implied by 
Gelfand-Gonchorav' s mathematical approach (making use of the 
same intricate quantity) to tomography.

  Then we have returned back to the mathematical methods of tomography. 
These methods provide us elegant mathematical tools for investigating 
the interrelations of these integral transforms in a unified view. 
First, these endowed us with an integral geometric view and motivation
for the discussions above. Second, these enabled us to discuss these 
mathematical methods with a view towards tomographical studies of 
Trkalian field models in nature. For this purpose, we have studied 
these mathematical methods using Trkalian fields. We have adopted 
a mathematical approach rather than a tomographical implementation.

  We have made use of four basic mathematical approaches of 
tomography due to Grangeat,\cite{G} Smith,\cite{S1} Tuy\cite{HKT} 
and Gelfand-Goncharov\cite{GG}. These methods are outflow 
of the fundamental relation of mathematical tomography. 
They are based on relations of the X-ray and Divergent 
beam transforms to the intricate quantities mentioned above: 
the Hilbert transform of the derivative of Radon transform. 

  These methods basically make use of the Radon inversion (for 
reconstruction). They lead us to new inversion formulas for the 
X-ray and Divergent beam transforms of Trkalian fields with a view 
towards tomographical applications. 

  In general, these simplify for Trkalian fields. The simplification
arises in the crucial intricate quantities mentioned above. This is 
basically due to the eigenvalue equation (\ref{transelfdualfield2}) 
and the Hilbert transform of the derivative for Radon transform of 
Trkalian fields in the Moses basis.

  The Grangeat approach leads to another simple, direct inversion
formula for the Divergent beam transform of Trkalian fields. 

  The Smith method reveals that the X-ray transform is in the form of 
a Minkowski-Funk transform of the intricate quantity mentioned above. 
This quantity reduces to the Radon transform for a Trkalian field, using 
the Moses basis. This provides the integral geometric view and motivation 
for the previous discussion of the X-ray transform of Trkalian fields. 
In this approach, the inversion formula can be expressed in terms of the 
Radon transform of Trkalian field. 

  The Tuy method enables us to study the Divergent beam transform in 
detail. It reveals that the Divergent beam transform is in the form 
of another closely related integral transform of a quantity related 
to the Radon transform. In this case, this quantity reduces to the spherical 
Curl transform, using the Moses basis. This leads us to the above mentioned
integral geometric view of the Divergent beam transform of Trkalian fields. 
In this case, the inversion formula can be expressed in terms of the 
spherical Curl transform of the field. 

  We have calculated the Divergent beam transform of the Lundquist field 
which is used to model solar magnetic clouds, benefitting the Tuy method.
This provides a mathematical \textit{R\"{o}ntgen} of these clouds.

  Meanwhile, the Gelfand-Goncharov approach leads to a direct 
inversion through the spherical mean that is mentioned above. 
This naturally makes use of the inverse transform that belongs 
to the above family of integral operators. The simplification 
is again based on the same intricate quantity: Hilbert transform 
of the derivative of Radon transform in the Moses basis.

  Briefly, the intricate quantity: Hilbert transform of the derivative 
of Radon transform which intrigued tomography simplifies for a Trkalian 
field in the Moses basis.

  The direct inversion formulas arising in Grangeat' s and 
Gelfand-Gonchorav' s approaches mathematically seem more feasible 
than the inversions in Smith' s and Tuy' s methods.

  These approaches provide different inversion formulas which may 
serve useful for designing reconstruction methods in tomographical studies 
of Trkalian field models in nature, depending on real physical situation.  
The author expects that the Moses basis which has led to a drastical 
simplification in the crucial quantity may also be of practical use 
in tomographical studies. We shall not discuss tomographical 
implementations of these inversion formulas.

    Furthermore, the Smith and Tuy methods mathematically enable us to define 
the X-ray and Divergent beam transforms of the Riesz potential (of order $2$) 
and Biot-Savart integrals. The X-ray transform of the Biot-Savart integral 
of a Trkalian field reduces to the X-ray transform of the field. The 
Radon,\cite{KS} X-ray and Divergent beam transforms of the Riesz potential 
(of order $2$) and Biot-Savart integrals lead to an integral geometric 
understanding of these integrals. However, a physical or tomographical 
discussion of these integrals is beyond the scope of this manuscript.

  In the second part of this manuscript we have discussed 
Trkalian fields using (mini-)twistors. The X-ray transform 
is a real analogue and a predecessor of Twistor 
theory.\cite{BEGM, BEGM1a, DW, D1, S, BE, LJM3a} The 
X-ray transform and the mini-twistors are both defined on the 
space $\mathbb{TS}^{2}\sim\mathbb{TCP}^{1}$ of oriented lines 
in $\mathbb{R}^{3}$. 

  We have discussed a mini-twistor representation, presenting 
a mini-twistor solution for the Trkalian fields equation. This 
is based on twistor solution of the (vector) Helmholtz equation 
which makes use of a time-harmonic reduction of the wave equation. 
A Trkalian field is given in terms of a null vector in $\mathbb{C}^{3}$ 
with an arbitrary holomorphic function of two variables and an exponential 
factor that results from the reduction. 

  The exponential factor contains the spatial part of an integrating 
factor for the time-harmonicity condition. The solution is of the same 
form containing the spatial part of any choosen integrating factor.
We have also used the general solution of this condition for writing 
our solution. 

  This solution can also be derived using a time-harmonic 
reduction of the twistor solution for electromagnetic fields 
in $(3+1)$ dimensions. 

  We are led to a time-harmonic extension of Trkalian fields, 
implicitly keeping this condition. This can be interpreted as 
a time-harmonic electromagnetic field.

  We have also presented examples of Debye potentials 
for CK type solutions using mini-twistors.

  This manuscript is aimed at studying the most basic integral geometric 
aspects and the mini-twistor representation of Trkalian fields. We have 
made use of the mathematical methods of tomography (but not the tomography).
These may serve useful for studying their physical properties in a realistic 
environment. 

  The Trkalian class of fields may also provide a simple and interesting 
example for studying the relation of ray transforms with Twistor theory.
However, twistor tomography is beyond the limitations of this manuscript.

\appendix

\section{RAY TRANSFORMS}

\subsection{X-ray transform of Lundquist Field}
\label{XrayLundquist}

  We write $\vect{x}=r\vect{e}_{r}(\phi)+z\vect{e}_{z}$, $r>0$ 
and $\vect{\theta}=v_{r}\vect{e}_{r}(\theta)+v_{z}\vect{e}_{z}$, $v_{r}>0$ 
in cylindrical coordinates. Then $\vect{x}^{\prime}=\vect{x}+s\vect{\theta}
=r^{\prime}\vect{e}_{r}(\phi^{\prime})+z^{\prime}\vect{e}_{z}$ where 
$r^{\prime}\cos \phi^{\prime}=r\cos \phi + sv_{r} \cos \theta$, 
$r^{\prime}\sin \phi^{\prime}=r\sin \phi + sv_{r} \sin \theta$, 
$r^{\prime 2}=r^{2}+s^{2}v_{r}^{2}+2rsv_{r}\cos (\theta-\phi)$, 
$z^{\prime}=z+sv_{z}$ and

\begin{eqnarray}
\vect{e}_{r}(\phi^{\prime}) &=& \cos \phi^{\prime}\vect{e}_{x}
+\sin \phi^{\prime}\vect{e}_{y}, 
\hspace*{20mm}
\vect{e}_{\phi^{\prime}} = -\sin \phi^{\prime}\vect{e}_{x}
+\cos \phi^{\prime}\vect{e}_{y} \\
&=& \frac{1}{r^{\prime}} \left[ r\vect{e}_{r}(\phi)
+sv_{r}\vect{e}_{r}(\theta) \right] 
\hspace*{20mm}
= \frac{1}{r^{\prime}} \left[ r\vect{e}_{\phi}
+sv_{r}\vect{e}_{\theta} \right]. \nonumber 
\end{eqnarray}

   Hence

\begin{eqnarray}
\vect{F}_{L}(\vect{x}+s\vect{\theta}) 
= F_{0} \left[ \lambda J_{1}(\lambda\nu r^{\prime})\vect{e}_{\phi^{\prime}} 
+J_{0}(\lambda\nu r^{\prime})\vect{e}_{z} \right]
= F_{0} \left[ \lambda r\frac{1}{r^{\prime}}J_{1}(\lambda\nu r^{\prime})
\vect{e}_{\phi}
+\lambda v_{r}\frac{s}{r^{\prime}}J_{1}(\lambda\nu r^{\prime})\vect{e}_{\theta}
+J_{0}(\lambda\nu r^{\prime})\vect{e}_{z} \right], 
\end{eqnarray}

\noindent and we have 

\begin{eqnarray}
\vect{\mathcal{X}}\vect{F}_{L}(\vect{\theta}, \vect{x})=F_{0} \left[ 
\lambda r\int_{-\infty}^{\infty} \frac{1}{r^{\prime}}J_{1}
(\lambda\nu r^{\prime})ds\vect{e}_{\phi} 
+\lambda v_{r}\int_{-\infty}^{\infty} \frac{s}{r^{\prime}}J_{1}
(\lambda\nu r^{\prime})ds\vect{e}_{\theta} 
+\int_{-\infty}^{\infty} J_{0}(\lambda\nu r^{\prime})ds\vect{e}_{z} \right],
\end{eqnarray} 

\noindent where $r^{\prime}(s)$. We define a new variable: 
$t=v_{r}s+r\cos(\theta-\phi)$ $\Rightarrow$ $ds=(1/v_{r})dt$, 
$r^{\prime}=\sqrt{t^{2}+u^{2}}$, $u=r\sin(\theta-\phi)$. 
Then

\begin{eqnarray} \label{integral}
\vect{\mathcal{X}}\vect{F}_{L}(\vect{\theta}, \vect{x}) 
&=&  F_{0} \frac{1}{v_{r}} \Bigg\{
\lambda r \int_{-\infty}^{\infty} \frac{1}{\sqrt{t^{2}+u^{2}}} 
J_{1}(\lambda\nu\sqrt{t^{2}+u^{2}})dt 
\left[ \vect{e}_{\phi}-\cos(\theta-\phi)\vect{e}_{\theta} \right] 
+ \lambda \int_{-\infty}^{\infty} \frac{t}{\sqrt{t^{2}+u^{2}}} 
J_{1}(\lambda\nu\sqrt{t^{2}+u^{2}})dt \vect{e}_{\theta} \nonumber \\ 
& & \hspace*{97mm} 
+\int_{-\infty}^{\infty} J_{0}(\lambda\nu \sqrt{t^{2}+u^{2}})
dt\vect{e}_{z} \Bigg\} \\
&=& 2 F_{0} \frac{1}{v_{r}} \Bigg\{
\lambda r \int_{0}^{\infty} \frac{1}{\sqrt{t^{2}+u^{2}}} 
J_{1}(\lambda\nu\sqrt{t^{2}+u^{2}})dt 
\left[ \vect{e}_{\phi}-\cos(\theta-\phi)\vect{e}_{\theta} \right]
+\int_{0}^{\infty} J_{0}(\lambda\nu \sqrt{t^{2}+u^{2}})dt\vect{e}_{z} \Bigg\}. 
\nonumber 
\end{eqnarray} 

\noindent We introduce another variable: $\sinh y=t/|u|$ $\Rightarrow$ 
$dt=|u|\cosh ydy$. This reduces to 

\begin{eqnarray}
\vect{\mathcal{X}}\vect{F}_{L}(\vect{\theta}, \vect{x}) 
= 2 F_{0} \frac{1}{v_{r}} \Bigg\{
\lambda u  \int_{0}^{\infty}J_{1}(2\omega\cosh y)dy \vect{e}_{r}(\theta)
+|u|\int_{0}^{\infty} J_{0}(2\omega\cosh y)\cosh ydy\vect{e}_{z} \Bigg\}, 
\end{eqnarray}

\noindent using $\vect{e}_{\phi}-\cos(\theta-\phi)\vect{e}_{\theta}
=\sin(\theta-\phi)\vect{e}_{r}(\theta)$, where $\omega=(1/2)\lambda\nu|u|>0$.
We can evaluate these integrals using  

\begin{eqnarray}
\int_{0}^{\infty}J_{m+n}(2z\cosh x) \cosh[(m-n)x]dx
=-\frac{\pi}{4} \left[ J_{m}(z)N_{n}(z)+J_{n}(z)N_{m}(z) \right], 
\hspace*{3mm} z>0
\end{eqnarray}

\noindent the formula 6.663(3) in Ref. \onlinecite{GR}. 
Here $J_{n}$ and $N_{m}$ are respectively Bessel functions 
of the first and second (Neumann) kind. We also\cite{F} need

\begin{eqnarray}
& & N_{-1/2}(z)=J_{1/2}(z), \hspace*{18mm} N_{1/2}(z)=-J_{-1/2}(z), \\
& & J_{1/2}(z)= \left( \frac{2}{\pi z} \right) ^{1/2} \sin z, \hspace*{10mm} 
J_{-1/2}(z)= \left( \frac{2}{\pi z} \right) ^{1/2} \cos z. \nonumber 
\end{eqnarray}

\noindent We respectively find

\begin{eqnarray}
\int_{0}^{\infty}J_{1}(2\omega\cosh y)dy 
= \frac{1}{2}\frac{1}{\omega} \sin 2\omega, 
\hspace*{10mm}
\int_{0}^{\infty}J_{0}(2\omega\cosh y)\cosh ydy 
= \frac{1}{2}\frac{1}{\omega} \cos 2\omega, 
\end{eqnarray}

\noindent for $m=1/2$, $n=1/2$ and $m=1/2$, $n=-1/2$.
This leads to

\begin{eqnarray} \label{XLe}
\vect{\mathcal{X}}\vect{F}_{L}(\vect{\theta}, \vect{x}) 
= F_{0} \frac{1}{v_{r}} \frac{1}{\omega}
\left[ \lambda u \sin 2\omega \vect{e}_{r}(\theta)
+|u| \cos 2\omega \vect{e}_{z} \right], 
\end{eqnarray}

\noindent that is equation (\ref{XLundquist}) upon rearranging.

\subsection{X-ray transform of a Trkalian Field}
\label{XraytrfTrkalian}

\subsubsection{Reduction of the integral} 
\label{reduceintegral}

  We can simplify (\ref{XFR1}) decomposing $\vect{\kappa}$ 
into two components which are respectively parallel and orthogonal 
to $\vect{\theta}$: $\vect{\kappa}=\vect{\kappa}_{\parallel}
+\vect{\kappa}_{\perp}=u\vect{\theta}+v\vect{e}_{r}(\phi)$. That 
is $\vect{\kappa}_{\parallel}=u\vect{\theta}$, $u=\cos \theta$ and 
$\vect{\kappa}_{\perp}=v\vect{e}_{r}(\phi)$, $v=\sqrt{1-u^{2}}=\sin \theta$ 
where $\vect{e}_{r}(\phi)$ is the unit radial vector parametrized by angle 
$\phi$ in the plane $\theta^{\perp}$ orthogonal to $\vect{\theta}$: 
$\vect{e}_{r}(\phi)\cdot\vect{\theta}=0$. Then 
$\vect{\kappa}\cdot\vect{x}=[u\vect{\theta}+v\vect{e}_{r}(\phi)]\cdot
\vect{x}$, $\delta(\vect{\kappa}\cdot\vect{\theta})=\delta(u)$ and 
$d\Omega_{\vect{\kappa}}=\sin \theta d\theta d\phi=-dud\phi$. Hence 
we can write (\ref{XFR1}) as

\begin{eqnarray} 
\vect{\mathcal{X}}\vect{F}_{\lambda}(\vect{\theta}, \vect{x}) 
= \frac{1}{4\pi}\lambda\nu \int_{C} \int_{u=-1}^{u=1}
\vect{F}_{\lambda}^{R} \big( [u\vect{\theta}+\sqrt{1-u^{2}}\vect{e}_{r}(\phi)]
\cdot\vect{x}, u\vect{\theta}+\sqrt{1-u^{2}}\vect{e}_{r}(\phi) \big) \delta(u) 
dud\phi,
\end{eqnarray}

\noindent where $C$ is the unit circle in the plane $\theta^{\perp}$ 
which corresponds to a great circle on $S_{\vect{\kappa}}^{2}$ 
(the intersection of the plane with the sphere). 
This reduces to 

\begin{eqnarray} \label{XFR2}
\vect{\mathcal{X}}\vect{F}_{\lambda}(\vect{\theta}, \vect{x}) 
= \frac{1}{4\pi}\lambda\nu \int_{C} \vect{F}_{\lambda}^{R} 
\big( \vect{e}_{r}(\phi)\cdot\vect{x}, \vect{e}_{r}(\phi) \big) d\phi 
= \frac{1}{(2\pi)^{1/2}}\frac{1}{g}\frac{1}{\lambda\nu}
\int_{C} e^{i\lambda\nu\vect{e}_{r}(\phi)\cdot\vect{x}}
\vect{Q}_{\lambda}(\vect{e}_{r}(\phi))s_{\lambda}(\lambda\nu\vect{e}_{r}(\phi))
d\phi, 
\end{eqnarray}

\noindent the integral along the great circle $C$ in $S_{\vect{\kappa}}^{2}$ 
determined by $\vect{\theta}$ in the transform space.  

  One can further try writing (\ref{XFR2}) in terms of rotation about 
the axis determined by $\vect{\theta}$ through angle $\phi$ in the plane 
$\theta^{\perp}$ since $\vect{Q}_{\lambda}(\vect{e}_{r}(\phi))$ can be 
regarded as an \textit{eigenfunction} of this rotation.\cite{M1}

\subsubsection{Fourier slice-projection theorem}
\label{Fouriersliceprojection}

  If we use the Curl expansion\cite{KS1, Ml1, Ml3} 
for a Trkalian field, we are led to

\begin{eqnarray}
\vect{{\mathcal{F}}}[\vect{F}_{\lambda}(\vect{x})](\vect{\xi})
&=& \frac{1}{(2\pi)^{3/2}}\int e^{-i\vect{\xi}\cdot\vect{x}} 
\vect{F}_{\lambda}(\vect{x}) d^{3}x 
=\frac{1}{(2\pi)^{3}}\frac{1}{g} 
\int e^{-i\vect{\xi}\cdot\vect{x}} 
\int e^{i\vect{k}\cdot\vect{x}} \vect{Q}_{\lambda}(\vect{k})
f_{\lambda}(\vect{k})d^{3}kd^{3}x \\ 
&=& \frac{1}{g} \vect{Q}_{\lambda}(\vect{\xi})
f_{\lambda}(\vect{\xi}) \nonumber \\
&=& \frac{1}{g} \vect{Q}_{\lambda}(\vect{\xi}) 
\frac{\delta(\xi-\lambda\nu)}{\xi^{2}}s_{\lambda}(\vect{\xi}), 
\hspace*{5mm} \xi=|\vect{\xi}|. \nonumber
\end{eqnarray}

\noindent If we substitute this in Fourier slice-projection 
theorem (\ref{Fourierslicproj}), we immediately find
 
\begin{eqnarray}
\vect{\mathcal{X}}\vect{F}_{\lambda}(\vect{\theta}, \vect{x})
&=& (2\pi)^{1/2} \vect{{\mathcal{F}}}^{-1}
\big\{ \vect{{\mathcal{F}}}[\vect{F}_{\lambda}(\vect{x})](\vect{\xi}) \big\} 
=\frac{1}{(2\pi)^{1/2}} \int_{\theta^{\perp}} 
e^{i\vect{\xi}\cdot\vect{x}} 
\vect{{\mathcal{F}}}[\vect{F}_{\lambda}(\vect{x})](\vect{\xi}) d^{2}\xi \\ 
&=& \frac{1}{(2\pi)^{1/2}}\frac{1}{g}\frac{1}{\lambda\nu} \int_{C} 
e^{i\lambda\nu\vect{e}_{r}(\phi)\cdot\vect{x}} 
\vect{Q}_{\lambda}(\vect{e}_{r}(\phi)) 
s_{\lambda}(\lambda\nu\vect{e}_{r}(\phi)) d\phi, \nonumber 
\end{eqnarray}

\noindent (\ref{XFR2}), using $\vect{\xi}=\xi\vect{e}_{r}(\phi) 
\, \in \theta^{\perp}$, $d^{2}\xi=\xi d\xi d\phi$ and 
$\vect{Q}_{\lambda}(\lambda\nu\vect{e}_{r}(\phi))
=\vect{Q}_{\lambda}(\vect{e}_{r}(\phi))$ since $\lambda\nu=|\nu|>0$.

\subsection{The integral operators: $\vect{\mathcal{A}}^{\alpha}
=\vect{\mathcal{U}}^{\alpha}+i\vect{\mathcal{V}}^{\alpha}$}
\label{transforms}

  The family of integrals $\vect{\mathcal{A}}^{\alpha}
=\vect{\mathcal{U}}^{\alpha}+i\vect{\mathcal{V}}^{\alpha}$, 
$\alpha \in \mathbb{C}$, for $Re(\alpha)>0$, ($n=2$) are 
given by 

\begin{eqnarray}
\vect{\mathcal{U}}^{\alpha}[\vect{G}(\vect{\kappa})](\vect{\theta})
&=& \frac{\Gamma((1-\alpha)/2)}{2\pi\Gamma(\alpha/2)} 
\int_{S^{2}_{\vect{\theta}}} \vect{G}(\vect{\kappa}) 
\frac{1}{|\vect{\theta}\cdot\vect{\kappa}|^{1-\alpha}} 
d\Omega_{\vect{\kappa}}, 
\hspace*{3mm} \alpha \neq 1, 3, 5, ... \\
\vect{\mathcal{V}}^{\alpha}[\vect{G}(\vect{\kappa})](\vect{\theta})
&=& \frac{\Gamma(1-\alpha/2)}{2\pi\Gamma((1+\alpha)/2)} 
\int_{S^{2}_{\vect{\theta}}} \vect{G}(\vect{\kappa}) 
\frac{1}{|\vect{\theta}\cdot\vect{\kappa}|^{1-\alpha}} 
sgn(\vect{\theta}\cdot\vect{\kappa})d\Omega_{\vect{\kappa}}, 
\hspace*{3mm} \alpha \neq 2, 4, 6, ... 
\nonumber
\end{eqnarray}

\noindent $\vect{G}(\vect{\kappa})\in C^{\infty}(\mathbb{S}^{2})$. See 
the Refs. \onlinecite{Rubin3, Rubin4} and the references therein.
The transform $\vect{\mathcal{U}}^{\alpha}$ ($\vect{\mathcal{V}}^{\alpha}$) 
represents the even (odd) part of $\vect{\mathcal{A}}^{\alpha}$ and 
annihilates odd (even) functions. For $Re(\alpha)\leq 0$, these are 
to be understood in the sense of analytic continuation. 
In the case $\alpha\longrightarrow 0$

\begin{eqnarray} 
\vect{\mathcal{U}}^{0}[\vect{G}(\vect{\kappa})]
=\frac{1}{2\pi^{1/2}}\vect{\mathcal{M}}[\vect{G}(\vect{\kappa})],
\end{eqnarray}

\noindent and $\vect{\mathcal{V}}^{1}$ is related to hemispherical 
transform.\cite{Rubin3, Rubin4}

  The inverse transforms\cite{Rubin3, Rubin4} 
are given as 

\begin{eqnarray} 
(\vect{\mathcal{U}}^{\alpha})^{-1}=\vect{\mathcal{U}}^{-1-\alpha}, 
\hspace*{5mm} 
(\vect{\mathcal{V}}^{\alpha})^{-1}=\vect{\mathcal{V}}^{-1-\alpha},
\end{eqnarray}

\noindent in the sense of analytic continuation for certain 
values of $\alpha$. This inversion of $\vect{\mathcal{U}}^{\alpha}$ 
was established by Semyanistyi who studied the connection of 
$\vect{\mathcal{U}}^{\alpha}$ with Fourier transform.\cite{VIS, VIS1} 

  We have\cite{Rubin3, Rubin4} the relation 

\begin{eqnarray} \label{Fourierhomogeneous}
\int_{R^{3}} \frac{\vect{G}(\vect{\kappa})}{k^{2+\alpha}}
e^{i\vect{k}\cdot\vect{\eta}} d^{3}k 
=c_{\alpha} \eta^{\alpha-1} A^{\alpha}[\vect{G}(\vect{\kappa})]
(\vect{\theta}, \vect{x}), 
\hspace*{5mm} 
c_{\alpha, n}=2^{1-\alpha}\pi^{3/2}
\end{eqnarray}

\noindent where $\vect{k}=k\vect{\kappa}$, $k=|\vect{k}|$, 
$\vect{\eta}=\eta\vect{\theta}$, $\eta=|\vect{\eta}|$. This 
can also be extended to all $\alpha\in \mathbb{C}$ by analytic continuation.

  Note $\Gamma(1)=1$, $\Gamma(-1/2)=-2\pi^{1/2}$ for our purpose. 

\subsubsection{Evaluation of an integral}
\label{evaluationintegral}  

  We decompose $\vect{\theta}$ and $\vect{\kappa}$ 
into two components which are respectively parallel and orthogonal 
to $\vect{\kappa}^{\prime}$: $\vect{\theta}=\vect{\theta}_{\parallel}
+\vect{\theta}_{\perp}=u\vect{\kappa}^{\prime}+v\vect{e}_{r}(\phi)$. That 
is $\vect{\theta}_{\parallel}=u\vect{\kappa}^{\prime}$, $u=\cos \theta$ and 
$\vect{\theta}_{\perp}=v\vect{e}_{r}(\phi)$, $v=\sqrt{1-u^{2}}=\sin \theta$ 
where $\vect{e}_{r}(\phi)$ is the unit radial vector parametrized by angle 
$\phi$, in the plane $\kappa^{\prime\perp}$ orthogonal to 
$\vect{\kappa}^{\prime}$: $\vect{e}_{r}(\phi)\cdot\vect{\kappa}^{\prime}=0$. 
Also $\vect{\kappa}=\vect{\kappa}_{\parallel}+\vect{\kappa}_{\perp}$. 
Then $\vect{\kappa}\cdot\vect{\theta}
=u\vect{\kappa}_{\parallel}\cdot\vect{\kappa}^{\prime}
+\sqrt{1-u^{2}}\vect{\kappa}_{\perp}\cdot\vect{e}_{r}(\phi)$, 
$\delta(\vect{\kappa}^{\prime}\cdot\vect{\theta})=\delta(u)$ 
and $d\Omega_{\vect{\theta}}=\sin \theta d\theta d\phi=-dud\phi$. 
The integral (\ref{integralimp}) becomes

\begin{eqnarray} 
I(\vect{\kappa}, \vect{\kappa}^{\prime}) 
= \int_{\phi=0}^{2\pi} \int_{u=-1}^{u=1}
\frac{\delta(u)}{|u\vect{\kappa}_{\parallel}\cdot\vect{\kappa}^{\prime}
+\sqrt{1-u^{2}}\vect{\kappa}_{\perp}\cdot\vect{e}_{r}(\phi)|^{2}}
dud\phi.
\end{eqnarray}

\noindent This reduces to 

\begin{eqnarray} \label{impint}
I(\vect{\kappa}, \vect{\kappa}^{\prime}) 
= \int_{\phi=0}^{2\pi} 
\frac{1}{[\vect{\kappa}_{\perp}\cdot\vect{e}_{r}(\phi)]^{2}}
d\phi.
\end{eqnarray}

\noindent If we use the planewave decomposition\cite{K, GS} 
of Dirac delta function 

\begin{eqnarray} \label{Diracplanewave}
\delta(\vect{x})=-\frac{1}{4\pi^{2}} 
\int_{S^{1}} \frac{1}{[\vect{x}\cdot\vect{v}(\psi)]^{2}} d\psi, 
\hspace*{5mm} |\vect{v}|=1
\end{eqnarray}

\noindent in the plane $\kappa^{\prime\perp}$, we find

\begin{eqnarray} 
I(\vect{\kappa}, \vect{\kappa}^{\prime}) 
= -4\pi^{2}\delta(\vect{\kappa}_{\perp}).
\end{eqnarray}

\noindent Thus we are led to

\begin{eqnarray} \label{impintegral}
I(\vect{\kappa}, \vect{\kappa}^{\prime}) 
=-4\pi^{2}[\delta(\vect{\kappa}-\vect{\kappa}^{\prime}) 
+\delta(\vect{\kappa}+\vect{\kappa}^{\prime})], 
\end{eqnarray}

\noindent using the fact that as 
$\vect{\kappa}_{\perp}=\vect{\kappa}-\vect{\kappa}_{\parallel}=0$, 
we have $\vect{\kappa}_{\parallel}=\pm\vect{\kappa}^{\prime}$ 
and hence the identity

\begin{eqnarray} \label{Identityimp}
\delta(\vect{\kappa}_{\perp})
=\delta(\vect{\kappa}-\vect{\kappa}_{\parallel})
=\delta(\vect{\kappa}-\vect{\kappa}^{\prime}) 
+\delta(\vect{\kappa}+\vect{\kappa}^{\prime}). 
\end{eqnarray}

\noindent [Simply: $\vect{\kappa}=\cos\alpha\vect{\kappa}^{\prime}
+\sin\alpha\vect{e}_{r}$ in the plane spanned by 
$\vect{\kappa}^{\prime}$ and $\vect{e}_{r}$, thus 
$\delta(\vect{\kappa}_{\perp})=\delta(\sin \alpha)=\delta(\alpha)
+\delta(\alpha-\pi)=\delta(\vect{\kappa}-\vect{\kappa}^{\prime})
+\delta(\vect{\kappa}+\vect{\kappa}^{\prime})$.]

\subsubsection{Example: Lundquist field}
\label{InvFunkSemLunquist} 

   If we substitute the X-ray transform (\ref{XLundquist}) 
[that is (\ref{XLe})] of the Lundquist field in (\ref{InvFunkSem}), 
we find

\begin{eqnarray} \label{InvFunkSemLundquistint}
\vect{F}_{L\lambda}^{R} 
(\vect{\kappa}\cdot\vect{x}, \vect{\kappa}) 
&=& -\frac{1}{\pi}F_{0}\frac{1}{\nu^{2}} 
\int_{\theta=0}^{2\pi}  
\big\{ \lambda\cos\theta \sin\left[ \lambda\nu r\sin(\theta-\phi) \right] 
\vect{e}_{x}+\lambda\sin\theta \sin\left[ \lambda\nu r\sin(\theta-\phi) \right] 
\vect{e}_{y} \big. \\
& & \hspace*{67mm} \big. 
+\cos \left[ \lambda\nu r\sin(\theta-\phi) \right] \vect{e}_{z} \big\}
I_{1}(\beta, \theta-\psi) d\theta, 
\nonumber
\end{eqnarray}

\noindent where 

\begin{eqnarray} \label{I1}
I_{1}(\beta, \theta-\psi) 
=\int_{\alpha=0}^{\pi}
\frac{1}{[\sin\alpha \sin\beta \cos(\theta-\psi)+\cos\alpha \cos\beta]^{2}}
d\alpha. 
\end{eqnarray}

\noindent Here $\vect{\theta}=v_{r}\vect{e}_{r}(\theta)+v_{z}\vect{e}_{z}$, 
$v_{r}=sin\alpha$, $v_{z}=\cos\alpha$ $\Rightarrow$ 
$d\Omega_{\vect{\theta}}=\sin\alpha d\alpha d\theta$ and 
$\vect{\kappa}=\kappa_{r}\vect{e}_{r}(\psi)+\kappa_{z}\vect{e}_{z}$, 
$\kappa_{r}=sin\beta$, $\kappa_{z}=\cos\beta$. 

  Consider the integral 

\begin{eqnarray}
I = \int_{\alpha=0}^{2\pi} \frac{1}{[\vect{y}\cdot\vect{v}(\alpha)]^{2}}d\alpha
= I_{1}+I_{2},
\end{eqnarray}

\noindent where $\vect{y}=\cos\beta\vect{e}_{x}+\sin\beta\cos(\theta-\psi)
\vect{e}_{y}$ and $\vect{v}(\alpha)=\cos\alpha\vect{e}_{x}+\sin\alpha
\vect{e}_{y}$, $|\vect{v}|=1$. Here $I_{1}$ is the integral (\ref{I1}) and 

\begin{eqnarray}
I_{2} &=& \int_{\alpha=\pi}^{2\pi} \frac{1}{[\vect{y}\cdot\vect{v}(\alpha)]^{2}} 
d\alpha.
\end{eqnarray}

\noindent We immediately find $I_{2}=I_{1}$
using: $\alpha^{\prime}=\alpha-\pi$. This 
leads to 

\begin{eqnarray}
I_{1} = \frac{1}{2} I 
= -2\pi^{2} \delta(\vect{y})
= -2\pi^{2} \delta(\cos\beta) \delta(\sin\beta\cos(\theta-\psi)), 
\end{eqnarray}

\noindent using (\ref{Diracplanewave}) and 
$\delta(\vect{y})=\delta(y_{1}\vect{e}_{x}+y_{2}\vect{e}_{y}) 
=\delta(y_{1})\delta(y_{2})$. This reduces to 

\begin{eqnarray}
I_{1} = -2\pi^{2} \delta(\cos\beta) \delta(\cos(\theta-\psi)) 
= -2\pi^{2} \delta(\kappa_{z}) \left[ \delta(\theta-\psi-\pi/2)
+\delta(\theta-\psi-3\pi/2) \right]. 
\end{eqnarray}

\noindent If we substitute this in (\ref{InvFunkSemLundquistint}), 
then we find $\vect{F}^{\mathcal{R}}_{L(\lambda=1)}
(\vect{\kappa}\cdot\vect{x}, \vect{\kappa})$,  
(\ref{RadonLundquist}) for $\lambda=1$ after 
a straightforward manipulation.

\section{MATHEMATICAL METHODS OF TOMOGRAPHY}

\subsection{The identity}
\label{Identity}

  The identity (\ref{essential}) follows as

\begin{eqnarray}
\int_{S^{2}_{\vect{\theta}}} \vect{\mathcal{D}}\vect{F}(\vect{\theta}, \vect{x})
h(\vect{\theta}\cdot\vect{b}) d\Omega_{\vect{\theta}}  
&=& \int_{S^{2}_{\vect{\theta}}} \int_{t=0}^{\infty} \vect{F}(\vect{x}+t\vect{\theta})
h(t\vect{\theta}\cdot\vect{b}) t^{2} dt d\Omega_{\vect{\theta}}, 
\hspace*{5mm} \vect{x}^{\prime}=t\vect{\theta}, \, 
\vect{x}^{\prime\prime}=\vect{x}+\vect{x}^{\prime} \\    
&=& \int \int_{-\infty}^{\infty} \vect{F}(\vect{x}^{\prime\prime}) 
\delta(p-\vect{b}\cdot\vect{x}^{\prime\prime}) 
h(p-\vect{b}\cdot\vect{x}) dp d^{3}x^{\prime\prime}
\nonumber \\ 
&=& \int_{-\infty}^{\infty} \vect{F}^{\mathcal{R}}(p, \vect{b}) 
h(p-\vect{b}\cdot\vect{x})dp, \nonumber 
\end{eqnarray}

\noindent similar to the derivation for scalar fields 
in Ref. \onlinecite{RK}, p. 277, Ref. \onlinecite{CD}.
We have only assumed $h(ap)=(1/a^{2})h(p)$, $a>0$.

\subsection{Grangeat' s method}
\label{Grangeat' s formula}

\subsubsection{Derivation}
\label{ProofGrangeat' s}

  The derivation 

\begin{eqnarray}
\frac{\partial}{\partial p}\vect{F}^{\mathcal{R}}
(p, \vect{\kappa})\Big|_{p=\vect{\kappa}\cdot\vect{x}} 
&=& \int \vect{F}(\vect{r}) \delta^{\prime}
(p-\vect{\kappa}\cdot\vect{r}) d^{3}r 
\Big|_{p=\vect{\kappa}\cdot\vect{x}} \\
&=& -\int \vect{F}(\vect{x}+\vect{r}^{\prime}) \delta^{\prime}
(\vect{\kappa}\cdot\vect{r}^{\prime}) d^{3}r^{\prime}, 
\hspace*{5mm} \vect{r}^{\prime}=\vect{r}-\vect{x}, \,
\delta^{\prime}(-t)=-\delta^{\prime}(t) 
\nonumber \\ 
&=& -\int_{S^{2}_{\vect{\theta}}}
\vect{\mathcal{D}}\vect{F}(\vect{\theta}, \vect{x})
\delta^{\prime} (\vect{\kappa}\cdot\vect{\theta}) 
d\Omega_{\theta},
\hspace*{5mm} 
\vect{r}^{\prime}=s\vect{\theta}, \, 
s=|\vect{r}^{\prime}|>0 \Rightarrow 
d^{3}r^{\prime}=s^{2}ds d\Omega_{\theta}, \,
\delta^{\prime}(st)
=\frac{1}{s^{2}} \delta^{\prime}(t)
\nonumber 
\end{eqnarray} 

\noindent follows the same reasoning\cite{TZG} for scalar fields.

  The identities\cite{L1} 

\begin{eqnarray} \label{identity1}
\int_{S^{2}_{\vect{\theta}}} \vect{F}(\vect{\theta})
\delta^{\prime} (\vect{\kappa}\cdot\vect{\theta}) 
d\Omega_{\theta}
= -\int_{S^{2}_{\vect{\theta}}\cap\kappa^{\perp}} 
\frac{\partial}{\partial \vect{\kappa}} 
\vect{F}(\vect{\theta}) d\vect{\theta} 
= -\int_{\psi=0}^{2\pi} \frac{\partial}{\partial q} 
\vect{F}(\vect{\theta}=q\vect{\kappa}+r\vect{v}(\psi))
\Big|_{q=0} d\psi,
\end{eqnarray}

\noindent and

\begin{eqnarray} \label{identity2}
\int_{S^{2}_{\vect{\theta}}} \vect{F}(\vect{x}\cdot\vect{\theta})
\delta^{\prime} (\vect{\kappa}\cdot\vect{\theta}) d\Omega_{\theta}
=-\vect{x}\cdot\vect{\kappa} \int_{S^{2}_{\vect{\theta}}\cap\kappa^{\perp}} 
\vect{F}^{\prime}(\vect{x}\cdot\vect{\theta}) d\vect{\theta},
\end{eqnarray}

\noindent are useful in handling these type of integrals.
Here ${\partial}/{\partial \vect{\kappa}}$ denotes directional 
derivative along $\vect{\kappa}$ [ $|\vect{\theta}|=1$, 
$|\vect{\kappa}|=1$ and $\vect{v}(\psi)$ takes values on 
the unit circle: $|\vect{v}|=1$ $\Rightarrow$ $q^{2}+r^{2}=1$, 
parametrized by angle $\psi$ in the plane $\kappa^{\perp}$ 
through the origin, Ref. \onlinecite{VP}, p. 71.] One 
can derive (\ref{identity2}) with a reasoning similar 
to the derivation of (\ref{identity1}). See Ref. 
\onlinecite{RK}, p. 277.

\subsubsection{Example}
\label{GrangeatTrkalian} 

   If we substitute $\vect{\mathcal{D}}\vect{F}(\vect{\theta}, \vect{x})$, 
(\ref{Dsimp}) in (\ref{Grangeat}), the term containing 
$\delta\delta^{\prime}$ vanishes and we find 

\begin{eqnarray} \label{Grangeatexample}
\frac{\partial}{\partial p}\vect{F}^{\mathcal{R}}
(p, \vect{\kappa}) \Big|_{p=\vect{\kappa}\cdot\vect{x}} 
=-i\frac{1}{k_{0}} e^{i k_{0}\vect{\kappa}_{0}\cdot\vect{x}} 
I^{\prime} \vect{F}_{0},
\end{eqnarray}

\noindent where

\begin{eqnarray} 
I^{\prime} = \int_{S^{2}_{\vect{\theta}}} 
\frac{1}{\vect{\kappa}_{0}\cdot\vect{\theta}}
\delta^{\prime}(\vect{\kappa}\cdot\vect{\theta}) d\Omega_{\vect{\theta}}
= \vect{\kappa}_{0}\cdot\vect{\kappa}   
\int_{S^{2}_{\vect{\theta}}\cap\kappa^{\perp}} 
\frac{1}{(\vect{\kappa}_{0}\cdot\vect{v})^{2}} d\psi
= \vect{\kappa}_{0}\cdot\vect{\kappa} 
I(\vect{\kappa}_{0}, \vect{\kappa}), 
\end{eqnarray}

\noindent (\ref{impint}). Here we have used the identity (\ref{identity1}) 
or (\ref{identity2}) and $(1/\vect{\kappa}_{0}\cdot\vect{\theta})^{\prime}
=-1/(\vect{\kappa}_{0}\cdot\vect{\theta})^{2}$ (in distributional\cite{K} 
sense). We decompose the vector $\vect{\kappa}_{0}=\vect{\kappa}_{0\parallel}
+\vect{\kappa}_{0\perp}$ into components which are respectively parallel and 
orthogonal to $\vect{\kappa}$ and $\vect{v}(\psi)\in\kappa^{\perp}$, 
$|\vect{v}|=1$. This yields

\begin{eqnarray}
I^{\prime} = -4\pi^{2} \vect{\kappa}_{0}\cdot\vect{\kappa} 
[\delta(\vect{\kappa}-\vect{\kappa}_{0}) 
+\delta(\vect{\kappa}+\vect{\kappa}_{0})]
= -4\pi^{2} [\delta(\vect{\kappa}-\vect{\kappa}_{0}) 
-\delta(\vect{\kappa}+\vect{\kappa}_{0})],
\end{eqnarray}

\noindent using the planewave decomposition of Dirac delta function 
(\ref{Diracplanewave}, \ref{impintegral}). If we substitute this in 
(\ref{Grangeatexample}), we find 

\begin{eqnarray}
\frac{\partial}{\partial p}\vect{F}^{\mathcal{R}}
(p, \vect{\kappa}) \Big|_{p=\vect{\kappa}\cdot\vect{x}} 
=4\pi^{2} i \frac{1}{k_{0}} e^{i k_{0}\vect{\kappa}_{0}\cdot\vect{x}} 
[\delta(\vect{\kappa}-\vect{\kappa}_{0}) 
-\delta(\vect{\kappa}+\vect{\kappa}_{0})] 
\vect{F}_{0},
\end{eqnarray}

\noindent (\ref{Radsimp}).

\subsubsection{Inversion for Trkalian fields}
\label{GrangetTrkalianinversion}

  If we substitute (\ref{DFR1},\ref {DFR2}) into 
the righthand side of (\ref{GrangetTrkalianinv}) 
and interchange the order of integrations, we find 

\begin{eqnarray}
\int_{S^{2}_{\vect{\theta}}} 
\vect{\theta} \bm{\times} 
\vect{\mathcal{D}}\vect{F}_{\lambda}
(\pm\vect{\theta}, \vect{x}) d\Omega_{\vect{\theta}}
=\frac{1}{(2\pi)^{1/2}}\frac{1}{g}\frac{1}{\lambda\nu} 
\int_{S_{\vect{\kappa}}^{2}} e^{i\lambda\nu\vect{\kappa}\cdot\vect{x}}
\vect{I}(\vect{\kappa}) \bm{\times} 
\vect{Q}_{\lambda}(\vect{\kappa})s_{\lambda}(\lambda\nu\vect{\kappa})
d\Omega_{\vect{\kappa}}, 
\end{eqnarray} 

\noindent where 

\begin{eqnarray}
\vect{I}(\vect{\kappa}) 
= \int_{S_{\vect{\theta}}^{2}} 
\vect{\theta} \delta^{\pm}
(\vect{\kappa}\cdot\vect{\theta})
d\Omega_{\vect{\theta}} 
= \pm\frac{1}{2\pi} i \vect{I}^{\prime}.
\end{eqnarray}

\noindent Here the first integral vanishes. We can evaluate the 
second integral $\vect{I}^{\prime}$ decomposing $\vect{\theta}$ 
into components which are respectively parallel and orthogonal 
to $\vect{\kappa}$. That is $\vect{\theta}=u\vect{\kappa}
+v\vect{e}_{r}(\phi)$, $u=\cos\theta$, $v=\sin\theta$ 
where $\vect{e}_{r}(\phi)$ is the unit radial vector 
parametrized by angle $\phi$ in the plane $\kappa^{\perp}$ 
orthogonal to $\vect{\kappa}$. Then $\vect{\kappa}\cdot
\vect{\theta}=u$ and $d\Omega_{\vect{\theta}}=-dud\phi$. 
Hence 

\begin{eqnarray} 
\vect{I}^{\prime} 
= \int_{S_{\vect{\theta}}^{2}} 
\frac{\vect{\theta}}{\vect{\kappa}\cdot\vect{\theta}}
d\Omega_{\vect{\theta}} 
= \int_{\phi=0}^{\phi=2\pi} \int_{u=-1}^{u=1} 
[u\vect{\kappa}+v\vect{e}_{r}(\phi)] \frac{1}{u} 
dud\phi
= 4\pi\vect{\kappa}, 
\end{eqnarray}

\noindent since the second integral vanishes, 
($\vect{e}_{r}=\cos\phi\vect{e}_{1}+\sin\phi\vect{e}_{2}$ 
in the plane $\kappa^{\perp}$). Thus we find 

\begin{eqnarray}
\int_{S^{2}_{\vect{\theta}}} 
\vect{\theta} \bm{\times} 
\vect{\mathcal{D}}\vect{F}_{\lambda}
(\pm\vect{\theta}, \vect{x}) d\Omega_{\vect{\theta}}
= \pm 4\pi \frac{1}{\nu} \vect{F}_{\lambda}(\vect{x}),
\end{eqnarray} 

\noindent using 
$\vect{\kappa}\bm{\times}\vect{Q}_{\lambda}(\vect{\kappa})
=-i\lambda\vect{Q}_{\lambda}(\vect{\kappa})$ and (\ref{invRadtrf}).

\subsection{Smith' s method}
\label{Smith}

  The equation (\ref{extxrayFourier}) follows as  

\begin{eqnarray} \label{extxrayFourierdetail}
\vect{\mathcal{G}}\vect{F}(\vect{\beta}, \vect{x}) 
&=& \vect{\mathcal{F}}[\vect{\mathfrak{g}}
\vect{F}(\vect{\alpha}, \vect{x})](\vect{\beta}, \vect{x}), 
\hspace*{5mm} \vect{\alpha}=\alpha\vect{\theta} 
\Rightarrow d^{3}\alpha=\alpha^{2} d\alpha d\Omega_{\vect{\theta}} \\
&=& \frac{1}{2} \frac{1}{(2\pi)^{3/2}} 
\int_{S^{2}_{\vect{\theta}}} \int_{\alpha=-\infty}^{\infty}    
\vect{\mathcal{X}}\vect{F}(\vect{\theta}, \vect{x})
|\alpha| e^{-i\alpha\vect{\theta}\cdot\vect{\beta}} 
d\alpha d\Omega_{\vect{\theta}} \nonumber \\
&=& \frac{1}{(2\pi)^{3/2}} \int_{S^{2}_{\vect{\theta}}} 
\vect{\mathcal{D}}\vect{F}(\vect{\theta}, \vect{x})
h(\vect{\theta}\cdot\vect{\beta}) d\Omega_{\vect{\theta}},
\nonumber 
\end{eqnarray}

\noindent where 

\begin{eqnarray} \label{identity3}
h(s) = \int_{\alpha=-\infty}^{\infty} |\alpha|e^{-i\alpha s}d\alpha
= (2\pi)^{1/2} i\partial_{s} \vect{\mathcal{F}}[sgn(\alpha)](s) 
= 2\partial_{s} \frac{1}{s} 
= -2 \frac{1}{s^{2}},
\hspace*{5mm} (s=\vect{\theta}\cdot\vect{\beta})
\end{eqnarray}

\noindent satisfies $h(as)=(1/a^{2})h(s)$, $a>0$ and
$h(-s)=h(s)$. Hence we obtain (\ref{extxrayFourier}), 
since $\vect{\beta}=\beta\vect{b}$, $s=\beta p$.

\subsection{Tuy' s method} 
\label{Tuy}

  The equation (\ref{extdbeamFourier}) follows as

\begin{eqnarray} 
\vect{\mathcal{G}}\vect{F}(\vect{\beta}, \vect{x}) 
&=& \vect{\mathcal{F}}[\vect{\mathfrak{g}}
\vect{F}(\vect{\alpha}, \vect{x})](\vect{\beta}, \vect{x}), 
\hspace*{5mm} \vect{\alpha}=\alpha\vect{\theta} 
\Rightarrow d^{3}\alpha=\alpha^{2} d\alpha d\Omega_{\vect{\theta}} \\
&=& \frac{1}{2} \frac{1}{(2\pi)^{3/2}} 
\int_{S^{2}_{\vect{\theta}}} \int_{\alpha=-\infty}^{\infty}    
\vect{\mathcal{D}}\vect{F}(\vect{\theta}, \vect{x})
(|\alpha|+\alpha) e^{-i\alpha\vect{\theta}\cdot\vect{\beta}} 
d\alpha d\Omega_{\vect{\theta}} \nonumber \\
&=& \frac{1}{(2\pi)^{3/2}} \int_{S^{2}_{\vect{\theta}}} 
\vect{\mathcal{D}}\vect{F}(\vect{\theta}, \vect{x})
f(\vect{\theta}\cdot\vect{\beta}) d\Omega_{\vect{\theta}}, 
\nonumber 
\end{eqnarray}

\noindent where 

\begin{eqnarray}
f(s) &=& \frac{1}{2} \int_{\alpha=-\infty}^{\infty} (|\alpha|+\alpha) 
e^{-i\alpha s} d\alpha
=(2\pi)^{1/2} 
\vect{\mathcal{F}}[\mathcal{R}(\alpha)](s) 
=\frac{1}{2} \int_{\alpha=-\infty}^{\infty} (|\alpha|-\alpha) 
e^{i\alpha s} d\alpha,
\hspace*{5mm} (s=\vect{\theta}\cdot\vect{\beta}) \\
&=& \frac{1}{2} [ h(s)-\frac{1}{i}k(s)]. \nonumber 
\end{eqnarray}

\noindent Here $\mathcal{R}(\alpha)=\alpha H(\alpha)$ is the Ramp 
function, $h(s)=2\partial_{s} 1/s$, (\ref{identity3}) and 

\begin{eqnarray}
k(s) = \partial_{s} \int_{\alpha=-\infty}^{\infty}  e^{i\alpha s}d\alpha 
= 2\pi \delta^{\prime}(s). 
\end{eqnarray}

\noindent Hence $f(s)=2\pi i \partial_{s} \delta^{-}(s)$.
This also satisfies $f(as)=(1/a^{2})f(s)$, $a>0$ and 
$f(-s)=-2\pi i \partial_{s} \delta^{+}(s)$. 
Thus we obtain (\ref{extdbeamFourier}), 
since $\vect{\beta}=\beta\vect{b}$, 
$s=\beta p$.

\subsection{Divergent beam transform of Lundquist field} 
\label{LundquistDivbeamTuy}

  If we substitute the Radon transform (\ref{RadonLundquist}) 
of the Lundquist field ($\lambda=1$) into the integral in 
(\ref{Tuymine3}) using $\vect{x}=r\vect{e}_{r}(\phi)+z\vect{e}_{z}$, 
$\vect{\theta}=v_{r}\vect{e}_{r}(\theta)+v_{z}\vect{e}_{z}$, 
and $\vect{b}=b_{r}\vect{e}_{r}(\psi)+b_{z}\vect{e}_{z}$ where 
$b_{r}=\sqrt{1-b^{2}_{z}}$, $d\Omega_{\vect{b}}=-db_{z}d\psi$, 
the integration over $b_{z}$ leads to 

\begin{eqnarray} \label{intytrf1}
I(\vect{\theta}, \vect{x})
&=& (2\pi)^{2}\vect{\mathcal{Y}}\vect{F}(\vect{\theta}, \vect{x}) \\ 
&=& -2\pi F_{0}\frac{1}{\nu}\frac{1}{v_{r}} \int_{\psi^{\prime}=0}^{2\pi} 
\left[ e^{i\nu r\cos(\psi^{\prime}+\theta-\phi)}\vect{L}(\psi^{\prime}+\theta)
e^{-i\nu r\cos(\psi^{\prime}+\theta-\phi)}\vect{L}^{\prime}(\psi^{\prime}+\theta) 
\right] \frac{1}{\cos\psi^{\prime}} d\psi^{\prime}. \nonumber
\end{eqnarray}

\noindent Here $\psi^{\prime}=\psi-\theta$ (and the integral over 
$\psi^{\prime}$ is insensitive to shift of limits of integration). 
This yields 

\begin{eqnarray} \label{intytrf2}
I(\vect{\theta}, \vect{x})
= -2\pi F_{0}\frac{1}{\nu}\frac{1}{v_{r}}
\left[ (I^{+}_{3}+I^{-}_{3})\vect{e}_{r}(\theta)-(I^{+}_{1}+I^{-}_{1})
\vect{e}_{\theta}-i(I^{+}_{2}-I^{-}_{2})\vect{e}_{z} \right],
\end{eqnarray}

\noindent where 

\begin{eqnarray}
& & I^{\pm}_{1}
= \int_{\psi^{\prime}=0}^{2\pi} e^{\pm i\nu r\cos(\psi^{\prime}+\theta-\phi)} 
d\psi^{\prime}, \hspace*{10mm} 
I^{\pm}_{2} 
= \int_{\psi^{\prime}=0}^{2\pi} e^{\pm i\nu r\cos(\psi^{\prime}+\theta-\phi)} 
\frac{1}{\cos\psi^{\prime}} d\psi^{\prime}, \\ 
& & \hspace*{59mm} 
I^{\pm}_{3} 
= \int_{\psi^{\prime}=0}^{2\pi} e^{\pm i\nu r\cos(\psi^{\prime}+\theta-\phi)} 
\tan\psi^{\prime} d\psi^{\prime}. \nonumber 
\end{eqnarray}

  We shall use the variable $\zeta=i\omega z$ where 
$z=e^{i\psi^{\prime}}$ and $\omega=e^{i(\theta-\phi)}$. Hence 
$i\cos(\psi^{\prime}+\theta-\phi)=(1/2)(\zeta-1/\zeta)$, 
$\cos \psi^{\prime}=(1/2)(1/i\omega)(1/\zeta)(\zeta^{2}-\omega^{2})$, 
$\tan \psi^{\prime}=-i(\zeta^{2}+\omega^{2})/(\zeta^{2}-\omega^{2})$,
and $d\psi^{\prime}=-i(1/\zeta)d\zeta$. These integrals become

\begin{eqnarray}
& & I^{\pm}_{1} 
=-i \int_{C^{\prime}} e^{\frac{1}{2}u^{\pm}(\zeta-1/\zeta)} 
\frac{1}{\zeta} d\zeta, 
\hspace*{10mm} 
I^{\pm}_{2}=2\omega \int_{C^{\prime}} e^{\frac{1}{2}u^{\pm}(\zeta-1/\zeta)} 
\frac{1}{\zeta^{2}-\omega^{2}} d\zeta, \\
& & \hspace*{55mm} 
I^{\pm}_{3}=-\int_{C^{\prime}} e^{\frac{1}{2}u^{\pm}(\zeta-1/\zeta)} 
\frac{1}{\zeta} \frac{\zeta^{2}+\omega^{2}}{\zeta^{2}-\omega^{2}} d\zeta, 
\nonumber 
\end{eqnarray}

\noindent where $u^{\pm}=\pm\nu r$ and $C^{\prime}$ is a unit circle 
around the origin (the integrals are insensitive to shift of limits 
of integration). Note the notation: the Cauchy principal value is meant 
in these contour integrals. 

    We shall make use of the generating function\cite{GNW} 
$e^{\frac{1}{2}u^{\pm} (\zeta-1/\zeta)}=\sum_{n=-\infty}^{\infty} 
\zeta^{n} J_{n}(u^{\pm})=J_{0}(u^{\pm})+\sum_{n=1}^{\infty} 
[\zeta^{n}+(-1)^{n}\zeta^{-n}] J_{n}(u^{\pm})$ 
for Bessel functions. 

  In $I^{\pm}_{1} $ there are no poles on $C^{\prime}$ and only 
the term of order $n=0$ in the generating function contributes 
to the integral

\begin{eqnarray} \label{intone}
I^{\pm}_{1} =2\pi J_{0}(\nu r). 
\end{eqnarray}

  If we substitute the generating function in $I^{\pm}_{2}$, 
we get 

\begin{eqnarray} 
I^{\pm}_{2}=2\omega \left\{ I^{\pm}_{2a} J_{0}(u^{\pm})
+\sum_{n=1}^{\infty} \left[ I^{\pm}_{2b}+(-1)^{n}I^{\pm}_{2c} \right] 
J_{n}(u^{\pm}) \right\}.
\end{eqnarray}

\noindent The contributions to the principal value integral $I^{\pm}_{2a}$ 
of the residues of simple poles at $\zeta=\pm\omega$ on $C^{\prime}$ cancel out

\begin{eqnarray}
I^{\pm}_{2a}=\int_{C^{\prime}} \frac{1}{\zeta^{2}-\omega^{2}}d\zeta
= \pi i Res[f=1/(\zeta^{2}-\omega^{2}), \zeta=\omega] 
+\pi i Res[f=1/(\zeta^{2}-\omega^{2}), \zeta=-\omega] 
= 0.
\end{eqnarray}

\noindent Meanwhile, for $n\geq 1$

\begin{eqnarray}
I^{\pm}_{2b}=\int_{C^{\prime}} \frac{\zeta^{n}}{\zeta^{2}-\omega^{2}}d\zeta
&=& \pi i Res[g=\zeta^{n}/(\zeta^{2}-\omega^{2}), \zeta=\omega] 
+\pi i Res[g=\zeta^{n}/(\zeta^{2}-\omega^{2}), \zeta=-\omega] \\ 
&=& \pi i \frac{1}{2} [1+(-1)^{n+1}] \omega^{n-1}, \nonumber
\end{eqnarray}

\noindent and

\begin{eqnarray}
I^{\pm}_{2c}=\int_{C^{\prime}} \frac{\zeta^{-n}}{\zeta^{2}-\omega^{2}}d\zeta
&=& 2\pi i Res[h=\zeta^{-n}/(\zeta^{2}-\omega^{2}), \zeta=0] 
+ \pi i Res[h=\zeta^{-n}/(\zeta^{2}-\omega^{2}), \zeta=\omega] \\
& & \hspace*{52mm} 
+\pi i Res[h=\zeta^{-n}/(\zeta^{2}-\omega^{2}), \zeta=-\omega] \nonumber \\
&=& -\pi i \frac{1}{2} [1+(-1)^{n+1}] \omega^{-(n+1)}. \nonumber
\end{eqnarray}

\noindent Thus

\begin{eqnarray} \label{inttwo}
I^{\pm}_{2}
&=& 2\pi i \omega \sum_{n=1}^{\infty} \frac{1}{2} [1+(-1)^{n+1}]  
\left[\omega^{n-1} -(-1)^{n} \omega^{-(n+1)} \right] J_{n}(u^{\pm}) 
= \pm 4\pi i \sum_{k=0}^{\infty} \cos [(2k+1)(\theta-\phi)]
J_{2k+1}(\nu r). 
\end{eqnarray}

  The integral $I^{\pm}_{3}$ reduces to 

\begin{eqnarray} 
I^{\pm}_{3}=- \left\{ I^{\pm}_{3a} J_{0}(u^{\pm})
+\sum_{n=1}^{\infty} \left[ I^{\pm}_{3b}+(-1)^{n}I^{\pm}_{3c} \right] 
J_{n}(u^{\pm}) \right\}. 
\end{eqnarray}

\noindent Here

\begin{eqnarray}
I^{\pm}_{3a}=\int_{C^{\prime}} 
\frac{1}{\zeta}\frac{\zeta^{2}+\omega^{2}}{\zeta^{2}-\omega^{2}}d\zeta
&=& 2\pi i Res[f=\zeta^{-1}(\zeta^{2}+\omega^{2})/(\zeta^{2}-\omega^{2}), 
\zeta=0] +\pi i Res[f=\zeta^{-1}(\zeta^{2}+\omega^{2})/(\zeta^{2}-\omega^{2}), 
\zeta=\omega] \nonumber \\
& & \hspace*{66mm}
+\pi i Res[f=\zeta^{-1}(\zeta^{2}+\omega^{2})/(\zeta^{2}-\omega^{2}), 
\zeta=-\omega] \nonumber \\
&=& 0, 
\end{eqnarray}

\noindent and since $n\geq 1$

\begin{eqnarray}
I^{\pm}_{3b}=\int_{C^{\prime}} 
\zeta^{n-1}\frac{\zeta^{2}+\omega^{2}}{\zeta^{2}-\omega^{2}}d\zeta
&=& +\pi i Res[g=\zeta^{n-1}(\zeta^{2}+\omega^{2})/(\zeta^{2}-\omega^{2}), 
\zeta=\omega]
+\pi i Res[g=\zeta^{n-1}(\zeta^{2}+\omega^{2})/(\zeta^{2}-\omega^{2}), 
\zeta=-\omega] \nonumber \\
&=& \pi i [1+(-1)^{n}] \omega^{n}, 
\end{eqnarray}

\noindent and 

\begin{eqnarray}
I^{\pm}_{3c}=\int_{C^{\prime}} 
\frac{1}{\zeta^{n+1}}\frac{\zeta^{2}+\omega^{2}}{\zeta^{2}-\omega^{2}}d\zeta
&=& 2\pi i Res[h=\zeta^{-(n+1)}(\zeta^{2}+\omega^{2})/(\zeta^{2}-\omega^{2}), 
\zeta=0] \\
& & +\pi i Res[h=\zeta^{-(n+1)}(\zeta^{2}+\omega^{2})/(\zeta^{2}-\omega^{2}), 
\zeta=\omega] \nonumber \\
& & +\pi i Res[h=\zeta^{-(n+1)}(\zeta^{2}+\omega^{2})/(\zeta^{2}-\omega^{2}), 
\zeta=-\omega]  
\nonumber \\
&=& -\pi i [1+(-1)^{n}] \omega^{-n}. \nonumber
\end{eqnarray}

\noindent Thus

\begin{eqnarray} \label{intthree}
I^{\pm}_{3}
= -\pi i \sum_{n=1}^{\infty} [1+(-1)^{n}]  
\left[\omega^{n} -(-1)^{n} \omega^{-n} \right] J_{n}(u^{\pm}) 
= 4\pi \sum_{k=1}^{\infty} \sin [2k(\theta-\phi)]
J_{2k}(\nu r). 
\end{eqnarray}

  Then we obtain (\ref{Lundquistdiftrf}) using 
(\ref{intone}, \ref{inttwo}, \ref{intthree})
in (\ref{intytrf1}, \ref{intytrf2}).

  We also need\cite{GNW}

\begin{eqnarray} \label{expansion}
\sin(z\sin\alpha)=2\sum_{k=0}^{\infty} \sin[(2k+1)\alpha] J_{2k+1}(z), 
\hspace{10mm} 
\cos(z\sin\alpha)=J_{0}(z)+2\sum_{k=1}^{\infty} \cos(2k\alpha) J_{2k}(z), 
\end{eqnarray}

\noindent for the Divergent beam transform (\ref{DivbeamLundquist}) 
of the Lundquist field.

\subsection{Gelfand-Goncharov' s method}
\label{GelfandGoncharov}

  If we substitute equation (\ref{GelfandGoncharovMoses}) 
in (\ref{invRadtrf}) and interchange the order of 
integrations, we find

\begin{eqnarray}
\vect{F}_{\lambda}(\vect{x}) 
=-\frac{1}{8\pi^{3}}\lambda\nu 
\int_{S^{2}_{\vect{\theta}}} \vect{\mathcal{D}}
\vect{F}_{\lambda}(\vect{\theta}, \vect{x}) 
I(\vect{\theta}) d\Omega_{\vect{\theta}},
\end{eqnarray}

\noindent where

\begin{eqnarray} \label{contint}
I(\vect{\theta})= \int_{S^{2}_{\vect{\kappa}}} 
\frac{1}{(\vect{\theta}\cdot\vect{\kappa})^{2}} 
d\Omega_{\vect{\kappa}}. 
\end{eqnarray}

\noindent If we decompose $\vect{\kappa}$ into components which 
are respectively parallel and orthogonal to $\vect{\theta}$: 
$\vect{\kappa}=\vect{\kappa}_{\parallel}+\vect{\kappa}_{\perp}$ 
($\vect{\kappa}_{\parallel}=u\vect{\theta}$, $u=\cos \theta$, 
and $d\Omega_{\vect{\kappa}}=\sin \theta d\theta d\phi=-dud\phi$), 
then we find 

\begin{eqnarray} 
I(\vect{\theta})=2\pi I,
\end{eqnarray}

\noindent where 

\begin{eqnarray} 
I=\int_{u=-1}^{1} \frac{1}{u^{2}} du. 
\end{eqnarray}

\noindent The principal value of this integral can 
be easily calculated using a contour integral: $I=-2$. 
This leads to (\ref{SphmeanDiv}).

\section{MINI-TWISTORS}

\subsection{Helmholtz equation: (mini-)twistor solution}
\label{Helmholtztwistor}

  We reproduce the following solution of Helmholtz equation from 
Ref. \onlinecite{WS} which is partially based on Hitchin' s notes.

  The solution\cite{WS} of wave equation 

\begin{eqnarray} \label{Wave}
\Box^{2} \psi=0, 
\end{eqnarray}

\noindent  $\Box^{2}=\partial^{2}/\partial t^{2}
-\nabla^{2}$, $\nabla^{2}=\partial^{2}/\partial x^{2}
+\partial^{2}/\partial y^{2}+\partial^{2}/\partial z^{2}$
is given by

\begin{eqnarray} \label{BatemanPenrose}
\psi(x, y, z, t) 
= \int_{C} g[\omega (t+z)+(x+iy), \omega(x-iy)+(t-z), \omega ] d\omega 
= \int_{C} g(p, q, \omega) d\omega, 
\end{eqnarray}

\noindent where $g$ is holomorphic in each of its arguments:
$p=x+iy+\omega(t+z)$, $q=t-z+\omega(x-iy)$ and $\omega$. 
We can easily verify (\ref{Wave}) using $\partial/\partial t
= \omega\partial/\partial p +\partial/\partial q$, 
$\partial/\partial x=\partial/\partial p 
+\omega \partial/\partial q$, $\partial/\partial y
=i \left( \partial/\partial p -\omega\partial/\partial q \right)$,
$\partial/\partial z=\omega\partial/\partial p-\partial/\partial q$.

  We can write the wave equation (\ref{Wave}) as $\partial^{2} 
\psi/\partial u \partial v-\partial^{2} \psi/\partial \zeta 
\partial \overline{\zeta}=0$ introducing new variables $u=t-z$, 
$v=t+z$, $\zeta=x+iy$, $\overline{\zeta}=x-iy$ and $\psi(x, y, z, t) 
\longrightarrow \psi(u, v, \zeta, \overline{\zeta})$. This is 
immediately satisfied since: $\partial/\partial u=\partial/\partial q$, 
$\partial/\partial v=\omega \partial/\partial p$, $\partial/\partial 
\zeta=\partial/\partial p$, $\partial/\partial \overline{\zeta}=\omega 
\partial/\partial q$. Then $p=\zeta+\omega v$, $q=\omega\overline{\zeta}+u$ 
in (\ref{BatemanPenrose}).

  For the Helmholtz equation,\cite{WS} we suppose 

\begin{eqnarray} 
\frac{\partial \psi}{\partial t} = -ik \psi.
\end{eqnarray}

\noindent This leads to $(\partial/\partial t) 
g[\omega (t+z)+(x+iy), \omega(x-iy)+(t-z), \omega ]=-ik g$
using (\ref{BatemanPenrose}), that is 

\begin{eqnarray} \label{eqn1}
\left( \omega \frac{\partial }{\partial p} 
+\frac{\partial }{\partial q} \right) g(p, q, \omega) =-ik g.
\end{eqnarray}

  A simple method for solving this equation is to use an integrating 
factor. For example we can use\cite{WS} an integrating factor that 
is a function of $q$ so as to remove the right-hand side 

\begin{eqnarray} \label{intfac1}
g(p, q, \omega)= e^{-ikq} h(p, q, \omega).
\end{eqnarray}

\noindent Then (\ref{eqn1}) becomes 

\begin{eqnarray} \label{eqn2}
\omega \frac{\partial h}{\partial p} 
+\frac{\partial h}{\partial q} =0.
\end{eqnarray}

\noindent The solution of this equation is given by $h(p, q, \omega) 
= H(\eta, \omega)$, where $H$ is a holomorphic function of two arguments: 
$\eta_{\vect{x}}(\omega)= p-\omega q = x+iy+2z\omega-(x-iy)\omega^{2}$
and $\omega$.\cite{WS}

  Then $g(p, q, \omega)= e^{-ikq} H(\eta_{\vect{x}}(\omega), \omega)$
and (\ref{BatemanPenrose}) yields

\begin{eqnarray} \label{timeharmonicextension}
\psi= e^{-ikt} \phi(x, y, z),
\end{eqnarray}

\noindent where 

\begin{eqnarray} \label{scalarHelmholtzSol}
\phi(x, y, z)= \int_{C} e^{-ik[\omega(x-iy)-z]} 
H(\eta_{\vect{x}}(\omega), \omega) d\omega,
\end{eqnarray}

\noindent satisfies the Helmholtz equation $\nabla^{2}\phi=-k^{2}\phi$.
This reduces to solution of Laplace equation for $k=0$. 

  We could also use another\cite{WS} integrating factor 

\begin{eqnarray} \label{intfac2a}
g(p, q, \omega)= e^{-i\frac{k}{2}(p/\omega+q)} H(p-\omega q, \omega).
\end{eqnarray}

\noindent This would yield 

\begin{eqnarray} \label{intfac2b}
\phi=\int_{C} e^{-i\frac{k}{2}[\omega(x-iy)+(x+iy)/\omega]}
H(\eta_{\vect{x}}(\omega), \omega) d\omega. 
\end{eqnarray} 

  Any solution of (\ref{eqn1}) can be written in this form.\cite{WS} 
Because introducing new variables: $\alpha=p-\omega q$, $\beta=p+\omega q$,
the equation (\ref{eqn1}) becomes 

\begin{eqnarray} \label{gensol}
\frac{\partial g}{\partial \beta}
=-i \frac{k}{2} \frac{1}{\omega} g,
\end{eqnarray} 

\noindent and the general solution of this equation is 
$g(\alpha, \beta, \omega)=e^{-i \frac{k}{2} \frac{1}{\omega} \beta} 
h(\alpha, \omega)$, $\alpha=\eta(\omega)$.

\subsection{Contour integrals for Bessel functions}
\label{Otherformulas}

  We need the following integrals

\begin{eqnarray} \label{Besselintegral1}
J_{m}(\nu r) \cos m\varphi = \frac{i^{m}}{2\pi} \int_{\theta=0}^{2\pi} 
e^{-i\nu (x\cos \theta + y\sin \theta)} \cos m\theta d\theta, 
\hspace*{2mm} J_{m}(\nu r) \sin m\varphi = \frac{i^{m}}{2\pi} 
\int_{\theta=0}^{2\pi} 
e^{-i\nu (x\cos \theta + y\sin \theta)} \sin m\theta d\theta, 
\hspace*{2mm} 
\end{eqnarray}

\noindent where $x=r\cos \varphi$, $y=r\sin \varphi$. 
These lead to 

\begin{eqnarray} \label{Besselintegral2}
J_{m}(\nu r) e^{im\varphi}=\frac{i^{m}}{2\pi} \int_{\theta=0}^{2\pi} 
e^{-i\nu (x\cos \theta + y\sin \theta)} e^{im\theta} d\theta.
\end{eqnarray}

\noindent Hence

\begin{eqnarray} \label{Besselintegral3}
& & J_{0}(\nu r)=\frac{1}{2\pi} \int_{\theta=0}^{2\pi} 
e^{-i\nu (x\cos \theta + y\sin \theta)} d\theta, 
\hspace*{8mm} 
J_{1}(\nu r) \cos \varphi=\frac{i}{2\pi} \int_{\theta=0}^{2\pi} 
e^{-i\nu (x\cos \theta + y\sin \theta)} \cos \theta d\theta, \\ 
& & \hspace*{67mm} J_{1}(\nu r) \sin \varphi=\frac{i}{2\pi} 
\int_{\theta=0}^{2\pi} e^{-i\nu (x\cos \theta + y\sin \theta)} 
\sin \theta d\theta. \nonumber 
\end{eqnarray}

\noindent See Ref. \onlinecite{W}. One can prove 
these, for example following Ref. \onlinecite{Ml3}.
We have: $J_{0}(-s)=J_{0}(s)$, $J_{1}(-s)=-J_{1}(s)$.

  Another expression\cite{MH} is 

\begin{eqnarray} \label{Besselintegral4}
J_{0}(\beta)=\frac{1}{2\pi} \int_{0}^{2\pi} e^{i\beta\sin u} du.
\end{eqnarray}

\subsection{Solution with Laurent series}
\label{SolutionLaurent}

  If we use $f=(1/2)[\omega(x-iy)+(x+iy)/\omega]$ with 
$\omega=i\omega^{\prime}$, ($k=\nu$) and choose $u=h(\omega^{\prime})$ 
which has a Laurent series: $h(\omega^{\prime})=1/{\omega^{\prime}}^{(n+1)}$,  
the equation (\ref{Trkaliantwistor1}) leads to

\begin{eqnarray} \label{sollau}
\vect{F}(\vect{x})=\int_{C} 
\left[ i\left( 1+{\omega^{\prime}}^{2} \right), \, 
-\left( 1-{\omega^{\prime}}^{2} \right), \, 
-2{\omega^{\prime}} \right] 
e^{-\frac{1}{2}\nu [(x+iy)/\omega^{\prime}-(x-iy)\omega^{\prime}]} 
\frac{1}{{\omega^{\prime}}^{(n+1)}} d\omega^{\prime}.
\end{eqnarray}

\noindent The generating function\cite{WS} 
$e^{\frac{1}{2}p(t-1/t)}=\sum_{m=-\infty}^{\infty} t^{m} J_{m}(p)$
for the Bessel functions yields

\begin{eqnarray}
e^{-\frac{1}{2}\nu [(x+iy)/\omega^{\prime}-(x-iy)\omega^{\prime}]}  
=\sum_{m=-\infty}^{\infty} e^{-im\varphi} {\omega^{\prime}}^{m} J_{m}(\nu r), 
\end{eqnarray}

\noindent with $p=-\nu r$, 
$t=-(x-iy)\omega^{\prime}/r=-e^{-i\varphi}\omega^{\prime}$, 
($x+iy=re^{i\varphi}$) and $J_{m}(-x)=(-1)^{m}J_{m}(x)$.
The integrals in (\ref{sollau}) can be evaluated using 

\begin{eqnarray}
\int_{C}\sum_{m=-\infty}^{\infty} c_{m} {\omega^{\prime}}^{m} 
{\omega^{\prime}}^{-(n+1)}d\omega^{\prime}=2\pi i c_{n}, 
\hspace*{3mm} c_{m}=e^{-im\varphi} J_{m} (\nu r), 
\end{eqnarray}

\noindent which is based on residues.\cite{WS} 
We find

\begin{eqnarray}
\vect{F}(\vect{x})= -2\pi e^{-in\varphi}   
\left( \left[ J_{n}(\nu r) + e^{i2\varphi} J_{n-2}(\nu r)\right], \, 
i \left[ J_{n}(\nu r) - e^{i2\varphi} J_{n-2}(\nu r)\right], \, 
2 i e^{i\varphi}J_{n-1}(\nu r) \right).
\end{eqnarray}

\noindent This reduces to 

\begin{eqnarray} \label{ccCKz}
\vect{F}(\vect{x})= 4\pi i e^{-im\varphi}  
\left[ im\frac{1}{\nu r} J_{m}(\nu r) \vect{e}_{r} 
+ J_{m}^{\prime}(\nu r) \vect{e}_{\varphi} 
- J_{m}(\nu r) \vect{e}_{z} \right], 
\hspace*{5mm} m=n-1
\end{eqnarray}

\noindent in cylindrical coordinates: 
$\vect{e}_{r}=\cos\varphi\vect{e}_{x}+\sin\varphi\vect{e}_{y}$,
$\vect{e}_{\varphi}=-\sin\varphi\vect{e}_{x}+\cos\varphi\vect{e}_{y}$, 
using\cite{AJ} the identities: $J_{n}(x)-J_{n-2}(x)=-2J_{n-1}^{\prime}(x)$, 
$x[J_{n}(x)+J_{n-2}(x)]=2(n-1)J_{n-1}(x)$, $(x=\nu r)$. This is a circular 
cylindrical CK\cite{CK} solution\cite{Y, TC} with no $z$ dependence, upto 
conventions.


\begin{thebibliography}{199}

\bibitem{KS} K. Saygili, J. Math. Phys. {\bf {51}}, 033513 (2010).  

\bibitem{VT} V. Trkal, \v{C}asopis pro p\v{e}stov\'{a}n\'{\i} matematiky 
a fysiky {\bf {48}}, 302 (1919); English translation by I. Gregora: 
Czechoslovak Jour. Phys. {\bf {44}}(2), 97 (1994).

\bibitem{AL} A. Lakhtakia, Czechoslovak Jour. Phys. {\bf {44}}(2), 89 (1994).

\bibitem{Ml1} M. A. MacLeod, J. Math. Phys. {\bf {36}}, 2951 (1995). 

\bibitem{Ml3} M. A. MacLeod, J. Math. Phys. {\bf {39}}, 1642 (1998).

\bibitem{KS1} K. Saygili, Int. J. Mod. Phys. A {\bf {23}}, 2015 (2008).

\bibitem{DJTS1} S. Deser, R. Jackiw, S. Templeton,
                Phys. Rev. Lett. {\bf {48}}, 975 (1982). 

\bibitem{DJTS2} S. Deser, R. Jackiw, S. Templeton, 
                Ann. Phys. {\bf{140}}, 372 (1982). 

\bibitem{DJTS3} J. F. Schonfeld, Nuc. Phys. B 
                {\bf {185}}, 157 (1981).  

\bibitem{ANS} A. N. Aliev, Y. Nutku, K. Saygili, Class. Quant. Grav.
              {\bf {17}}, 4111 (2000). 

\bibitem{KS2} K. Saygili, e-print arXiv: hep-th/0610307.

\bibitem{KS3} K. Saygili, Int. Jour. Mod. Phys. A {\bf {22}}, 2961 (2007).

\bibitem{BW} P. Baird, J. C. Wood, \textit{Harmonic Morphisms Between 
             Riemannian Manifolds} (Oxford University Press, New York, 2003).

\bibitem{PW1} R. Pantilie, J. C. Wood, Asian J. Math. 
              {\bf {6}}(2), 337 (2002).

\bibitem{P1} R. Pantilie, Comm. Anal. Geo. {\bf {10}}, 779 (2002).

\bibitem{SL} S. Lundquist, Arc. Fys. {\bf {2}}, 361 (1950).  

\bibitem{JR} J. Radon, Berichte \"{u}ber die Verhandlungen der 
             K\"{o}niglich-S\"{a}chsische Akademie der Wissenschaften  
             zu Leipzig, \\ 
             Mathematisch-Physische Klasse {\bf {69}}, 262 (1917),

\bibitem{John} F. John, Duke Math. Jour. {\bf {4}}, 300 (1938).

\bibitem{M1} H. E. Moses, Siam J. Appl. Math. {\bf {21}}, 114 (1971).

\bibitem{CDG} J. Cantarella, D. DeTurck, H. Gluck, J. Math. Phys.
              {\bf {42}}, 876 (2001). 

\bibitem{P} R. J. Parsley, ``The Biot-Savart operator and 
            electrodynamics on bounded subdomains of the 
            three-sphere,'' Ph.D. Thesis, \\ 
            University of Pennsylvania, 2004; 
http://users.wfu.edu/parslerj/research/dissertation.parsley.pdf.

\bibitem{GGG} I. M. Gelfand, S. G. Gindikin, M. I. Graev, \textit{Selected 
              Topics in Integral Geometry, Translations of Mathematical
              Monographs}, Vol. 220 (American Mathematical Society, 
              Providence, 2003).

\bibitem{M} H. Minkowski, Mathematics Sbornik {\bf {25}}, 505 (1904).

\bibitem{PF} P. G. Funk, Mathematische Annalen, Band {\bf {74}}, 278 (1913).

\bibitem{FBG} F. B. Gonzalez, \textit{John' s Equation and the Plane-to Line 
              Transform on $R^{3}$,} p. 1 in \textit{Harmonic Analysis and 
              Integral Geometry}, Research Notes in Mathematics, Vol. 422, 
              edited by M. Picardello (Chapman \& Hall/CRC, Florida, 2001).

\bibitem{Rubin3} B. Rubin, Jour. D' Analyse Math\'ematique
                 {\bf {77}}, 105 (1999). 

\bibitem{Rubin4} B. Rubin, Frac. Cal. App. Anal. {\bf {6}}(1), 25 (2003).

\bibitem{G} P. Grangeat, \textit{Mathematical framework of cone beam 3D 
            reconstruction via the first derivative of the Radon transform,} 
            p. 66 in \textit{Mathematical Methods in Tomography}, Lecture 
            Notes in Mathematics, Vol. 1497, edited by G. T. Herman, 
            A. K. Louis, F. Nattarer (Springer Verlag, Berlin, 1991).

\bibitem{S1} B. D. Smith, IEEE Transactions on Med. Im. 
             MI-{\bf {4}}(1), 14 (1985).

\bibitem{HKT} H. K. Tuy, SIAM J. Appl. Math. {\bf {43}}(3), 546 (1983).

\bibitem{GG} I. M. Gelfand, A. B. Goncharov, Dokl. {\bf {290}}(3), (1986); 
             English translation: Soviet Math. Dokl. 
             {\bf {34}}, 373 (1987). 

\bibitem{HSSW} C. Hamaker, K. T. Smith, D. C. Solmon, S. l. Wagner, 
               Rocky Mountain Jour. Math. {\bf {10}}(1), 253 
               (Winter 1980).

\bibitem{NW} F. Natterer, F. W\"ubbeling, \textit{Mathematical Methods 
             in Image Reconstruction} (SIAM, Philadelphia, 2001). 

\bibitem{RP4} R. Penrose, W. Rindler, \textit{Spinors and space-time}, 
              Vol. 1 (Cambridge University Press, Cambridge, 1993).  

\bibitem{RP3} R. Penrose, W. Rindler, \textit{Spinors and space-time}, 
              Vol. 2 (Cambridge University Press, Cambridge, 1993).  

\bibitem{W} E. T. Whittaker, Mathematische Annalen {\bf {57}}, 333 (1903),

\bibitem{Bateman} H. Bateman, Proc. London Math. Soc. {\bf {1}}, 451 (1904). 

\bibitem{BEGM} T. N. Bailey, M. G. Eastwood, R. Gover, L. J. Mason, 
               J. Korean Math. Soc. {\bf {40}}, 577 (2003). 

\bibitem{BEGM1a} T. N. Bailey, M. G. Eastwood, R. Gover, L. J. Mason, 
                 Math. Proc. Cambridge Phil. Soc. {\bf {125}}, 67 (1999).

\bibitem{DW} M. Dunajski, S. West, \textit{Anti-self-dual conformal structures 
             in neutral signature,} p. 113 in \textit{Recent Developments in 
             Pseudo-Riemannian Geometry}, edited by D. V. Alekseevsky, 
             H. Baum (European Mathematical Society, Zurich, 2008).

\bibitem{D1} M. Dunajski, J. Phys. A: Math. Theor. {\bf {42}}, 404004 (2009).

\bibitem{S} G. Sparling, Phil. Trans. R. Soc. London A 
            {\bf {356}}, 3041 (1998).

\bibitem{BE} T. N. Bailey, M. G. Eastwood, \textit{Twistor results for 
             integral transforms,} p. 77 in \textit{Radon transforms and 
             tomography}, Contemp. Math., Vol. 278, edited by E. T. Quinto, 
             L. Ehrenpreis, A. Faridani, F. Gonzalez, E. Grinberg
             (American Mathematical Society, Providence, 2001).

\bibitem{LJM3a} L. J. Mason, J. reine angew. Math. {\bf {597}}, 105 (2006).

\bibitem{H1} N. J. Hitchin, Commun. Math. Phys. {\bf {83}}, 579 (1982). 

\bibitem{WS} W. T. Shaw, \textit{Complex Analysis with Mathematica} 
             (Cambridge University Press, Cambridge, 2006)

\bibitem{CK} S. Chandrasekhar, P. C. Kendall, Astrophys. Jour. 
             {\bf {126}}, 457 (1957).

\bibitem{RK} A. G. Ramm, A. I. Katsevich, \textit{The Radon Transform 
             and Local Tomography} (CRC Press, Florida, 1996).

\bibitem{N} F. Natterer, \textit{The Mathematics of 
            Computerized Tomography} (SIAM, Philadelphia, 2001).

\bibitem{VP} V. Palamodov, \textit{Reconstructive Integral Geometry}
             (Birkh\"auser, Basel, 2004).

\bibitem{VP1} V. P. Palamadov, \textit{Inversion formulas for the 
              three-dimensional ray transform,} p. 53 in \textit{Mathematical 
              Methods in Tomography}, Lecture Notes in Mathematics, 
              Vol. 1497, edited by G. T. Herman, A. K. Louis, F. Nattarer
              (Springer Verlag, Berlin, 1991).

\bibitem{GS} I. M. Gelfand, G. E. Shilov, \textit{Generalized Functions, 
             Volume 1, Properties and Operations} 
             (Academic Press, London, 1964).

\bibitem{K} R. P. Kanwal, \textit{Generalized Functions, Theory and 
            Applications} (Birkhauser, Boston, 2004).
                                                                
\bibitem{VIS} V. I. Semyanistyi, Soviet Math. Dokl. {\bf {2}}, 59 (1961). 
                 
\bibitem{VIS1} V. I. Semyanistyi, Trudy Sem. Vektor. Tenzor. Anal. 
               {\bf {12}}, 397 (1963). 

\bibitem{T} P. Tooft, ``The Radon Transform, Theory and Implementation,'' 
            Ph.D. thesis, Technical University of Denmark, 1996. 

\bibitem{PM} P. Maass, Inv. Prob. {\bf {3}}, 729 (1987).

\bibitem{Markoe} A. Markoe, \textit{Analytic Tomography, 
                 Encyclopedia of Mathematics and Its Applications}, 
                 Vol. 106 (Cambridge University Press, Cambridge, 2006). 

\bibitem{CD} R. Clack, M. Defrise, J. Opt. Soc. Am. A
             {\bf {11}}(2), 580 (1994).

\bibitem{XY} X. Yang, ``Geometry of Cone-beam Reconstruction,'' 
            Ph.D. thesis, Massachusetts Institute of Technology, 2002.

\bibitem{TZG} K. Taguchi, G. L. Zeng, G. T. Gullberg, 
              Phys. Med. Bio. {\bf {46}}, N127 (2001).

\bibitem{YL} X.-H. Yan, R. M. Leahy, IEEE Trans. Med. Imag. 
             {\bf {10}}(3), 462 (1991).

\bibitem{GNW} G. N. Watson, \textit{A Treatise on the Theory 
              of Bessel Functions} 2nd ed. (Cambridge University 
              Press, Cambridge, 1944). 

\bibitem{MKM} M. K. Murray, \textit{Twistor Theory}, 
              http://www.maths.adelaide.edu.au/michael.murray/twistors.pdf.

\bibitem{WS1} W. T. Shaw, e-print Arxiv: 1005.4184v1 [physics.flu-dyn].

\bibitem{PB} P. Baird, \textit{An Introduction to Twistors}, 
http://www.math.jussieu.fr/$\sim$helein/encyclopaedia/baird-twistors.pdf.

\bibitem{PB1} P. Baird, Phys. Let. A {\bf {179}}, 279 (1993).

\bibitem{MKM1} M. K. Murray, e-print Arxiv: math-ph/0101035v1.

\bibitem{CGHKM} D. Chiou, O. J. Ganor, Y. P. Hong, B. S. Kim, I. Mitra, 
                Phys. Rev. D {\bf {71}}, 125016 (2005). 

\bibitem{Y} Z. Yoshida, J. Math. Phys. {\bf {33}}, 1252 (1992).

\bibitem{TC} G. F. Torres del Castillo, J. Math. Phys. 
             {\bf {35}}, 499 (1994).

\bibitem{RB1} M. N. Rosenbluth, M. N. Bussac, Nuc. Fus. 
              {\bf {19}}(4), 489 (1979).
                      
\bibitem{JAS} J. A. Stratton, \textit{Electromagnetic Theory}
              (John Wiley \& Sons, New Jersey, 2007).

\bibitem{NPFRCBC} A. A. Neves, L. A. Padilha, A. Fontes, E. Rodriguez, 
                  C. H. B. Cruz, L. C. Barbosa, C. L. Cesar, J. Phys. A: 
                  Math. Gen. {\bf {39}}, L293 (2006). 

\bibitem{VD} V. V. Dodonov, J. Phys. A: Math. Theor. {\bf {40}}, 14329 (2007). 

\bibitem{CS} P. J. Cregg, P. Svedlindh, J. Phys. A: Math. Theor. 
             {\bf {40}}, 14029 (2007).

\bibitem{SK} S. Koumandos, Int. J. Math. \& Math. Sci. 
            {\bf {2007}}, 73750 (2007). 

\bibitem{E} A. Erdelyi (ed.) \textit{Bateman Manuscript Project: 
            Higher Transcendental Functions}, Vol. II 
            (McGraw-Hill, New York, 1953).

\bibitem{GR} I. S. Gradshteyn, I. M. Ryzhik, \textit{Table of Integrals, 
             Series, and Products} 5th ed. (Academic Press, Boston, 1994).
  
\bibitem{F} G. B. Folland, \textit{Fourier Analysis and Its Applications}
            (Wadsworth \& Brooks/Cole Advanced Books \& Software, California, 
            1992).

\bibitem{L1} A. K. Louis, \textit{Developments of Algorithms in Computerized
            Tomography,} p. 25 in \textit{The Radon Transform, Inverse 
            Problems, and Tomography},  Proceedings of Symposia in Applied 
            Mathematics, Vol. 63, edited by G. Olafsson, E. T. Quinto 
            (American Mathematical Society, Providence, 2006).
                      
\bibitem{MH} J. E. Marsden, M. J. Hoffman, \textit{Basic Complex Analysis} 
             (W. H. Freeman and Company, New York, 1999).

\bibitem{AJ} A. Jeffrey, \textit{Applied Partial Differential Equations, 
             An Introduction} (Academic Press, Massachusettts, 2003).


\end{thebibliography}
\end{document}